\documentclass[
	bibliography=totoc, 
	listof=totoc,      
]{scrbook}              
 \pdfoutput=1

\usepackage[utf8]{inputenc}
\usepackage[ngerman,english]{babel}

\usepackage{tikz}
    \usetikzlibrary{external, arrows, positioning, trees, shapes.arrows} 
\usepackage{auxhook}

\usepackage[
        a4paper,
        top=35mm,
        bottom=40mm,
        inner=35mm,
        outer=40mm,
        bindingoffset=5mm,
        marginparsep=5mm,
        marginparwidth=40mm,
    ]{geometry}

\usepackage[
        headsepline,
        footsepline,
        draft=false     
    ]{scrlayer-scrpage}

\clearpairofpagestyles
\ohead{\headmark}
\automark[chapter]{chapter}
\automark*[section]{}
\ofoot*{\pagemark}
\usepackage{float}
\usepackage[
        font={normal,small,sl,color=black!80},
        labelfont={normal,bf}
    ]{caption}

\usepackage[
        backend=biber,
        style=apa,
        minnames=1,
        maxnames=2,
        urldate=long
    ]{biblatex}

\DeclareSourcemap{
	\maps{
		\map[overwrite]{
			\step[fieldsource=urldate,
			match=\regexp{([0-9]{2})\.([0-9]{2})\.([0-9]{4})},
			replace={$3-$2-$1},
			final]
		}
	}
}

\usepackage{blindtext} 
\usepackage{csquotes}
\usepackage{graphicx}
\usepackage{marvosym} 
\graphicspath{ {./images/} }
\usepackage{enumitem}
\usepackage{amsmath}
\usepackage{amstext}
\usepackage{subcaption}
\usepackage{MnSymbol}

\usepackage[noabbrev,capitalise]{cleveref}

\usepackage{multirow}
\usepackage{colortbl}

\usepackage[acronym]{glossaries}
\makeglossaries
\newacronym{API}{API}{An \textbf{A}pplication \textbf{P}rogramming \textbf{I}nterface provides a  protocol allowing standardized communication between different components of software systems (server and client or user and web service, e.g.)}
\newacronym{DTC}{DTC}{\textbf{D}irect \textbf{T}o \textbf{C}ustomer marketing describes the parctice of advertisingproducts of services directly to customers, in an 1-to-1 fashion. Thus, no broadcast media like banner ads, newspaper advertisements or TV commercials are needed as potential customers are directly addressed. Direct phone calls, mailing or targeted advertising are means to do so.}
\newacronym{EDD}{EDD}{The \textbf{E}uroStemCell \textbf{D}ata \textbf{D}onation is the project, this thesis was situated in. It was an effort by Anna Couturier and AALAB to crowdsource data collection of responses from ISEs with a browser plugin and analyze the results with respect to questionable advertising of stem cell treatments.}
\newacronym{ISE}{ISE}{In this thesis, the term \textbf{I}ntegrated \textbf{S}earch \textbf{E}ngine is used to refer to monolithic but multifaceted information systems that combine search utility with the capabilities of an advertising exchange. Some of these platforms (like e.g. Google) unify subsystems dedicated with web services like web search, ad networking, ad-exchange and ad-delivery as well as data exchange. Though they still offer the services separately, they enable customers and users to use the services with a single account and benefit from the horizontal integration.}
\newacronym{SEO}{SEO}{\textbf{S}earch \textbf{E}ngine \textbf{O}ptimization describes the activity of optimizing websites to receive high rankings in search engines and thus a prominent position in their listing~\autocite{Siepermann.2019b}. There are \textit{black hat} SEO tactics, that are disapproved by search providers and \textit{white hat} techniques that they encourage to ensure high quality ranking and relevance among the results~\autocite{Pasquale.2008}}
\newacronym{SCT}{SCT}{\textbf{S}tem \textbf{C}ell \textbf{T}reatment or Stem Cell Treatments (SCTs) are medical practices that use stem cells to cure particular diseases. As the domain is an evolving area of research, the are only few applications known to be effective. They include therapies of blood, immunesystem, skin and eye conditions.)}
\newacronym{SERP}{SERP}{The \textbf{S}earch \textbf{E}ngine \textbf{R}esult \textbf{P}age is delivered by a web search engine upon request. It displays the ranked search results and possibly includes advertisements.}
\newacronym{VPS}{VPS}{The \textbf{V}irtual \textbf{P}rivate \textbf{S}erver were introduced to the study to prvide control data from an unbiased and clean browser profile. They are servers running on a virtual machine, administered by remote access.}

    \title{Abusive Advertising: Scrutinizing socially relevant algorithms in a Black Box analysis to examine their impact on vulnerable patient groups in the health sector}

    \author{Martin Reber}

    \date{\today}

    \newcommand{\examinerA}{Prof. Dr. Katharina Zweig}
    \newcommand{\examinerB}{Tobias Krafft}

    \newcommand{\thesisType}{Master Thesis}

\makeatletter
\hypersetup{
    pdftitle = {\@title},
    pdfauthor= {\@author}
}
\makeatother

\begin{document}

\frontmatter


\newgeometry{margin=1in}
\begin{titlepage}
    \makeatletter
        \centering
        \includegraphics{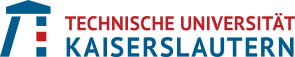} \par \vspace*{\fill}
        {\scshape\LARGE \@title} \par \vspace*{\fill}
        {\bfseries \thesisType} \par\vspace{1cm}
        {by} \par\vspace{1cm}
        {\Large\itshape \@author} \par\vspace{1cm}
        {March 2, 2020} \par\vspace*{\fill}
        {Technische Universität Kaiserslautern,\\
        Department of Computer Science,\\
        67653 Kaiserslautern,\\
        Germany} \par\vspace*{\fill}
        {
        \begin{tabular}{rl}
        Examiner: & \examinerA\\
        		  & \examinerB
        \end{tabular}
        }
    \makeatother
\end{titlepage}
\restoregeometry


\section*{Eigenständigkeitserklärung}

Hiermit versichere ich, dass ich die von mir vorgelegte Arbeit mit dem Thema \makeatletter\@title\makeatother selbstständig verfasst habe, dass ich die verwendeten Quellen und Hilfsmittel vollständig angegeben habe und dass ich die Stellen der Arbeit - einschließlich Tabellen und Abbildungen -, die anderen Werken oder dem Internet im Wortlaut oder dem Sinn nach entnommen sind unter Angabe der Quelle als Entlehnung kenntlich gemacht habe.

\vspace{0.5cm}

Kaiserslautern, den 2.3.2020

\vspace{2cm}
\begin{flushleft}
\includegraphics[width=0.2\linewidth]{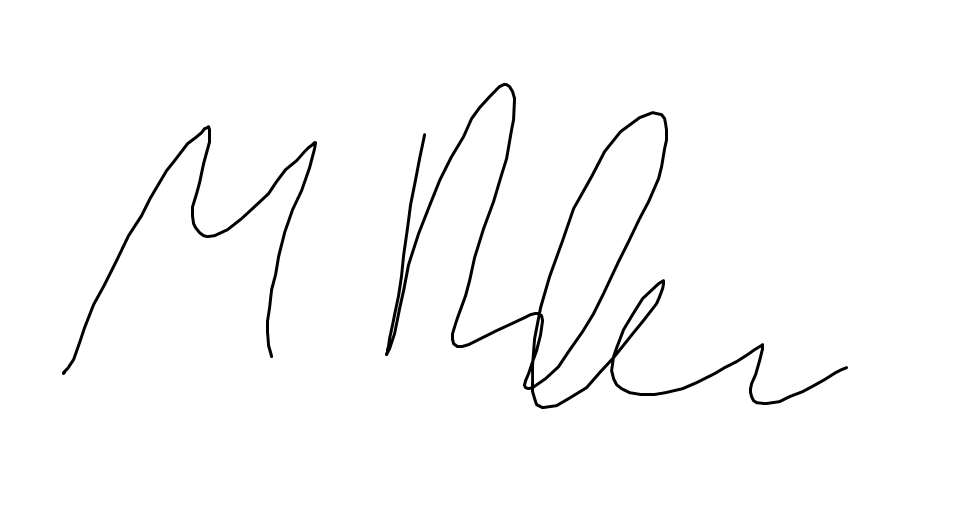}
\end{flushleft}
\begin{tabular}{@{}l@{}}\hline
\makeatletter\@author\makeatother
\end{tabular}


\clearpage
\pdfbookmark{Abstract}{Abstract}
\section*{Abstract} 
\begin{center}
\begin{minipage}[t]{0.7\textwidth}
The targeted direct-to-customer marketing of unapproved stem cell treatments by a questionable online industry is directed at vulnerable users who search the Internet in the hope of a cure. This behavior especially poses a threat to individuals who find themselves in hopeless and desperate phases in their lives. They might show low reluctance to try therapies that solely promise a cure but are not scientifically proven to do so. In the worst case, they suffer serious side-effects.

Therefore, this thesis examines the display of advertisements of unapproved stem cell treatments for Parkinson's Disease, Multiple Sclerosis, Diabetes on Google's results page. The company announced a policy change in September 2019 that was meant to prohibit and ban the practices in question. However, there was evidence that those ads were still being delivered.

A browser extension for Firefox and Chrome was developed and distributed to conduct a crowdsourced Black Box analysis. It was delivered to volunteers and virtual machines in Australia, Canada, the USA and the UK. Data on search results, advertisements and top stories was collected and analyzed. The results showed that there still is questionable advertising even though Google announced to purge it from its platform.
\end{minipage}
\end{center}

\vfill

\section*{Zusammenfassung}
\begin{center}
\begin{minipage}[t]{0.7\textwidth}
\begin{otherlanguage}{ngerman}
Die Direktvermarktung von nicht zugelassenen Stammzellbehandlungen von der fragwürdigen Online-Industrie dahinter zielt auf Patienten, die das Internet in der Hoffnung auf Heilung durchsuchen. Dieses Verhalten stellt eine besondere Gefahr für Menschen dar, die sich in verzweifelten Phasen in ihrem Leben befinden. Sie könnten wenig Zurückhaltung zeigen, die beworbenen Therapien auszuprobieren, die zwar eine Heilung versprechen, diese aber nicht durch anerkannte klinische Tests belegen können. Im schlimmsten Fall erwarten die Patienten schwerwiegende Nebenwirkungen.

Daher untersucht diese Theis die oben genannten Werbeanzeigen auf der Ergebnisseite der Online-Suchmaschine Google nach einer Änderung der Platformrichtlinien im September 2019. Besonders ging es dabei um Anzeigen bezüglich Behandlugen von Parkinson, Multipler Sklerose und Diabetes, die in den Verhaltensregeln ausdrücklich verboten wurden.

Browsererweiterungen für Firefox und Chrome wurden entwickelt und verteilt, um damit eine \textit{crowdsourced} Black Box Analyse durchzuführen. Freiwillige Teilnehmer und virtuelle Maschinen in Australien, Kanada, den USA und Großbritannien wurden rekrutiert. Es wurden Daten zu Suchergebnissen, Werbung und Schlagzeilen auf der Ergebnisseite von Google gesammelt. Die Analyse derer ergab, dass es trotz des expliziten Verbots dieser Praktiken noch immer fragwürdige Werbung gab.
\end{otherlanguage}
\end{minipage}
\end{center}

\vfill

\thispagestyle{empty}


\cleardoublepage                    
\pdfbookmark[0]{\contentsname}{toc} 
\tableofcontents                    


\listoffigures

\printglossary[type=\acronymtype]

\mainmatter
\chapter{Introduction} 
Digitalization has changed mankind in many ways. One technology that has grown to be indispensable is the search engine. They serve as an entry point to the WWW separating the websites most relevant to a user from noise. Most search engines provide their service to Internet users at no monetary cost. They finance their operation through advertisements (or short: ads) displayed along with search results on the search engine's result page (SERP). They call this sponsored or affiliated search. These integrated search engines (ISEs) combine search utility with the capabilities of an advertising exchange and thus connect advertisers, content providers and searchers.
However, most ISEs are privately operated Internet platforms. They rose to be powerful intermediaries that take the role of algorithmic gatekeepers. Not only do they control the flow of communication between users and content providers. On their platform, they also organize ad distribution and direct attention. Concurrently, they get to retain transaction data of all involved participants, e.g. user data, website content, ad efficacy, conversion costs of businesses.

In some domains, misconfiguration of algorithms has only minor consequences, like irrelevant search results or dysfunctional technical components. When it comes to medicine and health though, digitalization is probably going to have the \enquote{most immediate and profound personal and social consequences}~\parencite[p.368]{Petersen.2019}.
ISEs and their advertising partners combine various personal data to explore \enquote{the most intimate aspects of our selves}~\parencite[p.368]{Petersen.2019}. In the realm of health, they can have immediate effect on the well-being of citizens and their surroundings.

Although society is heavily affected by privately-operated Internet-based platforms it cannot assess the functionality and safety of those software systems. A Guardian journalist describes this situation as \enquote{operating on blind, ignorant, misplaced trust}~\parencite{Goldacre.2014} and adds that choices in algorithm design are generally being made without citizens noticing.
To counter-balance problematic and business-driven development of algorithms, the concept of \enquote{\textit{algorithmic accountability}} arose.
It describes the aspiration to scrutinize the mechanisms of opaque algorithms and understand how and why they produce a certain output. It also demands for institutions to be held responsible for the algorithms they produce~\autocite{USACM.2017}. This can be achieved with means like the Black Box analysis portrayed in this thesis. In this context, a \textit{Black Box} denotes an \enquote{opaque technical device of which only the inputs and outputs are known}~\parencite[83]{Bucher.2016}.

\section{Motivation}
\begin{figure}
	\centering
	\includegraphics[width=0.7\linewidth]{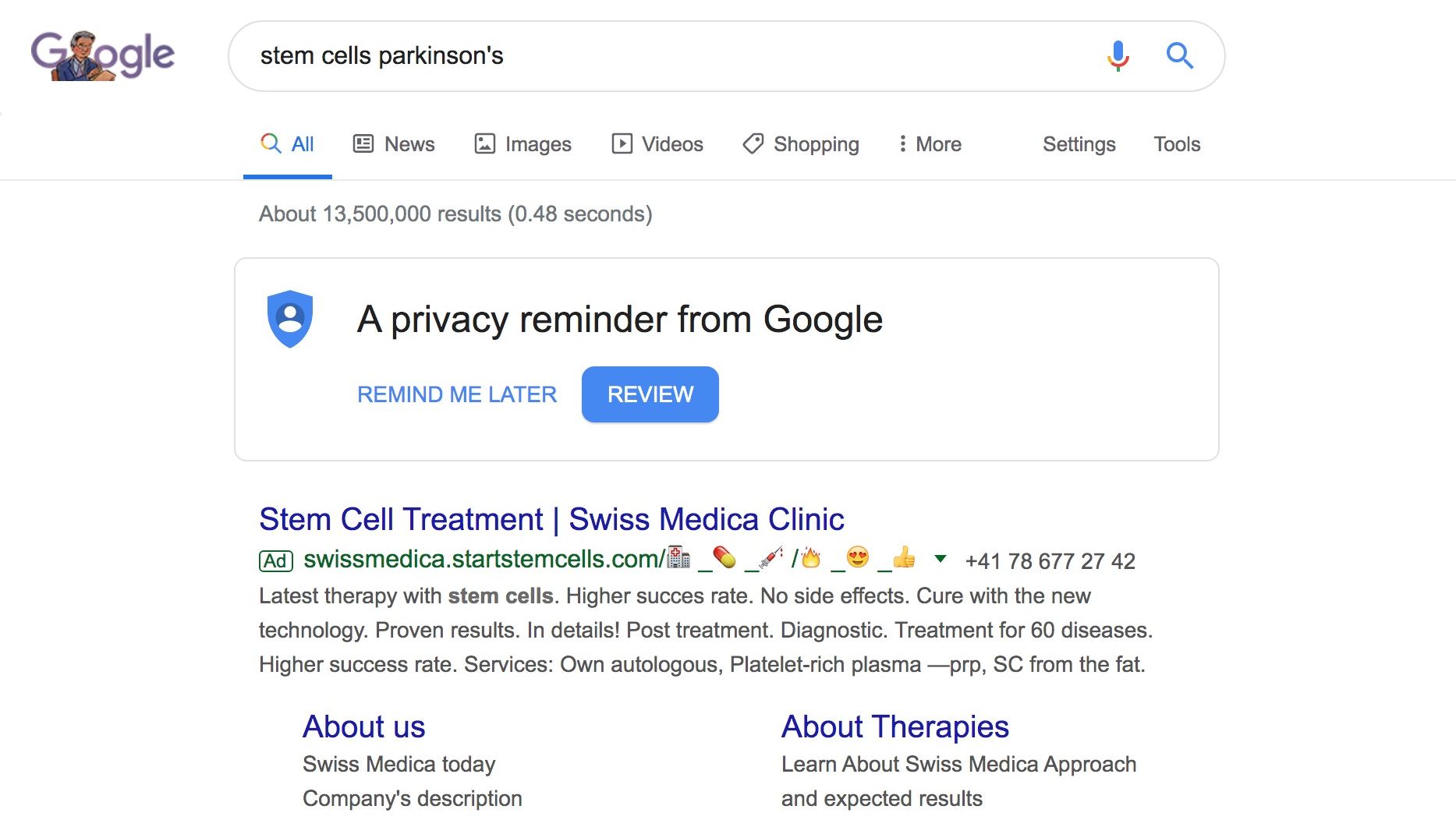}
	\caption[Sample SERP before policy change II]{Single ad of questionable stem cell treatment provider (30.9.2019) before the policy change, courtesy of Anna Couturier}
	\label{fig:screenshotresultswissmedica}
\end{figure}

The motivation of this project has come from the work of Anna Couturier who holds the dual role of PhD researcher in Science, Technology and Innovation Studies at the University of Edinburgh and Digital Manager at EuroStemCell\footnote{
	EuroStemCell is an organization dedicated to educate European citizens about stem cells (Website: \url{https://www.Eurostemcell.org/about-EuroStemCell})}.
In this secondary role, she has observed the impact of targeted advertisement and Google as an intermediary on inquiries made to the EuroStemCell project by patients and carers looking for information about stem cell treatments and serious conditions and diseases online. This master's thesis contributes to a deeper analysis of stem cell treatments and digital health information as part of a collaboration between the University of Edinburgh\footnote{Website: \url{https://www.ed.ac.uk/}} and the Algorithm Accountability Lab(AALAB)\footnote{
	The AALAB strives to establish ethics in programming, especially in socially sensitive applications, automated decision making systems (ADMs) and artificial intelligence (AI)}.
At the project's completion, the findings will be handed over to a number of patient organizations including the Anne Rowling Clinic\footnote{Website: \url{https://www.annerowlingclinic.org/}}, Parkinson's UK\footnote{\url{https://www.parkinsons.org.uk/}}, the Centre for Regenerative Medicine\footnote{\url{www.crm.ed.ac.uk}} and the Australian Stem Cell Network\footnote{\url{http://www.stemcellsaustralia.edu.au/}}. 
EuroStemCell fosters an interdisciplinary network of scientists and patient groups to research and communicate the subjects surrounding stem cells. They fill the role as a professional medical organization to \enquote{counteract the un-controlled and premature commercialization of stem cell interventions.}~\parencite{Weiss.2018}.
From these tight partnerships (and academic literature alike), evidence arose that patients diagnosed with Parkinson's Disease or Multiple Sclerosis were exposed to questionable advertisement when searching the web on Google. They were questionable and problematic in a sense that they advertised scientifically unproven stem cell treatments (SCT)\footnote{
	To my knowledge, stem cell-related treatments are only clinically tested and empirically proved to be helpful for diseases concerning the blood and immune system with advances in the area of skin and cornea (eye)~\autocite{FDA.2019,Eurostemcell.2020}.
	Thus, in this thesis, by \enquote{questionable SCT} I denote those treatments that use stem cell-related practices that are \emph{NOT} yet approved by medical authorities. These questionable procedures are not yet approved and possibly dangerous.}
~\autocite{ISSCR.2019,Enserink.2006} to affected Internet users that might be looking for a cure to their disease. \cref{fig:screenshotresultswissmedica} shows examples of problematic ads.
\begin{figure}
	\centering
	\includegraphics[width=0.7\linewidth]{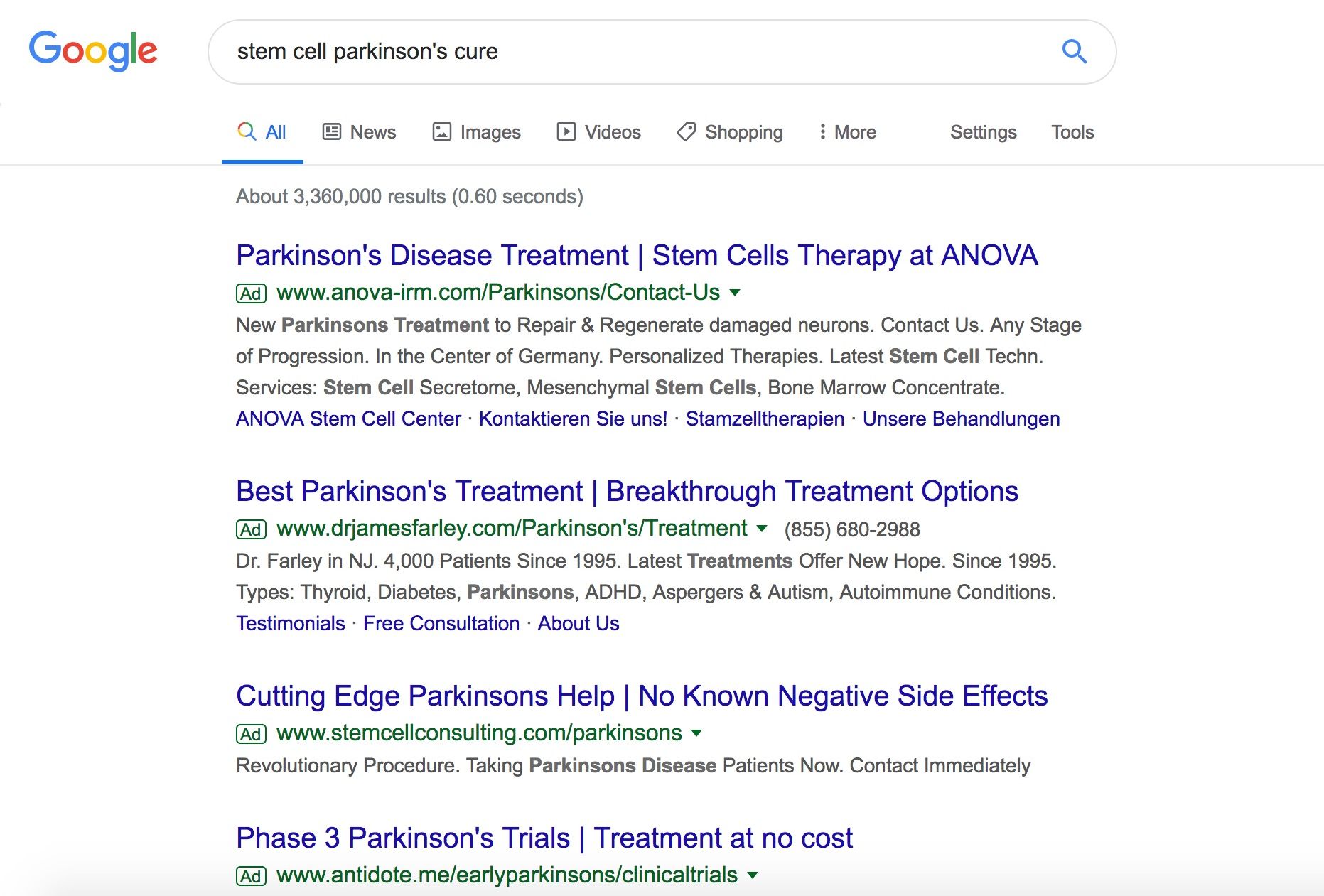}
	\caption[Sample SERP before policy change I]{Sample screenshots of advertisements presented at a typical Google search result page (SERP) (30.9.2019) before the policy change, courtesy of Anna Couturier}
	\label{fig:screenshotsearchads}
\end{figure}
The motivation of this thesis was to examine whether vulnerable user groups (patients of severe diseases) were specifically targeted by advertisement on the Google search engine result page (SERP). This research would have been especially concerned with the promotion of unproven stem cell treatments for Parkinson's disease, Multiple Sclerosis and Diabetes (Type I and II) on Google's web search platform. This is important as the presumably targeted users represent a vulnerable group whose exploitation can have severe consequences.
We picked Google because it is a popular integrated search engine (ISE) with large market share~\autocite{Ratcliff.2019}. Most importantly, anecdotal evidence suggested that the problematic phenomena appeared on Google, which supposedly handles promotion of unproven treatments very strictly according to their advertising policy. They claim to ban all ads concerned with speculative and experimental medical treatments, specifically including stem cell therapy~\autocite{Google.2019j,GoogleAdsHelp.2019}.

This thesis presents a browser plugin associated with a client-server software system to crawl the Google SERP and store results, ads and top stories in a database for further analysis. The goal was to find out whether the problematic advertisements were still being delivered over Google's ISE.
Google's announcement to implement adaptive measures from the beginning of October 2019 pressured this project to engineer a lightweight, flexible and practical solution on the fast track. Thus, many potentials for improvement could not be considered (see \cref{sec:lessons}).

A qualitative analysis of advertisements on Google's SERP is presented and a general assessment of the Black Box approach is conducted.
The analyses showed that there still is questionable advertising of unapproved SCT practices. Furthermore, they showed that a variety of actors compete for attention in the advertising ecosystem surrounding SCT.

The assessment of our Black Box approach produced interesting insights for future work concerning methodology and requirements of such analyses.

\section{Outline}
\cref{ch:fund} introduces fundamentals that are required to follow the reasoning in this work. In \cref{sec:info} and \ref{sec:models} the notions of information and modelling are explored. \cref{ssec:algo} discusses algorithms in general to deduce how ISEs operate and whether a programming artifact can be responsible for its outcomes. Systems theory will be explored to draw models of communication and the socio-technical system of web-search in \cref{sec:com} and \cref{sec:sts}, respectively. Because this thesis elaborates on the workings of Integrated Search Engines and web-advertisement, \cref{sec:dataeco} and \ref{sec:se} elaborate on the respective topics. Then the socio-technical system will be applied to web search to explain the interactions of ISEs and their users in \cref{sec:appl_sts}.

In the second part, \cref{ch:relwork} describes the context in which this work was embedded and examines the notion of digitalized health in \cref{sec:digihealth}. \cref{sec:algacc} further discusses how algorithms can be assessed regarding their accountability. The chapter closes with a closer look at Black Box analysis as means to analyze undisclosed algorithms in information systems (\cref{sec:blackbox}) and approaches to their governance (\cref{sec:gov}).
Finally, \cref{ch:datadonation} describes the \textit{EuroStemCell Data Donation} that took place in 2019. Its goal was to verify the impact of policy changes Google initiated after anecdotal evidence about ads on unproved therapies arose\footnote{see \autocite{GoogleAdsHelp.2019} for Google's announcement.}.

\chapter{Fundamentals}\label{ch:fund}
This chapter describes the required key concepts on which the following parts of this thesis are based on. Because the conceptions of these terms differ greatly depending on the domain, there is a need to define and contrast some of them.
First, this thesis describes multidimensionality of information in \cref{sec:info}. Then, it will be situated in the communication process in \cref{sec:com}. This lays foundations for the introduction of socially relevant algorithms in \cref{ssec:algo}. It makes use of several examples to show the considerable impact of these artifacts of information technology on our society. Based on this, technical and social systems and ultimately, the socio-technical system are referenced in \cref{sec:sts}. The derivation of the concept of socio-technical system is based on \autocite{Kienle.2014}.
To illustrate how the theoretical model of socio-technical systems finds its real-world application in \cref{sec:appl_sts}, the advertising ecosystem as well as integrated search engines are reviewed in \cref{sec:dataeco} and \cref{sec:se}, respectively.

\section{Information}\label{sec:info}
Information is defined as meaningful part of a message or a set of symbols\footnote{
		From German:\enquote{Derjenige Anteil einer Nachricht, der für den Empfänger einen Wert besitzt}~\autocite{Siepermann.2019}}~\autocite{Meadow.1997}.
This distinguishes it from \textit{data}, which has \enquote{little or no meaning to a recipient}~\autocite[p.701]{Meadow.1997} and underlines how the notion of information is strongly dependent on the recipient's context. In \autocite{Meadow.1997}, Meadow and Yuan claim that there cannot be information overload through too much data. They argue that data is only considered informative if it was received and comprehended. They also require it to ultimately change the \textit{knowledge state} \footnote{\enquote{Knowledge is the accumulation and integration of information received and processed by a recipient}~\autocite[701]{Meadow.1997}} of the recipient.

This reflects a search engine's capability to crawl the web (composed of data) and extract only those webpages that it deems worthy to present to a user (information). The users on the other hand perceive the results as potential information\footnote{
	Meadow offers three interpretations of data, one of which explains it as \enquote{set of symbols in which the individual symbols have potential for meaning but may not be meaningful to a given recipient}~\autocite[704]{Meadow.1997}.}
and subjectively judge which data gets their attention.
Meadow assumes recipients determine relevance regarding understandability, redundancy and alignment with subjective beliefs~\autocite{Meadow.1997}. Madden lists geographical, cultural and social as well as educational and professional (area of interest and level of experience) factors\autocite{Madden.2000} that contribute to perceived relevance.

However, this thesis follows Belkin's argumentation in \autocite{Belkin.1978} concerning the usefulness of a definition of information. He argues that by dropping the compulsion to define, one is enabled to choose a useful interpretation that caters ones needs. Consequently, he suggests accepting diverse concepts as a way of looking at a phenomenon rather than squeezing all applications into one definition. 
Hence, the following paragraph depicts how scholars summarize the conceptualizations of information. According to McCreadie and Rice \autocite{McCreadi.1999}, information can be:
\begin{itemize}
	\item \textbf{Representation of knowledge:} Information stored on a medium (e.g. a website or a database),
	\item \textbf{Data in an environment:} signals obtained from the environment, including unintentional communication,
	\item \textbf{Resource or commodity:} \enquote{A message, a commodity, something that can be produced, purchased, replicated, distributed, sold, traded, manipulated, passed along, controlled}~\autocite[p.47]{McCreadi.1999},
	\item \textbf{Part of process of communication:} Assumes meaning originates from people, not from words, hence context plays an important role.
\end{itemize}

The concepts above help to describe the manifold manifestations of transmitted information in the communication processes of the socio-technical system. This thesis is mainly concerned with information as part of a process of communication\footnote{
	However, all other forms are present as well. The creative of ads constitutes a representation of knowledge, be it legitimate or questionable. Data in the environment of the human-computer interaction are constantly being extracted through tracking and exploited by analysis services. Usage data and user profiles are regularly traded as a commodity at low per-unit prices(see ~\cref{sec:dataeco})}.
The suggestive nature of SCT-related advertising only unfolds in the context of patients or carers desperately searching for support. The informational character of the promotional message arises from the subjective relevance for users affected by a medical condition. The following section deals with the specifics of communication and \cref{sec:appl_sts} relates the methodology of information concepts to the web search and advertising context.

\section{Models}\label{sec:models}

In the following, Weisberg's elaborations on modeling in \autocite{Weisberg.2013} are described. This is required to understand the premises on which the following models are constructed on. He distinguishes physical, mathematical and computational models. They are \enquote{potential representations of a target system}~\autocite[171]{Weisberg.2013} that differ in their representational capability. Each model consists of a structure and its construal  (interpretation). The latter defines the assignments of real entities to structural elements and the intended scope. The scope limits the model's expressiveness to some specific aspects of a phenomenon. Finally, the fidelity criteria describe the standards by which a model's  representational qualities can be judged.
However, this work only presents \textit{descriptions} of models\footnote{
	Even though I present the \enquote{models} of communication and socio-technical systems I am aware that they are rather models' descriptions than a model themselves. However, I use the term \enquote{model} to refer to their respective descriptions for the sake of readability.}.
They are distinct from the models themselves and from the target system. The target system is constructed by the modeler through abstraction of a real-world phenomenon. This abstraction intentionally reduces complexity while preserving similarity with respect to a certain subject of interest. It does so by reducing it to the most relevant aspects. 

Due to vagueness or ignorance, these descriptions may specify more than one distinct model or a family of models~\autocite[172]{Weisberg.2013}. This is important to acknowledge, because the web search and advertising ecosystem is a highly complex and opaque agglomeration of a multitude of actors. Thus, a modeler must find a balance between simplification and explanatory power of a model. 
\section{Socially Relevant Algorithms}

This thesis highlights the importance to scrutinize \textit{Socially Relevant Algorithms} (SRAs) like the ones deployed in web search and advertising systems. It is required to understand the basic categories of algorithms and how they can express bias.

\subsection{Algorithms}\label{ssec:algo}

Integrated search engines like Google's platform are operated by algorithms.
By \textit{integrated search engine} we denote an information system that combines search engine and ad exchange\footnote{For a detailed description, see \cref{ssec:webads} for web advertisement and \cref{app:functsearch} for search engines, respectively.)}
An \textit{information system} consists of humans and machines that create information and who are interrelated through communication processes which generally describes a computer-assisted system designed for a special purpose~\autocite{Gabriel.2016}. Zwass for example describes it as \enquote{an integrated set of components for collecting, storing, and processing data and for providing information, knowledge, and digital products}~\parencite{Zwass.2016}.

An algorithm is a finite set of rules that yield a sequence of well-defined instructions which need to be followed to solve a class of problems or produce a distinct outcome from an input in finite time\autocite{Introna.2016,Knuth.1968}\footnote{
	Knuth also lists \textit{effectiveness} as an equally important future. However, he deems an algorithm effective when it can be fulfilled by a human using pen and paper only. A definition that does not realistically hold with today's advanced algorithms. Another notable fact is that Knuth means \textit{reasonable duration}, when he speaks of finite time. Acceptable running times for algorithms naturally are a moving target due to technological advancement.}.
It consists of a logic component that describes the domain-specific problem and data structures and a control component dedicated to the problem-solving strategy~\autocite{Kowalski.1979}. This allows to separate the efficiency-centered control from the functional logic. The latter is solely concerned with functional aspects. For example, to ask the right question, modeling a suitable representation, use appropriate data and find an adequate solution. This work addresses the logic components of ISEs and examines it all along the development axis. This is where companies and developers make conscious decisions about how an algorithms is designed and how outcomes are computed.

Algorithms can be arbitrarily complicated. They range from simple algebraic calculations via computational heuristics to applications of artificial intelligence (AI)\footnote{
	The interest of AI is in \enquote{the synthesis and analysis of computational agents that act intelligently}~\parencite{Poole.2010}. Some scholars prefer the term \textit{computational intelligence} to emphasize that the agency is based on computation\autocite{Poole.1998}.}.
Trivial algorithms dedicated to simple algebraic calculations, sorting or other unsophisticated operations are not deemed socially relevant. Only if their outcomes have repercussions on individual humans or society as a whole, their actions must be evaluated from a societal perspective.
Admittedly, this is a fuzzy distinction as socially relevant algorithms can be composed of other trivial algorithms. Additionally, the above description strongly depends on the deployment context.
Nevertheless, several algorithm classes are at risk to discriminate those affected (even unintentionally through their respective choice of criteria, training data, semantics, and interpretation~\autocite{Diakopoulos.2014}). Discrimination can occur through an advertiser’s malicious intent, the targeting process or the targeted audience (the eventual outcome)~\autocite{Speicher.2018}. This can make users subject to bias, manipulation, constrained freedom, surveillance, discrimination, commercial or political influence, or loss of sovereignty~\autocite{Gillespie.2014,Saurwein.2017}.
They are distinguished by~\autocite{WorldWideWebFoundation.2017} according to the way they process information. Below, the categories are listed along with the respective pitfalls.
\begin{description}
	\item[Prioritization] Rank or score entities based on certain characteristics. The choice of these characteristics and the underlying values and norms have immediate impact on the order of results which could falsify the original intention.
	\item[Classification] Categorize an entity and assign it to a group due to its features. A faulty classifier might wrongfully label an entity with severe consequences.
	\item[Association] Establish relationships between entities. They are deduced semantically, through similarity or connotation, thus not necessarily reasonable or real.
	\item[Filtering] Exercising choice about what to consider relevant, possibly without revealing the criteria this decision is based on and applying possibly biased filters.
\end{description}

According to~\autocite{Poole.2010}, algorithms are agents because they \emph{act} in an environment. Going by Max Weber's definition, to act means internal or external \enquote{doing} that is premised on subjective purpose or deliberate intention~\autocite{ArbeitsgruppeSoziologie.1978}\footnote{
	Herein, the terms agent and actor are used interchangeably to describe subjects or entities that act.}.

Algorithms that power Internet-based platforms like Google's platform fulfill Poole's requirements to be \textit{intelligent agents}~\autocite{Poole.2010} which are derived from Turing's approach to intelligence in~\autocite{Turing.2009}.
His notion explains intelligence by behavior. Following Skinner, \textit{behavior} is any externally observable action (or \textit{doing}) by an organism if it happens with reference to its environment~\autocite{Skinner.1938}\footnote{
	Unfortunately, this does not allow to observe and scrutinize technical components by their behavior as they are not alive in the original sense. Nonetheless, a technical system can be seen as an (non-biological) organism that is comprised of different organs (its components) and pursues a certain goal.}.
Baum notes how behavior is generally aimed at a goal and a result of deliberate choice of  actions that considers future consequences~\autocite{Baum.2013}. Because actions are intended behavior~\autocite{Kienle.2014}, inanimate entities like algorithms are capable to behave within the boundaries of their defined actions.
On top of that, technical components and the repercussions of their actions affect both social and technical entities in the socio technical system. Thus, they have a strong relational aspect when they facilitate communication processes~(see \cref{sec:sts}). Thus, they can be seen as agents of collective agency~(in \cref{ssec:algo}).

This explains why Google's platform qualifies as intelligent agent by fulfilling Poole's requirements~\autocite{Poole.2010}. It emphasizes the capability of their algorithms to act appropriately with respect to circumstance and goal. Furthermore, the intelligent algorithmic actors is flexible pertaining to resources (computational space and time) and learning experiences.

The paragraphs above explained how ISE's algorithms construct an information system that is designed for a certain purpose and to interact with humans through distinct technical components or computational artifacts, namely algorithms. 
Introna claims that \enquote{[a]lgorithmic action has become a significant form of action (actor) in contemporary society}~\parencite[37]{Introna.2016}.

In \cref{sec:algacc} algorithms and particularly those that act intelligently are described as agents that are able to perform self-sufficiently in their environment, Nonetheless, they cannot be perceived as self-sufficient moral agents of their doing.
They can be judged by the decisions that were made along the chain of instructions and the output they generate which constitutes their actions.

\subsection{Social Relevance}\label{ssec:sra}

Socially Relevant Algorithms constitute the technical components in \textit{socio-technical systems} (STS, see \cref{sec:sts}). They have a significant impact on a social system and can mostly be found in human-computer interaction, for example, when a human searches the World Wide Web (WWW) using a search engine and is computationally targeted with advertising. Here, human and computer engage in mutual communication.
The idea to evaluate algorithms as part of a greater system is a perspective that expands the boundaries of computer science beyond the realms of bare construction of computers and algorithms design. 
It addresses accountability and  responsibility concerning the development, implementation and use of algorithms that play a significant role in socio-technical systems. It addresses long-term effects and emergent behavior as well as a wider scope of stakeholders.
Sometimes, the outcomes of these algorithms are accompanied by discrimination, induce manipulation or express other unwanted side-effects. Basically, all classes of algorithms as denoted in \cref{ssec:algo} can suffer from biases.
Below, SRAs are listed that showed significant impact on either individuals or society, some problematic or at least questionable others merely thought-provoking\footnote{
	This list intends to give a rough overview of SRAs to demonstrate their widespread application in all sorts of domains of everyday live. The compilation includes scientific as well as journalistic sources and is in no means exhaustive.}.
\begin{itemize}
	\item	Scoring credit risk~\autocite{Citron.2014}, recidivism~\autocite{Larson.2016} and social behavior~\autocite{Kuhnreich.2017,Stanley.2015}	
	\item	Nation-wide face recognition~\autocite{Chen.2017} and predictive policing~\autocite{Peteranderl.2017}
	\item 	Fake News~\autocite{Albright.2017} and emotional manipulation in social networks~\autocite{Kramer.2014}\footnote{This study was widely criticized by popular media \parencite{Chambers.2014,Grohol.2018} and academia alike \parencite{Shaw.2016,Jouhki.2016} for including uninformed participants on a large scale and basing marginal findings on antique research methods.}
	\item	Racial ad delivery~\autocite{Angwin.2016,Sweeney.2013} and sexist recruitment~\autocite{Reuters.2018}
	\item 	Home automation~\autocite{Peterson.2020} and automotive software~\autocite{Gelles.2015,Koscher.2010}
	\item	Dubious autoplay feeds~\autocite{MaxFisher.2019,MaxFisher.2019b,Maheshwari.2017}
	\item 	Art performances~\autocite{Weckert.2020}\footnote{Weckert created a virtual traffic jam on \textit{Google Maps} by pulling  a wagon full of cell phones down an empty road.}
	
\end{itemize}
Another critical application domain is the search engine. Search engines act as the entry portal to the WWW, creating comprehensiveness in humongous mass of websites out there and satisfy users' information need. Users confidently trust a search engine to answer their query and rank results by true relevance \autocite{Pan.2007}. They allow an algorithm to deem some information more worthy than other. Thus, researchers claim that search engines have the power to
shape public opinion~\autocite{Zittrain.2014},
disseminate conspiracy theories~\autocite{Ballatore.2015},
redeﬁne history~\autocite{Grimmelmann.2008},
perpetuate negative stereotypes~\autocite{Baker.2013,Kay.2015},
manipulate individual users~\autocite{Epstein.2015,Epstein.2013}\footnote{Although \autocite{Epstein.2013,Epstein.2015} are widely referenced, the studies are equally harsh criticized due to their miscalculations, exaggeration and sensational claims, for example in \autocite{AlgorithmWatch.2017}.}
and discriminate based on race~\autocite{Sweeney.2013,Angwin.2016}
and gender~\autocite{Kay.2015,Otterbacher.2017,AdamGale.2015}.

An attempt to describe these effects is made by Gillespie in \autocite{Gillespie.2014}. He distinguishes six dimensions of algorithmic impact on society. \textit{Patterns of inclusion}, the \textit{evaluation of relevance} and the \textit{promise of algorithmic objectivity} all relate to the functionality of search engines as an unbiased information provider that delivers relevant answers from an objective selection of knowledge to users. The \textit{cycles of anticipation}, \textit{entanglement with practice} and \textit{production of calculated publics} describes how algorithms analyze and target users and how these inferences and the users' respective expectations rebound to society. Further down, we will see how the practices of integrated search engines like Google are subject to all of them.

These alarming consequences are not necessarily intended by their developers but usually emerge as unwanted side effects, unexpectedly and through interaction with society.
Algorithms may indirectly disadvantage users in ways that are not necessarily illegal or intended by their developers. Once they exercise socially problematic behavior, they should be scrutinized by the public~\autocite{Sandvig.2014}.

This thesis defines Socially Relevant Algorithms as algorithms that have an immediate effect on a social system through their close coupling with social processes (communication). They are part of a socio-technical system\footnote{See \cref{sec:sts}}, where they constitute the technical part.

\section{Communication}\label{sec:com}
To examine the characteristics of interaction between humans and computers, this chapter contemplates different models of communication. 

In the course of years, several models have gained popularity. This chapter discusses three popular models of interaction \footnote{
	Watzlawick et al. define \textit{interaction} as mutual exchange of messages between two or more persons~\autocite{Watzlawick.2007}.}
to describe the interactions in the process of communication between a human and a technical agent.
First, both extremes of the human-computer spectrum will be explored.
Shannon's model of tech-focused communication in the context of electrical communication engineering is to be contrasted with Watzlawick's approach of human psychology. Lastly, a context-conscious model by Kienle will be evaluated.
The comparison should illustrate why the subject of human-computer interaction present in web search and advertising requires a specific approach to communication. In order to fully explain the nature of the web search and advertising ecosystem, any arbitrary model might be insufficient. Consequently, an appropriate model must be able to reflect the system's properties and means of interaction. The following critique is mainly based on \autocite{Kienle.2014} with specific examples by the author of this thesis to illustrate the inapplicability or fitness of the respective model's characteristics to web search. 

\subsection{Shannon's technical Model of Communication}

\begin{figure}[htbp]
	\includegraphics[width=0.9\textwidth]{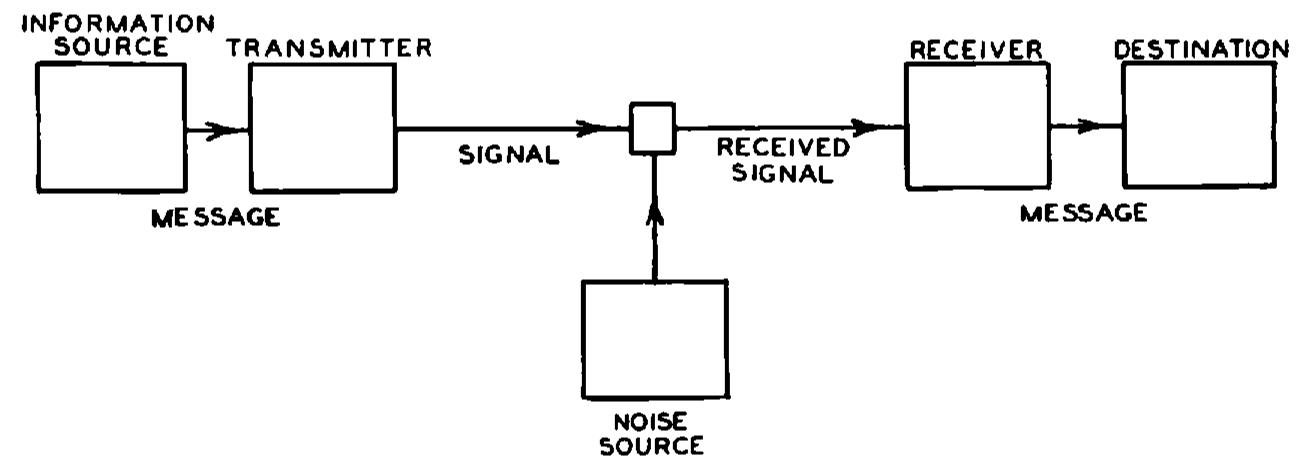}
	\caption[Shannon's Communication Model]{Schematic diagram of a general communication system, illustration from~\autocite[381]{Shannon.1948}}
	\label{fig:shannon}
\end{figure}

\cref{fig:shannon} shows a communication model dedicated to describing the exchange of information between two partners\footnote{
	Herein, the \textit{partners} denote agents involved in mutual communication.}
via telegraph or any wired connection. A source submits a message to a transmitter that encodes the message and sends a signal on a channel. During transmission, it may be affected by a noise source. The possibly corrupted message is then received and decoded by a receiver, which typically applies the inverse function of that done by the transmitter. After reconstructing the original message, it is delivered to its destination.\autocite{Shannon.1948}

This model falls short of many aspects that are essential for human-computer interaction in web search. \autocite{Kienle.2014} enumerates the following shortcomings that are then adapted to the web search context.
First, Shannon reduces the content of the message to its syntax only. The value of received information only depends on not-yet transmitted signals. The more of a message has been received, the lower is the informational value of residual signals. This is a wrong assumption in the context of web search. Even though searchers may have reviewed numerous results, the single most relevant result or advertisement that they eventually accept has higher informational value than the preceding signals.
Shannon further assumes that all messages are equivalently important for all destinations. This falls short of describing a web search scenario, where searchers have a unique background or context and expect a custom answer for a specific question.
Users only consider the subjective value of advertisements and search results.
Nonetheless, in this thesis the notion of sender, message and receiver is retained to describe the agent who initiates the communication, the transferred information and the addressed person.

\subsection{Watzlawick's psychological Communication Model}
Other models emphasize inter-human communication and add an empathic aspect. Watzlawick et al. present their psychological approach in \autocite{Watzlawick.2007}. His model is concerned with two human communication partners that are situated in vicinity to each other (possibly in one room). He interprets the entirety of behavior as means to transmit a multipartite message. The model is strictly restricted to observable actions and its trajectory depends on the subjective interpretation of the course of actions.

Watzlawick et al. formulated five axioms based on their experience as therapists in \autocite[53-70]{Watzlawick.2007}. Below, they are enumerated and subject to discussion with respect to their applicability to the human-computer interaction of web search.
\begin{description}
	\item[1. Axiom] Non-communication is impossible.
	\item[2. Axiom] All communication includes a content and a relational aspect such that the latter determines the former, which forms a meta-communication.
	\item[3. Axiom] The nature of a relationship is determined by the succession of communication perceived by the parties or their \textit{interpunction} thereof
	\item[4. Axiom] Human communication utilizes analogue and digital modes.
	\item[5. Axiom] Course of inter-human communication is either symmetrical or complementary.\footnote{
		Translated from German from \autocite[53-70]{Watzlawick.2007}}
\end{description}

For the \textbf{first axiom} to hold, Watzlawick presumes the analogous human communicators to be in one room. Obviously, this cannot be guaranteed with Internet-based service. Furthermore, remote communication offers many ways to not communicate, most of which pertain to not initiating a communication process online (not sending a message, not clicking a button). Through the technical communication channel (the Internet), the intention of non-communication remains shrouded and cannot be evaluated unlike other than with a passive agent in human-to-human interaction.

As to the \textbf{second axiom}, Determining a relationship between conversational partners over Internet-based services is difficult. Internet intermediaries such as platforms, search engines and ad exchanges complicate finding the true source of information. Imbalance of power over the communication channel (which is dictated by the platform) and a disparate state of knowledge about the respective partner usually leave users in the dark about the workings of the communication and the intentions of their counterpart.

With respect to static websites like most search engine result pages, the \textbf{third axiom} cannot be applied, too. Usually Internet-based services respond to user queries, the interchange accurately logged in files of the web server. There is no ambiguity to the course of communication. Users often perceive the Internet-based service's response to a user query, as a direct answer. However, the user query alone is not the only input to a search engine for example. Its algorithm organizes a plethora of information about the user and leverages background information in a way that the users can never be sure, when their communication with the platform provider actually started. As most users are unaware of the unobtrusive and constant tracking, testing and adapting of web-services, they are also ignorant about the entirety of exchanges in a communication

The depiction of analogue modes of communication as non-verbal can be sustained as claimed in the \textbf{fourth axiom}. However, gestures and facial expressions do no (yet) play a role in web search. Instead, background information in the form of data about a user and knowledge about context influence the communication.
Kienle and Kunau note that this can explain reduced communicative capability of interacting with and via technical systems\autocite[59]{Kienle.2014}.
Watzlawick argues that especially in human-computer interaction, it is important to provide \textit{meta-information} along with a message so the communication partners can negotiate their relationship and the interpretation of the message\autocite[55]{Watzlawick.2007}.

Interestingly, the \textbf{last axiom} allows a two tiered interpretation. At first, the role of interrogator (users) and respondent (search engine) are very clear and fulfill the requirements of complementary communication with mutual reinforcements of this distinct relationship. Nonetheless, one could see the search engine providers' learning strategies as a symmetrical approach to the question-answer-dialog. By learning more  about the users, their intentions and the context in which a query is formulated, there is a notion of reciprocal learning, though on different levels and orchestrated with a distinct intention. The users' learning is objective-oriented with respect to their information need, the focus of the search engine's learning however is subject-based and on the users and its own means to serve them.
In conclusion, even though this model is well suited to illustrate direct human communication, yet again it cannot be applied to human-computer communication without flaws.

\subsection{Context-oriented Communication Model}
The context-oriented communication model by Kienle \parencite[22-27]{Kienle.2003} depicts communication differently. It is no longer an unidirectional automatic process pushing a message from a sender to a receiver. Now, all involved parties are responsible for a common understanding~\autocite{Clark.1991}. Kienle adds that, the involved parties mutually refer to or react to each other's messages~\autocite[17]{Kienle.2003}. It is based on the notion of \textit{social action} by Luhmann. He describes it as action whose intention includes the supposed or expected attitudes of other people who are involved in the communication~\autocite[129]{ArbeitsgruppeSoziologie.1978}.
Kienle calls these assumptions \textit{context} and assigns to it the part of an environment that affects individuals' actions during interaction and facilitates mutual understanding~\autocite[22]{Kienle.2003}.
Kienle's model interchangeably assigns the roles of sender and receiver to the communication partners. Her model allows for switched positions and for technical entities to participate as long as they can fulfill the tasks involved in the process.

\begin{figure}[htbp]
	\centering
	\includegraphics[width=.7\textwidth]{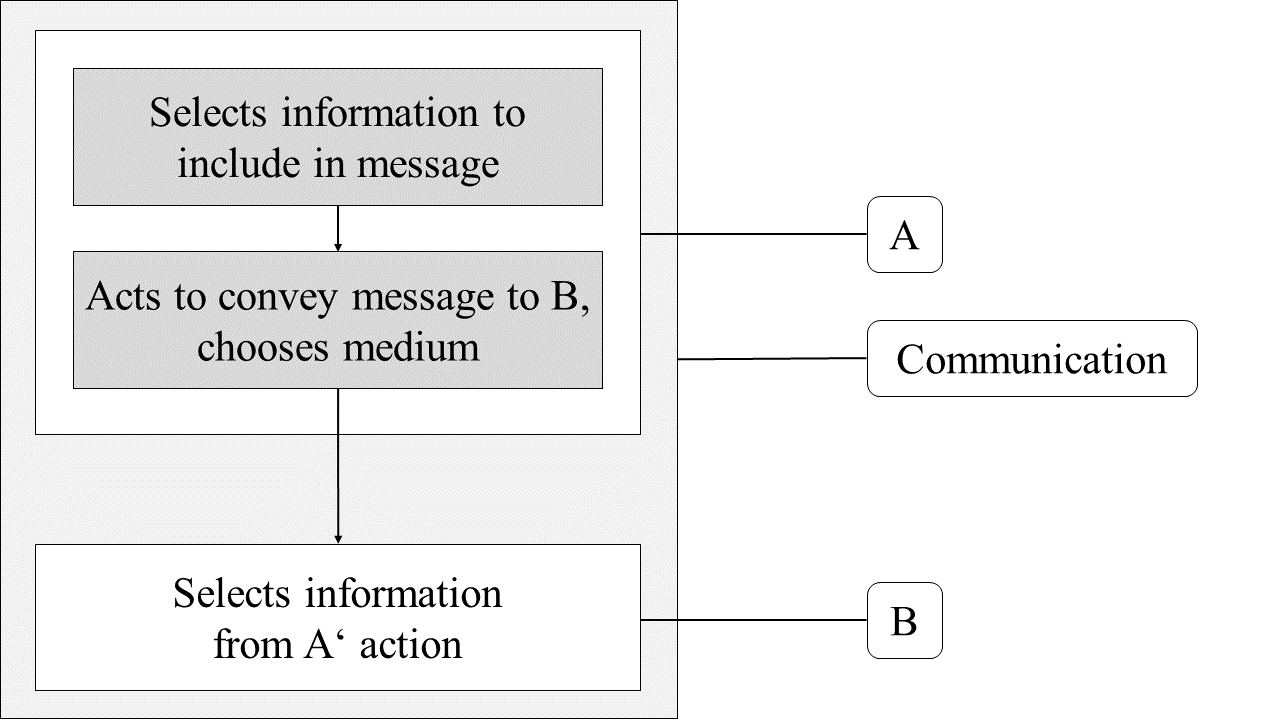}
	\caption[Luhmann's triple selection]{Luhmann's triple selection in social actions, from \autocite[71]{Kienle.2014}}
	\label{fig:triplesel}
\end{figure}

Luhmann derives his model from the idea that humans can only cope with complexity through selection ~\autocite[48]{Luhmann.1984}. Thus, he proposes a communication process that passes through several selections (see \cref{fig:triplesel}. First, the information to be communicated is selected among many alternatives, secondly the form of transmission is chosen (the kind of message), then the recipient evaluates how to understand the message. Eventually, the persons addressed select how the new \enquote{information difference} affects their behavior~\autocite[194ff]{Luhmann.1984}.
We see these steps in web search as well. A search engine selects only a fraction of available information and specifically choses a personalized subset thereof to answer the users' queries. Then, the results are presented in the most meaningful way. Based on their subjective assessment, users accept a relevant result, reformulate queries or reject the output, exhibiting a degree of satisfaction. Finally, they may or may not act by clicking on an organic or paid search result.

\begin{figure}[htbp]
	\includegraphics[width=\textwidth]{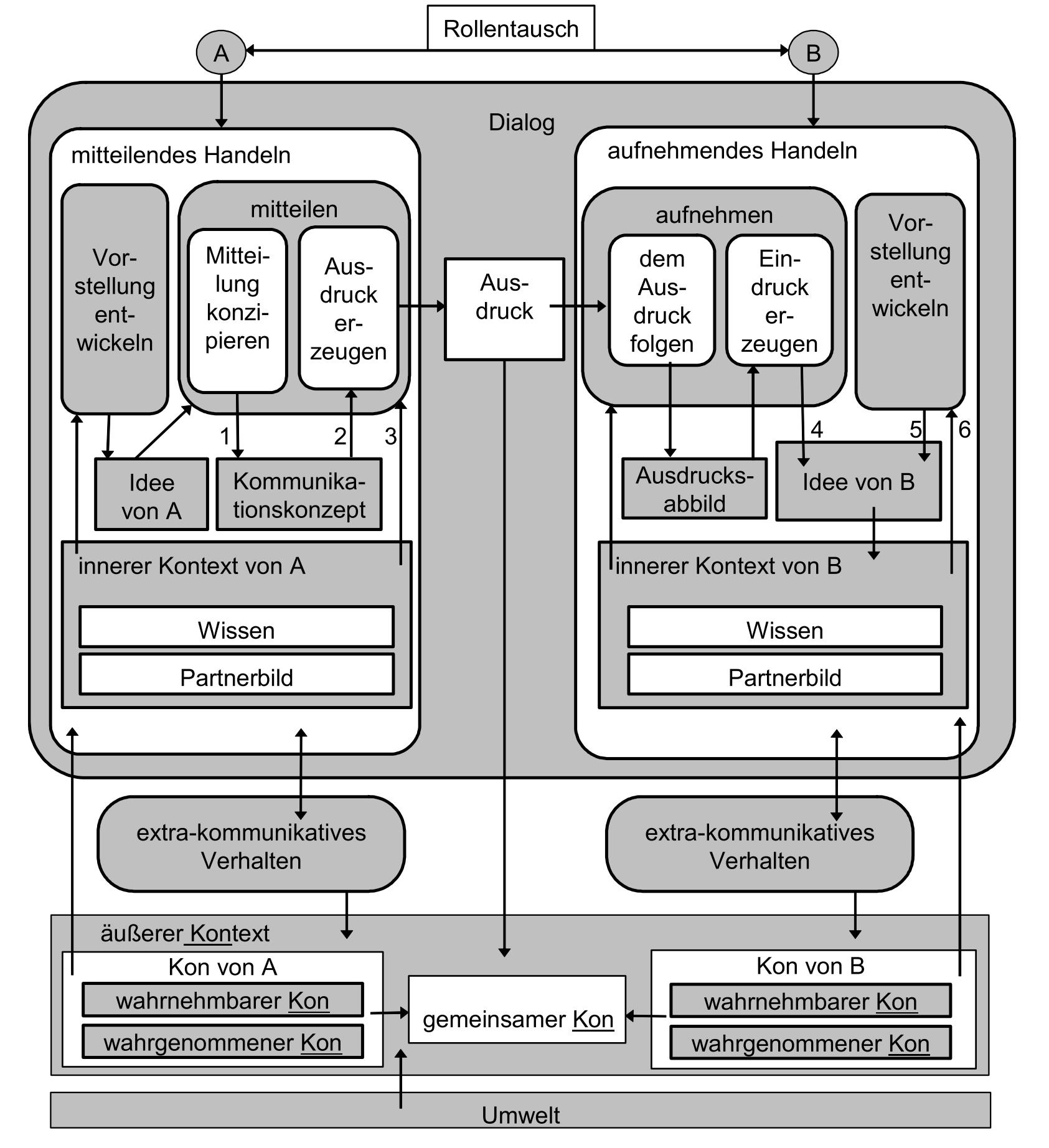}
	\caption[Kienle \& Kunau's Communication Model]{Context-oriented communication model (Kontext-orientiertes Kommunikationsmodell), from~\autocite[S.~35]{Kienle.2003}}
\end{figure}

In Kienle's model in \autocite{Kienle.2003} and \autocite{Kienle.2014}, these selections also takes place, though they are influenced by internal and external context of the agents.
According to her, internal context includes knowledge, emotions and assumptions (especially about the partner).

While the internal context is invisible for the counterpart, the external context is shared. It is based on common perceptions and experiences. as well as mutual beliefs.
Extra-communicational behavior is adapted to the context and enriches the verbal (direct) communication. In this model, context has a significant function.\\
First, shared context supports success monitoring with respect to the intended outcome of the communication. \\
Second, the explicit message can omit information that can be inferred from context. In the end, a consistent inner context about the counterpart's attitude and common belief or shared assumptions about an outer context are the premises for successful communication.

The development of search engine capabilities featured in \cref{app:functsearch} exhibit a tendency of to concentrate on user intent and context. Apparently, the ambiguity of textual queries degrades the result quality like verbal-only communication without contextual knowledge. Thus, it is vital for a technical agent to identify the respective human's context, attitudes and intentions to fully grasp the nature of the communication and answer accordingly. Both the inner and the outer context are explored through data-supported user modeling and predictive analysis. 

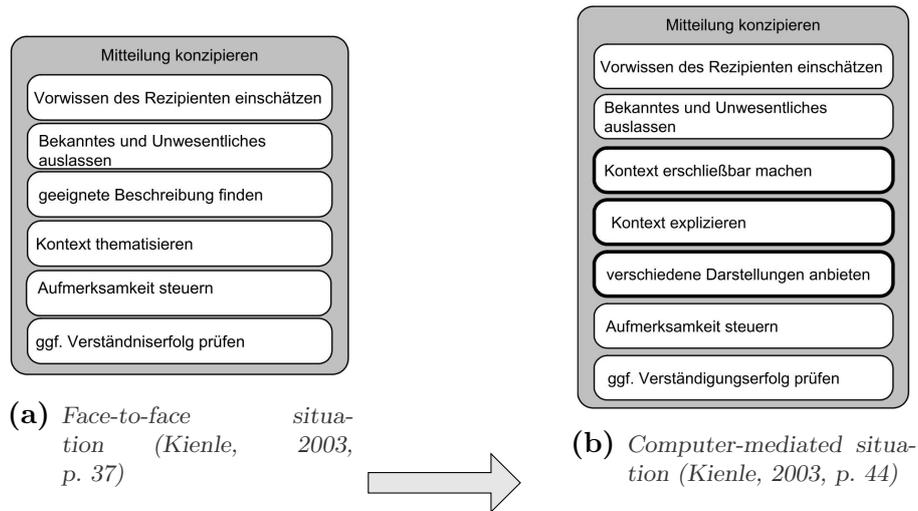
\begin{figure}
	\centering
\begin{subfigure}[b]{0.35\textwidth}
	\centering
	\includegraphics[width=\linewidth]{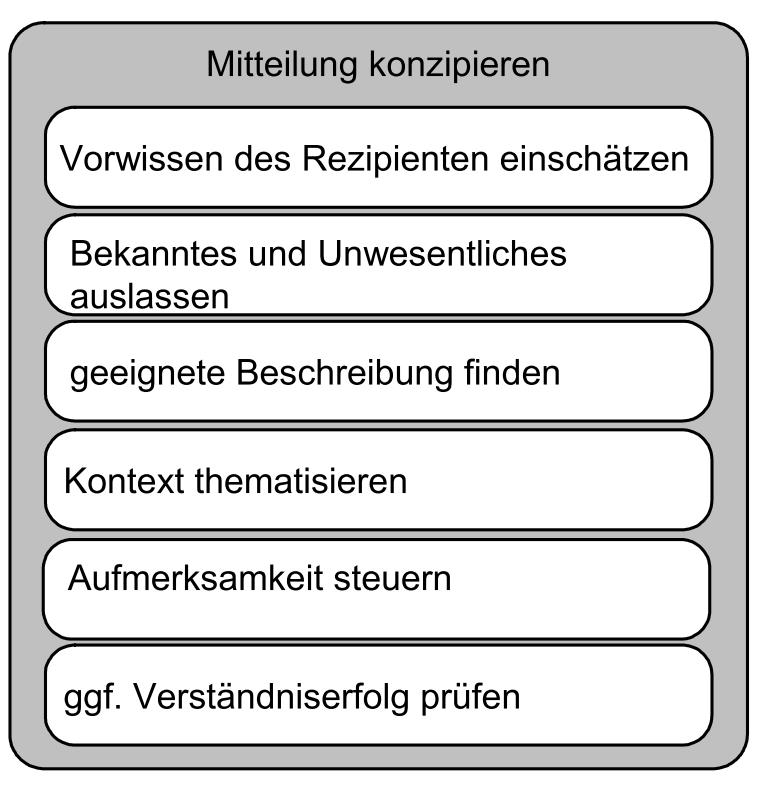}
	\caption{Face-to-face situation~\autocite[37]{Kienle.2003}}
	\label{fig:send_act_1}
\end{subfigure}
\begin{subfigure}{0.2\textwidth}
	 \begin{tikzpicture}
	\useasboundingbox (-2,0);
	\node[single arrow,draw=black,fill=black!10,minimum height=2cm] at (-1,0) {};
	\end{tikzpicture}
\end{subfigure}
	\begin{subfigure}[b]{0.35\textwidth}	
	\centering
	\includegraphics[width=\linewidth]{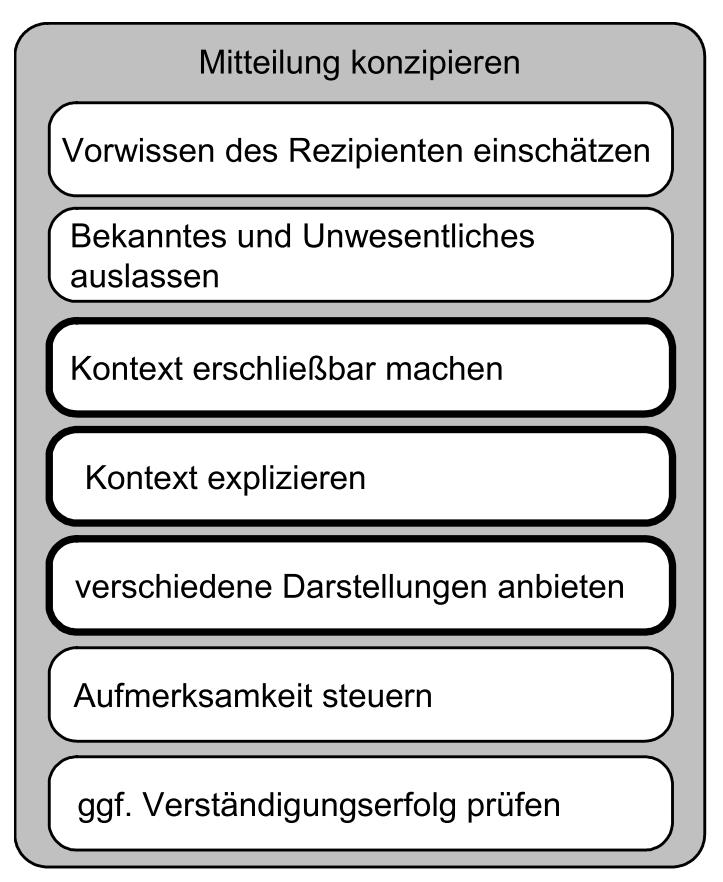}
	\caption{Computer-mediated situation~\autocite[44]{Kienle.2003}}
	\label{fig:send_act_2}
	\end{subfigure}
	\centering
	\caption[Kienle's sender activities]{Sender activities in the context-oriented communication model, from \autocite{Kienle.2003}}
\end{figure}

Kienle's consideration of the sender's activities is especially interesting regarding web search. In the face-to-face situation depicted in \cref{fig:send_act_1} many of the activities listed can be effortlessly applied to ISEs. They try to evaluate and estimate a searcher's background, intentions and knowledge through profiling and computational models (see \cref{app:functsearch} and \cref{sec:dataeco}). ISEs also exclude irrelevant advertisements and search results through selection and personalized ranking on the ad exchange. Then the algorithms determine appropriate descriptions and provide different forms of presentation through \textit{Knowledge Graphs} and \textit{infoboxes}\footnote{See \cref{app:functsearch} for an introduction to both}. After that, they steer attention through a structured search result page and ads on the bottom or top of the SERP.

Finally, an ISE validates success with click-through analysis~\autocite{Joachims.2007,Granka.2004}. They only fail at making context deducible. The context-attributes used in the selection and delivery process remain disclosed. Thus, a receiver might find a message relevant and useful. But users can never fully grasp why a subset of ads or results is shown. Nowadays, ISEs make this context explicit, at least pertaining to advertisements when they give reasons as to why an advertisement was shown. Google gives users some information on why they see a certain ad. Naturally, these explanations are only vague~\autocite{Google.2020c}.

With a computational agent as an intermediary, the communication process changes. A technical system transmits the message, reducing the choices available in the selection of medium (see the second step in Luhmann's three-fold selection, \cref{fig:triplesel}). The communication situation cannot be immediately experienced since there is usually a significant distance between sender and receiver. Context blurs or perishes making interaction more tedious. Now, communication partners must consider the limited means of expression. Extra-communicative behavior can no longer be directly observed, and the partners cannot necessarily assume a shared context. Thus, context information has to be made explicit if it contains useful information for a recipient.
This entails a change in senders' activities, as seen in the transformation from \cref{fig:send_act_1} to \cref{fig:send_act_2}. In computer-mediated communication instead of implicitly referring to context, context has to be made explicit to a degree that it supports the sender's intentions and the receiver's ability to understand and accept information. Kienle supposes to use different illustrations and cues to facilitate comprehension.\autocite{Kienle.2003}~\autocite{Kienle.2014}

Based on the argumentation above, this thesis understands communication as Kienle defines it and illustrates it in her model: \textit{Interaction encoded in symbols, regarding the mutual context}\footnote{From German: \enquote{Durch Zeichen vermittelte Interaktion, wobeauf gemeinsamen Kontext Bezug genommen wird.}~\autocite[20]{Kienle.2003}}.
We need this extended perspective on communication to comprehend the interaction between users, advertisers and integrated search engines\footnote{
	There are different ways to apply the the model of context-oriented communication to the web search ecosystem. In alternative approaches technical agents take the role of a mediator. Although this would better represent the actual flow of information, it would not significantly support this work's analysis.}.
Below, the nature of the communication's content is discussed, and \cref{sec:se} explains how search engines achieve context awareness without engaging in face-to-face communication.

\section{Socio-technical Systems}\label{sec:sts}

The notion of socio-technical systems represents the idea that social, psychological and technical factors can be tightly connected in a way that they can only be understood in combination as an integrated whole~\autocite[p.81]{Kienle.2014}. It originated from a very analogue mining context \autocite{Trist.1954} and was applied to modern software engineering~\autocite{Sommerville.2016} using the systems theory below\footnote{The line of argumentation is drawn from \autocite{Kienle.2014}}.
In the discipline of informatics or computer science this change of mind emphasizes, that not only the design and implementation of algorithms should be of concern, but also their impact on individuals and society as a whole. It is reflected by the constantly changing efforts of the discipline to self-define. 
\citeauthor{Coy.2013} shows how the trajectory of definitions changes over the years and shows how there is a growing conscience for applications and implications of algorithms~\autocite{Coy.2013}. \citeauthor{Coy.2013} quotes Wilfried Brauer twice over a decade, showing the scope of informatics grew from data processing by means of digital computers towards \enquote{theory, methodology, analysis and construction, application (and) consequences of deployment}~\parencite[p.489]{Coy.2013}\footnote{Translated from German~\autocite[p.489]{Coy.2013}}.

Kneer and Nassehi define a system as \enquote{the entirety of a set of entities and their mutual relations}~\parencite[p.25]{Kneer.1993}\footnote{
	Alongside \textit{entities}, this thesis interchangeably refers to the constituent parts of a system as \textit{elements}.}.
Anything not included in a system's definition is called \textit{environment}~\autocite{Kienle.2014}.
Sommerville extends this definition with a purpose the system is dedicated to. From his Software Engineering perspective, he adds that the components of a system cooperate to deliver a set of services to a user~\autocite[p.556]{Sommerville.2016}.

\subsection{Technical System}
Following the definition above, technical systems consist of interrelated technical components. They constitute the entities. 
Luhmann describes those components as coupling of causal elements~\autocite[370]{Luhmann.2000} which may include human behavior, if it happens in an automatic and determined manner and not through arbitrary decisions~\autocite[p.370]{Luhmann.2000}.
This reflects the connectedness of the discrete computational instructions that drive an algorithm. 
Further, he argues that technical systems are  \textit{allopoetic}\footnote{
	\textit{allo}, Greek for \enquote{different, other} and \textit{poiesis} for \enquote{An act or process of creation}, see \url{https://en.wiktionary.org/wiki/allopoiesis}}.
This means, they were constructed by an external force and are not self-sufficient. Thus, they cannot reproduce or renew themselves which means they are autonomous but not autarkic. They rely on external resources (like energy, replacement parts or activation though signals) what makes them non-autarkic. However, they autonomously carry out their operations in a self-determined manner. They halt operation when they receive no further input from their environment. Thus, Luhmann concludes that technical systems are externally controlled and organized~\autocite{Luhmann.2000}.
This applies to algorithms in so far as they are created from the outside through programmers and they rely on hardware and energy to operate. They perform their predetermined actions according to their instructions. They do not compute for the sake of computation but to enact their creator's intentions through \textit{performativity} (\cref{sec:algacc}) which denotes the outcomes that emerge from an algorithm's deployment rather than the written code. 

Following Kienle and Kunau, technical systems are deemed faulty, if not they do not behave as intended by its constructors~\autocite{Kienle.2014}. Here, we can observe a possible discrepancy between the purpose-directed actions and the eventual outcomes of an algorithm. The latter can deviate from the expected results even though a technical agent only performs as intended by its developers. This opens the space for discussion about what separates intentional functionality from undesired side-effects that algorithms can produce in a socio-technical system.

\subsection{Social System}
According to Luhmann, not humans but communications constitute a social system~\autocite[65]{Kneer.1993}.
Thus, Kneer and Nassehi define the social system as systems that recursively generate communication from communication in a continuous manner until the system perishes. Its constituent elements are communications, that reference each other. The relations describe the kind of dependence between them\autocite[p.80]{Kneer.1993}.

In contrast to technical systems, social systems are autopoietic~\autocite{Luhmann.2000}\footnote{
	Greek for \enquote{self-produced, self-organized}, see \url{https://en.wiktionary.org/wiki/autopoiesis}}.
They are self-sufficient as they proliferate through succeeding operations from the elements within. Only if newly created communication can reasonably connect to existing communication, the social system lives on~\autocite{Klymenko.2012}.
Additionally, they are self-describing in a sense that they constitute themselves through differentiation from their respective environment. This \textit{operational closedness} ensures that social systems develop their own structure based on intrinsic operations alone. These operations are not determined by the system's environment, but by a selective choice of environmental influences, at the system's discretion. Hereby, the social system can compose its own structure by selectively reacting to an arbitrarily complex environment. It observes the environment and creates its identity by distinguishing between inside and outside in its communications~\autocite{Luhmann.1998,Mayr.2012}
By determining what communication is acceptable within the system, it can differentiate between system, other systems and environment.
This emergent behavior is a result of the three-fold selection process in the creation of communication by Luhmann~\autocite{Luhmann.1984}(see \cref{sec:com}).
Through this self-description, the system can be observed, described and analyzed from the outside~\autocite{Kunau.2006,Kienle.2014}.
This way, subsystem can arrive at functional differentiation~\autocite{Mayr.2012}.
Similar to its technical counterpart, social systems are autonomous but not autarkic. Even though they sustain themselves through recursive communication (which makes them autonomous), they are not immune to impulses from the outside (their environment) and are subject to boundary conditions. Nonetheless, the social system sovereignly decides on how to incorporate impulses from the outside~\autocite{Kienle.2014}.

Hence, Luhmann deduces that society itself must be the ultimate social system, including the entirety of all social communication~\autocite[p.555]{Luhmann.1984}.
Klymenko points out that, according to Luhmann, this super-system can be partitioned into subsystems with their respective environments. Through self-description these fragments can distinguish themselves from other subsystems. Consequently, systems can recursively consist of interrelated systems. This allows us to treat society as an amalgamation of multiple social subsystems, each with its own communications. Today, this separation happens on a functional basis, so those sub-societies are shaped by their specific form of communication~\autocite{Klymenko.2012}.
Drawing from this distinction, we can make out the social system of online advertising that is comprised of the subsystems of users, advertiser and search engine providers.

\subsection{Kienle and Kunau's Socio-Technical System}
To describe and analyze social systems that sustain a tight relationship with a technical system, Kienle and Kunau came up with a new definition to merge both. According to them~\autocite[p.97]{Kienle.2014}, a social system constitutes a socio-technical system (STS) if:
\begin{enumerate}
	\item 	The technical system supports the social system's communication processes,
	\item 	There is mutual influence,
	\begin{enumerate}[label=\alph*]
		\item The technical system influences the social system,
		\item The social system shapes the technical system,
	\end{enumerate}
	\item The technical system becomes part of the social system's self-description.
\end{enumerate}

This underlines the interrelation of both. Now the social system is actively designing and constructing the technical system. This, in turn, is weaved into the communication processes that sustain the social system. Eventually, it becomes indispensable, so the social system integrates it into its self-description.

The model characteristics with respect to Weisberg's \enquote{model of models} can be described as follows. The structure of the STS is composed of a technical and a social system. Furthermore, in comprises communication processes of the social system that are affected by influences of the technical system. Additionally, there are creative and manipulating actions towards the technical system. Its intended scope is to explain a specific phenomenon that requires to involve both technical and social agents. It allows to analyze the mutual interferences and the technical adoptions that are integrated in a social systems self-description. The (unspoken) fidelity criteria is the capability of the modeler to somehow restrict the boundaries of said systems and narrow the narrow the significant variables.

In \cref{sec:appl_sts}, this model will be applied to the web advertising ecosystem. It will describe how the social system of advertiser, users, engineers and society as a whole interact through the technical system and integrate it in their self-description.
\newpage
\section{Data Economy and advertising}\label{sec:dataeco}

\begin{quote}
	\enquote{\textit{The predominant economic model behind most Internet services is to offer the service for free, attract users, collect information about and monitor these users, and monetize this information.}}~\autocite{Mikians.2012}
\end{quote}

Some Internet platforms exploit basic human needs like socializing with others, information seeking and communication to hoard personal data and capitalize on the analysis of this information~\autocite{Petersen.2019}.
A soon as customers are profiled and recognized online, they can be targeted with personalized advertising and search results \autocite{Google.2019b} in real time \autocite{Steel.2010}.
\enquote{[I]f an ad network is able to accurately target users, we can deduce that the ad network is able to determine user characteristics}~\autocite[1]{Guha.2010}, Guha concludes.

The sections below describe the methods and merits as well as a critique of data collection and targeted advertising. In the context of this work, it is important to understand them as the foundation for modern online advertising. Some problems that emerge from the technical systems in web search and advertising have their roots here.

\subsection{Web-tracking and data collection}
Today, information that was seemingly meaningless alone is enriched through the amalgamation of data from different sources.
There seems to be no such thing as useless data. According to \autocite{FTC.2014}, some of those companies have 3000 data segments for almost all U.S. consumers. Data brokers buy and sell information in packages that include overhead which was not ordered in the first place but are part of the deal~\autocite{FTC.2014}.
Some user data segments are sold off for less than \$0.0005 on average, because user data is so widely available~\autocite{Olejnik.2013}\footnote{Study was conducted 2013, so prices may have changed. Nonetheless, as there are more networked entities today, the amount of data most likely increased, with the price of an individual bit of information consequently decreasing.}.
This data is collected through web tracking.
There are two kinds of online tracking. Stateful technologies use cookies, cache, HTML5 properties and session IDs to identify users. Stateless technologies or \textit{fingerprinting} on the other hand, combine properties of hardware, operating system, browser and the configuration thereof to identify a user~\autocite{Laperdrix.2019}. While active fingerprinting is performed by scripts and plugins and therefore can be inhibited by prohibiting their execution, passive fingerprinting can be derived from network traffic and thus remains unseen and untouched by the user~\parencite[421]{Mayer.2012}

Web tracking enables companies to reveal users' demographics \autocite{Hu.2007}, location, purchasing decisions and interests as well as some sensitive information about them like health conditions, political or religious views \autocite{Bi.2013}, sexual orientation \autocite{Mistree.2009} and relationship status \autocite{Backstrom.2014} through their online activities \autocite{Mayer.2012}.
This amalgamation of data from various sources allows companies specialized in data-collection, -analysis and -fusion to derive PII from at first user-neutral data~\autocite{Krishnamurthy.2009b}.
Sparse individualized data like browsing histories or product ratings are sufficient to de-anonymize users in an approach presented in~\autocite{Narayanan.2008}. Browsing behavior also suffices to learn about a user's demographics~\autocite{Goel.2012}.
This even includes offline behavior such as movement, speech or geolocation~\autocite{Lane.2011}~\autocite{Lu.2012}.
Technological progress benefits this development. Social media entices users to unveil intimate details about themselves, digital communication can be crawled, mobile technology reveals geospatial data~\autocite{Yuan.2012}.
Machine Learning renders manual or explicit classification superfluous. Consequently, user profiles are no longer composed by query similarity and classified in groups of equal interest and purchase decisions. Modern approaches derive clusters and semantic relationships from user behavior~\autocite{Wu.2009}.

Profiling\footnote{
	Tufekci distinguishes \textit{profiling} and \textit{modeling}. According to him, profiling only aggregates data about individuals and categorizes them, whereas modelling infers attributes and intentions beyond the former knowledge with the help of data and computational methods~\autocite{Tufekci.2014}. However, as this distinction is not at the heart of this work, it uses the terms interchangeably.}
can generate problematic categories and unwanted side-effects that allow discrimination or questionable targeting of users. Angwin et al. showed how Facebook allowed to target \enquote{jew haters} \autocite{Angwin.2017} or exclude users by race \autocite{Angwin.2016}. Speicher et al. scrutinized different targeting methods used by advertising-based platforms and found three major methods: attribute-based targeting\footnote{
	Attribute based: Determining the target audience by selection of attributes that users must express},
PII-based (custom) audience targeting\footnote{
	PII-based targeting: Specifying distinct users by their PII}
and look-alike audience targeting\footnote{
	Look-alike targeting: Targeting an audience similar to an existing sample customer base, also known as \textit{remarketing}}~\autocite{Speicher.2018}. 
These methods can discriminate users or groups of users. Speicher et al. further showed that selectable categories on Facebook correlate with sensitive attributes of users (like ethnicity)~\autocite{Speicher.2018}. Further, through the use of look-alike audiences bias is propagated to the selection of new subjects. Lastly, the wide availability of personal data and the efficacy of combination and analysis thereof facilitates discrimination by PII.

Because data brokers and the industries that tap into their resources are generally customer-oriented but not consumer-oriented, it remains laborious for individuals to inquire about their data, their origin, sourcing techniques and usage~\autocite{Marwick.2014,FTC.2014}.
In 2015, Datta et al. found that users could not review all data that was used by Google to create their profile. Furthermore, protected attributes carrying sensitive personal information were used in the profiling process. This potentially exposes users to discrimination and deters the ability to comprehend the reason behind ad choices~\autocite{Datta.2015}.
The opacity may lead to distrust with respect to \textit{unaccountable data sources}~\autocite{Pasquale.2008} that leaves users in the dark about the origin of a computation. This can lead to users losing confidence in an algorithmic system.

Today, agency is passed onto privacy policies, Terms \& Conditions and the like to a degree that they deteriorate to \enquote{defaults}~\autocite{Introna.2016}. Apparently, they are usually skipped or skimmed and only read if coerced to~\autocite{Steinfeld.2016}. This questions the concepts of fair conduct and informed consent in this interconnected socio-technical system.
Ambiguous and misleading privacy policies further the collection as they are incomprehensible to an average Internet user and grant vaguely defined rights to first- and third- parties~\autocite{Reidenberg.2015}.
Eventually, giving truly informed consent to data collection may be hampered by the design of the decision process as people's capabilities with respect to memory load and concentration are challenged~\autocite{Veltri.2017}.
In \autocite{FTC.2014}, authorities voice recommendations that would allow citizens to easily identify data brokers that trade their data and require those businesses to disclose if and how they deduce from raw data. Specifically the categories or profiles that they attach to a consumer should be revealed, so concerned users can scrutinize and correct this information.
In this sense, EU's GDPR (General Data Protection Regulation)\autocite{EuropeanParliament.2016} allows subjects of data collection at least theoretically to demand details about the data stored on them.

Through third-party tracking technologies that are embedded into websites, \textit{personally identifiable information} (PII) is transferred to entities other than the first-party website a user originally intended to visit~\autocite{Krishnamurthy.2009}.
Tracking providers' services span across a wide variety of first-party websites\footnote{\autocite{Krishnamurthy.2009b} showed that 70\% of first-party websites were supported by the Top-10 tracking providers in 2008 already.}. Hence, they are able to aggregate usage data from multiple sources to create a user profile~\autocite{Krishnamurthy.2006} that allows inferences about the personality of a user~\autocite{Lambiotte.2014}.
Acquisitions and technological advance realize a \enquote{potential of significant growth in aggregate data}~\autocite[548]{Krishnamurthy.2007}, for example, when Google acquired \textit{DoubleClick} in 2007 \autocite{Google.2007}.
This \textit{diffusion} or \textit{leakage} of PII~\autocite{Krishnamurthy.2009b} leads to an imbalance of power as users cannot easily examine the usage of their data\footnote{Researchers add: \enquote{Aggregator nodes in possession of information that can be tracked to individual users could potentially use it in a manner that violates the legitimate privacy expectations of users}~\autocite[1]{Krishnamurthy.2006}.}.

This shows how intermediary platforms agglomerate data sources and collection utilities to enhance their services and horizontally integrate technologies that allow them to analyze and target specific users. Through the complex tracking networks and advanced analysis methods an information asymmetry arises~\autocite{Tufekci.2014b}. Users are unaware or resigned towards the collection and have to understanding of the tracking imposed on them. Online companies, however, can construct a rich representation of users. As a consequence, some users try to protect themselves from tracking.

\subsection{Tracking protection}
Whenever a citizen leaves a digital footprint, it can be added to their path. Avoidance is practically impossible due to the high degree of digitalization and the technological divide that separates tech companies from the average user's capabilities to fend off attempts of tracking.
On top of that, organizational structures in advertising ecosystems are hard to decipher, which makes blocking malicious content cumbersome~\autocite{Krishnamurthy.2006}.
Researchers suggest that efficacy of tracking protection techniques are inversely correlated with page quality or browsing experience strongly impede web browsing experience\autocite{Krishnamurthy.2007}.
The better the protection, the more features are unavailable ant the less comfortable the web browsing experience~\autocite{Krishnamurthy.2007}.
A study from 2010 showed that the vast majority of tested browsers could uniquely be identified, even after a fingerprint has changed~\autocite{Eckersley.2010}\footnote{Ironically, adding protection measures to a browser can help to identify an individual client~\autocite{Eckersley.2010}.}.

Still, there are some technical and behavioral measures that can reduce the dissemination of personal identifying information, for example using the TOR browser\footnote{https://www.torproject.org/} or the NoScript browser extension\footnote{\url{https://addons.mozilla.org/de/firefox/addon/noscript/}} as well as various tools to block ads~\autocite{Eckersley.2010}~\autocite{Krishnamurthy.2007}.
Ultimately, some scholars discuss obfuscation and misleading actions like entering ambiguous and false data as a last resort to privacy~\autocite{Brunton.2011}.
Because they see the free web's business model at stake, some scholars suggest tools like \textit{MyAdChoices}. This browser extensions detects behavioral advertising and allows fine-grained control over what information is shared with advertisers~\autocite{ParraArnau.05.02.2016}. Toch summarizes different approaches to preserve both privacy and online advertising including but not limited to aggregated profiles, client-side distribution of PII or supply-side user controls~\autocite{Toch.2012}.

The paragraphs above showed how tracking protection can actually facilitate tracking. One way or another, some users can be identified through associated data and a profile is compiled. Then, they can be subject to targeted or personalized advertising.

\subsection{Targeted advertising}

Marketing does no longer serve a large audience but can be tailored to individuals by deducing knowledge about them, that they were not necessarily willing to expose~\autocite{Tufekci.2014}.
Behavioral targeting of ads is increasing their click-through rate significantly compared to non-targeting controls an addresses similar users of a distinct audience~\autocite{Yan.2009}. It also enhances persuasion and motivates purchases~\autocite{Matz.2017}.

In 2010 already, Gauzente suggests that most of the Internet users are aware of sponsored ads on SERPs, with an increasing tendency~\autocite{Gauzente.2010}. Moreover she finds that a positive attitude towards them improves click-through-rate.
Users feelings towards targeted behavioral advertising and the heavy use of user data to identify customers and audiences are still manifold, undecided and ambiguous. They oppose persistent tracking, intrusive analysis and overly personal advertising, yet expect time-relevant and interest based advertisement~\autocite{Ur.2012}~\autocite{Ruckenstein.2019}.
Schumann et al. suggest that users may accept targeted advertising due to either perceived utility they of a website or an act of reciprocity with respect to the free service they receive. In doing so, they balance the negative loss of sensitive information against the benefits of the transaction~\autocite{Schumann.2014}.
Users may have different mental models of the Internet and its threats to privacy, however they do not express an increased effort to protect against privacy invasion if they are more literate~\autocite{Kang.2015}\footnote{
	The study in~\autocite{Kang.2015} was conducted in an university setting mostly with participants in their 20s (students). However, that is already worrisome.}.

This underlines how citizens are generally aware of data collection and targeting but mostly resign with respect to those practices~\autocite{Hargittai.2016}.
In accordance with that, \autocite{Kim.2019} claims that transparency about data collection practices increases user acceptance if they are deemed acceptable.

Google disallows misconduct on their platform and enumerates prohibited practices on its \textit{Advertising Policy Help} website~\autocite{Google.2019d}\footnote{Forbidden practices include omitting relevant information (payment model, legal or financial details), promoting unavailable offers (products not in stock, inactive deals), misleading content (specified above), unclear relevance (ads unrelated to the search keyword) and unacceptable business practices (fraud, unduly conduct of business).\autocite{Google.2019d}}.
In the context of this work the prohibition of misleading content is most interesting.
Google outlaws false statements about qualifications and claims that promise unrealistic results.
These two rules inhibit most of the practices documented in \cref{sec:digihealth}.

\section{Integrated Search Engines: Google as an advertisement enabler}\label{sec:se}
\begin{quote}
	\textit{Our mission is to organize the world’s information and make it universally accessible and useful.}--Google in 2020~\autocite{Google.2019c}
\end{quote}
\begin{quote}
	\textit{[W]e expect that advertising funded search engines will be inherently biased towards the advertisers and
	away from the needs of the consumers.}--Sergej Brin in 1999~\autocite{Brin.1999}
\end{quote}
Mission statements like the one above show the aspiration of ISE operators to make sense of the world wide web and put the chaos in order. 
Market leaders in this field have arrived at monopolistic scale with the capability to serve billions of users\footnote{
	\textit{Users} of web search engines are people who query the information system to return online content (websites, facts, media) that is relevant to their question.}
at once and satisfy an inexhaustible thirst for knowledge~\autocite{Ratcliff.2019,statcounter.2019}\footnote{
	According to~\autocite{comscore.2019}, Google has had a market share of 98.3\% in Germany and 62.5\% in the US search engine market in October 2019 serving 60 million unique users in Germany and processed 10,718 million search queries in the US~\autocite{comscore.2019}. Other sources claim, Google received at least two trillion inquiries in 2016~\autocite{Sullivan.2016}. Thus, some observers appoint Google the default search engine~\autocite{Editorial.2010}.}.

The following paragraphs deal with the functionality of ISEs and the role they play in a modern society\footnote{
	\cref{app:functsearch} dives deeper into the mechanisms of collection, indexing, ranking and serving that a search engine (SE) provides and gives a detailed overview on SE capabilities.}.
It further models the online advertising ecosystem and describes the different ways of advertisers to connect to users in \cref{ssec:webads} and \cref{ssec:bizmodels}.

The selection choices of this information selection process are subject to academic discussion concerned with the power of intermediary platforms to act as editors and the demand of accountability thereof \autocite{Granka.2010,Introna.2016,Edelman.2011,Grimmelmann.2010,Bracha.2008}. An excerpt of these works is discussed in \cref{ssec:iserole}.

This helps to understand how the decisions of an intermediary like Google have significant impact on advertisers and users alike.

\subsection{Integrated Search Engines}\label{ssec:se}
According to Battelle, \enquote{[$\cdots$] a search engine connects words you enter (queries) to a database it has created of Web pages (an index) [$\cdots$][and] then produces a list of URLs (and summaries of content) it believes are most relevant for your query}~\parencite{Battelle.2005}\footnote{In this context, relevance (or being relevant) is defined as being able to satisfy the needs of the user\autocite{MerriamWebster.28.12.2019} or being related to an event or subject\autocite{CambridgeDictionary.2019}.}.
This leads to a four-step model of search composed of formulation, action (search), review and refinement that has been established by Shneiderman et al. as early as 1954 and applied to many web search engines today~\autocite{Shneiderman.1997}\footnote{
	Additionally, Broder includes query refinement that enables users to iteratively enhance their satisfaction with results through reformulation and modification of the original query~\autocite{Broder.2002}. Similar to Battelle and other researchers he assumes that users take search engines as an aid to satisfy their information need.}

Broder categorizes intentions to search the web into navigational, informational and transactional approaches~\autocite{Broder.2002}\footnote{
	\textit{Navigational} queries describe the urge to access a specific site.\\	
	\textit{Informational} search includes directed and undirected questions, advice-seeking and requests pertaining to listing and locating on- and offline entities~\autocite{Rose.2004}.\\	
	\textit{Transactional} search is concerned with interaction and the intent to \enquote{perform some web-mediated activity}~\parencite[5]{Broder.2002}.}.
According to a study from 2007 in \autocite{Jansen.2008}, over 80\% of queries identify as informational. In \autocite{Rose.2004} Rose and Levinson suggest that navigational queries represent a minority of web search. Furthermore, they introduced the \textit{resource} category to replace the transactional one. This should contain all intentions to find non-informational content online (downloads, recipes, entertainment, aids to offline tasks such as purchases).
This is reflected by Ashkan et al.'s introduction of horizontal categories distinguishing \textit{commercial} from \textit{non-commercial} query interests~\autocite{Ashkan.2009} after scholars learned that frequent queries often originate from the intention to purchase something~\autocite{Dai.2006}.
Later research suggests that search intentions and strategies significantly vary between demographic groups and regional affiliation \autocite{Weber.2011} or gender and task \autocite{Lorigo.2006}.
This shows how users mainly engage with search engines when they perceive an information need or require resources to base their decisions on. On top of that, if they consistently search a topic, they are likely pondering a purchase decision. This can be interpreted as a willingness to spend money.

Google continuously advances and furthered its search engine capabilities through a plethora of updates, features and patents all in order to improve its algorithms and thus user satisfaction~\autocite{Slawski.2019}.
According to market observers, Google rolls out updates multiple times a day to enhance its service and adapt to changes in search behavior~\autocite{Illyes.2017,MozResources.2019}\footnote{
	Moz, a Search Engine Optimization provider writes, without indicating a source: \enquote{Each year, Google makes hundreds of changes to search. In 2018, they reported an incredible 3,234 updates — an average of almost 9 per day, and more than 8 times the number of updates in 2009.}~\autocite{MozResources.2019}}
Observers note that they usually are dedicated to optimize the search engine for user-oriented quality content, fend off malicious attempts of SEO, understand a searcher's context and intentions and expand the variety of queries that can be processed~\autocite{Vinoth.2017}\footnote{Origin unclear, information is available verbatim on various sites}.

Throughout this evolution, a paradigm shift has been and still is observable.
The search engine matured from only working with bare keyword association to processing conversational queries~\autocite{Sullivan.2013,Slawski.2018}.
Semantic analysis and context play an important role now ~\autocite{Broder.2002,Halevy.28.10.2014,Pasca.2.11.2012}. Furthermore, an intricate knowledge repository, fueled by ontologies~\autocite{Menzel.2010,Semturs.6.6.2015} and enriched by the users themselves is employed to make sense of at first incomprehensible queries.
Additionally, personalization of search results based on a user's background, search history and interaction with results seems to play an important role \autocite{Balog.2019,Brukman.6.12.2013,Zamir.13.7.2004,Lawrence.2010} up to the point where some people express their fear of a \enquote{closed-in effect}, that is  figuratively named \enquote{The Filter Bubble}~\autocite{Pariser.2011}.

\begin{figure}[htbp]
	\centering
	\includegraphics[width=0.7\linewidth]{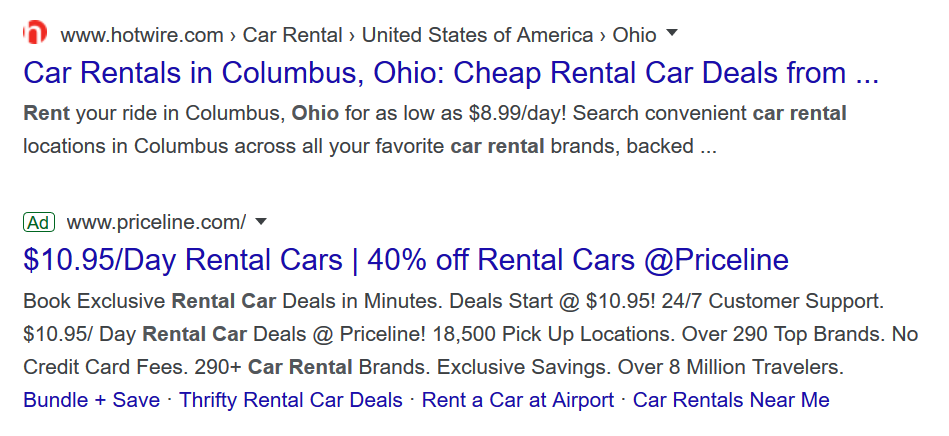}
	\caption[Organic vs. sponsored search]{Examples for ads on Google: Above, an organic search result, below a promotional one, denoted by the green marker on the top left}
	\label{fig:googlesearchcarad}
\end{figure}

To fund their operations, search engines often display promotional results along with their organic\footnote{\enquote{organic} denotes unpaid results on the search engine result page that are listed due to their relevance to the search query.\autocite{Google.2019m}} search results. They are similar styled but marked as advertisement.

\subsection{Web Advertisement}\label{ssec:webads}

Advertisements in its basic understanding refers to \enquote{drawing attention to something}~\autocite[2]{Dyer.2008}. This does not necessarily mean a product but can also address an idea, value belief or opinion, for example the claim of a therapy's superior efficacy~\autocite{Dyer.2008}.

The field of online advertising makes use of \enquote{Information Retrieval, Machine Learning, Data Mining and Analytic, Statistics, Economics, and even Psychology to predict and understand user behavior}~\parencite[1]{Yuan.2012}.
It quickly matured from merely displaying static promotional web banners in the mid- 90s to integrated networks which automatically deliver personalized multi-media advertisements in present days~\autocite{Rashtchy.2007}.
The advantages of online advertising over traditional formats are clear: pricing (cost control through different pricing models~\autocite{Yuan.2012}), optimization (variety of media, real-time display and measurability), reach (virtually unlimited advertising space, no geographical borders, tap into arbitrary demographics) and precisely targeted ads (targeting customers based on arbitrary attributes). Also, search marketing allows to \enquote{brand} search terms\footnote{
	\enquote{Search enables advertisers to associate a brand with a term, even a term that is traditionally associated with other companies or industries}~\autocite[184]{Rashtchy.2007}.}
.~\autocite{Rashtchy.2007}.
This can enable advertisers in the health sector to establish legitimacy through association of their brand with popular search terms (e.g. the name for a clinic appears in the top results after searching for \textit{stem cell treatments}).

The advancement of the Internet as a medium for communication, e-commerce and information allow ISEs to seize a strategic role in connecting advertisers with customers. They guide searchers to their goals and, by the way, place promotion preferably associated with the information need, search intent, product or service that is being searched. Similar to the results, the selected advertisements are to be as relevant to the specific user as possible.
Users that express an information need are more likely to engage with advertising relevant to their cause~\autocite{Yuan.2012}.

\subsection{Business Models}\label{ssec:bizmodels}
Online ads usually consist of a title, creative (text or media), an URL and a landing page~\autocite{Yuan.2012} whereas the latter two do not necessarily have to match exactly.

Scholars and professionals alike speak of \textit{push and pull} or \textit{search and display} advertising~\autocite{Rashtchy.2007}. The former targets searchers and ushers them to a specific webpage that addresses their information need. The latter is displayed along the web experience and may interrupt the browsing experience, thus annoy a user~\autocite{Rashtchy.2007}.

According to Mayer, six business models compose the online advertisement landscape. Advertising companies, hosting platforms, frontend services, analytics services, social networks and content providers cooperate in arbitrary combinations to deliver promotional messages to Internet users~\autocite{Mayer.2012}\footnote{
	Hosting services provide utility to easily set up websites while frontend services publish content or support extended functionality via JavaScript libraries or APIs, respectively.
	Content providers and social networks publish content, media or widgets to increase user engagement and collect usage data through tracking. This data is usually monetized in targeted advertising.}.

\begin{figure}[htbp]
	\includegraphics[width=\textwidth]{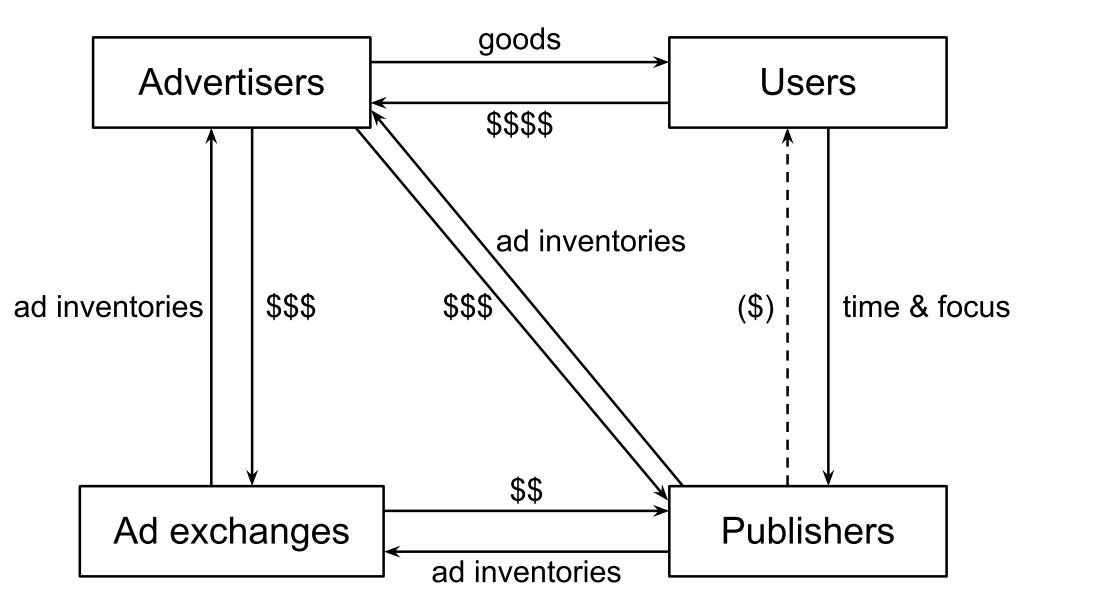}
	\caption[Yuan et al.'s online ad ecosystem]{Online Advertisement Ecosystem, by Yuan, Abidin et al~\autocite{Yuan.2012}}
	\label{fig:ad_ecosys}
\end{figure}

With respect to web search Yuan et al. boil this down to a 4-party model to simplify the workings and reduce its constituent parts to the most relevant functional entities.
The paragraph below describes this and illustrates how the system of online advertising is composed of ad exchanges, advertisers, publishers and users in \cref{fig:ad_ecosys}.
The descriptions are drawn from \autocite{Yuan.2012}.

\begin{description}
	\item[Publishers] offer advertisement space (the \textit{inventory slots}) on the service they offer to gain revenue. A search engine may opt to show promotional results in a designated space on its SERP. Thus, Yuan et al. argue that, conceptually, ISEs qualify to be a publisher in this model.
	\item[Ad exchanges] handle the negotiation of ad delivery and auctioning of inventory slots. As a broker focused on supply and demand it computes the matching based on keywords and query terms, website content and user data, respectively~\autocite{Google.2019t}. Since these networking agents act as intermediaries, systematic targeting of ads is possible, either based on website specifics (target group, topic, location) or user characteristics~\autocite{Mayer.2012}. Yuan et al. distinguish between supply-side or demand-side networks, combination of both and data exchanges. However they note, that the lines between them blur as an all-in-one approach popularizes. Nonetheless, data exchanges play a distinct role in delivering user data for behavioral targeting~\autocite{Yuan.2012}. The more an ad exchange can make sense of the relations between keywords in terms of similarity and relevance and the more it learns about users' search context, the more valuable the service it can provide.
	
	Google is a strong player in this field with 70\% market share\footnote{based on ad revenue} in the online advertisement business~\autocite{Graham.2019}.
	
	\item[Advertisers] are eager to promote their service or product. They bid on inventory slots through the ad exchange. The efficacy of their ads strongly varies with position, context and the number of other ads on the website. Hence, the price varies based on these metrics and the fit of bid phrase and query term or popularity of the keyword. They choose which promotional content to deliver, set up campaign goals, select a billing method and review the ads' performance.
	\item[Users] access websites to satisfy their information need. The results they receive from search engines are individually tailored and purely based on relevance. However, which ad they receive, depends on multiple factors. Quality of the match between advertisements and query keywords, bid prices and expected revenue ratios computed by the ad exchange influence the choice~\autocite{Yuan.2012}.
	Advertisement delivery can also be steered via signals emitted by a search user~\autocite{Shah.2019}. 
\end{description}

\cref{fig:ad_ecosys} by Yuan et al. shows how the four participants are related. There is a flow of cash between the commercial players in exchange for inventory slots. Users generally are compensated with value with respect to their information need as they receive online services, which usually are free (compared to traditional paper advertising, where magazines must be purchased). They in turn return to the promoted services or products with a commercial interest or even purchase intention.
The interactions between users, search engine and ad network providers and advertiser constitute the communication in the social system of the STS of web search.
The different perspectives on Google's role therein are discussed in the next part.

\begin{figure}
	\centering
	\begin{tikzpicture}[node distance = 7em, auto,
	every node/.style={text centered},
	every edge/.style={>=latex,draw},
	every edge quotes/.style = {align=center,draw=none}
	];
	
	\node (ad) {Advertisers};
	\node (aa) [right of = ad] {Ad Agency};
	\node (an)[right of = aa] {Ad Network};
	\node (pu) [right of = an]{Publisher};
	\node (us) [right of = pu]{User};
	\node (ax) [above of = an]{Ad Exchange};
	\path[->, every node/.style={align=center}]
	(ad) edge (aa)
	(ad) edge[bend left=20] (an)
	(ad) edge[bend left=20] (pu)
	(ad) edge[bend left=20] (an)
	(ad) edge[thick] (ax)
	(aa) edge (an)
	(aa) edge[bend right=20] (pu)
	(an) edge (pu)
	(pu) edge (us)
	(ax) edge (us);
	\end{tikzpicture}
	\caption[Muthukrishan's ad paths]{Ad paths, by Muthukrishnan~\autocite[2]{Muthukrishnan.2009}, extended by author (addition of ad exchange as a new intermediary)}
	\label{fig:adpath}
\end{figure}
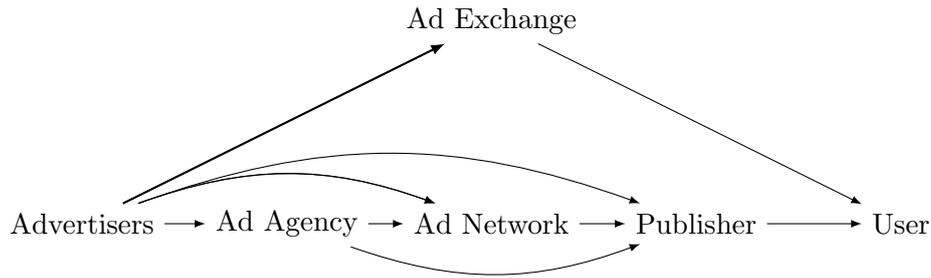

There are several different methods to place advertisements on a website (see~\cref{fig:adpath}).
In the traditional \textit{Direct Buy} pricing model advertisers buy a distinct slot on the first-party website of a publisher to place their promotion~\autocite{Mayer.2012}. Usually, these ads categorizes as \textit{Branding Ads} with long-term contracts for distinct slots and no targeting differentiation~\autocite{Yuan.2012}.
Yuan et al. describe other business models in web advertisement. For example, publisher networks or ad agencies / advertiser networks operate supply- or demand-side platforms. They act as intermediaries and facilitate their members' or customers' advertisement business. In doing so, they organize the entirety of the inventory slots or ads of their customers~\autocite{Yuan.2012,Muthukrishnan.2009}.
From this duality, \textit{ad exchanges} like Google AdSense emerged.
They manage different kinds of ads, including \textit{sponsored search} ads and \textit{contextual ads}. In the first case, ads are matched to users based on keywords (query terms, content on the relevant websites, e.g.) and user-PII and displayed among the search results. The latter describes ads that are targeted based on context (domain, user intent, e.g.) and PII with flexible localization on a publisher's website. These types of ads can further be differentiated by delivery method, trading place, competition method, pricing model and automation~\autocite{Yuan.2012}\footnote{
	A summary of differentiations found in \autocite{Yuan.2012} is described below:
	\begin{description}
		\item[Delivery method] Forward contract or on the spot
		\item[Trading place] Over the counter or on a transparent market (auction)
		\item[Competition method] 1st price negotiation, reservation or auction or 2nd price auction in a real-time bidding or pre-set bidding competition
		\item[Pricing models] Flat-rated (per time), or cost-wise (per click, mille, action or conversion)
		\item[Automation] Manual (mostly in negotiation and campaign planning) or automated (real-time bidding)
	\end{description}}

The principal goal of the Internet-based advertising system is to find \enquote{the best match} in terms of both relevance and revenue between a specific user in given context and set of available ads through computation.

Muthukrishnan describes the business model of ad exchanges like Google AdSense in \autocite[2]{Muthukrishnan.2009} as follows. A user $u$ visits a website $w$ that allots space to ads. The publisher $p(w)$ requests an ad from the ad exchange $E$ and also denotes a minimum price $p$ for the inventory slot. In this model, it is assumed, that $p(w)$ knows $u$'s characteristics and shares this information with $E$. The ad exchange provider furthermore knows about the \textit{ad configuration}\footnote{
	The ad configuration includes localization, dimensions, media type of an inventory slot and other conditions determined by the publishers. It is guided by presumptions about ad efficacy and user engagement drawn from empirical data~\autocite{Muthukrishnan.2009b}.}
\autocite{Muthukrishnan.30.6.2009} on the target page. Additionally, it can crawl content on $w$ to make inferences. Then, $E$ requests ads from ad networks $a_1,\cdots,a_m$. It may disclose some information $E(u)$ and $E(w)$ about $u$ and $w$ along with the minimum price to each of them. This could include PII of $u$ or topics of $w$. An ad network may return a bid $b_i>p$ and an ad $d_i$ of one of its customers to display on the slot. In a competition method (see above) determined by the exchange, the inventory slot is sold to the winner who can now serve its ad on the publisher's website to the user if it fits the configuration. This is called an \textit{impression}. The winners are notified of their success (and possibly, the losers, too).  All of this happens in a matter of milliseconds.\autocite{Muthukrishnan.2009}
Google extends the above model by using \textit{AdRank}, a measure that influences the position an advertisement can attain. It is influenced by the respective bid, ad-content and landing page quality, competing other ads, search context, relevance and performance. In an auction scenario, the AdRank determines an ads success in an auction and its position on the SERP~\autocite{Google.2019o}.
The dynamically computed \textit{AdRank threshold} is a score set by Google to determine the minimum price of a specific inventory slot and the rejection level for ads competing for the slot~\autocite{Google.2019n}.

Google serves both side of the market. The platform \textit{AdSense}\footnote{https://www.google.com/adsense} enables publishers to sell inventory slots on their respective sites via Google ad exchange. \textit{Google Ads}\footnote{\url{https://ads.google.com/}} on the other side allows advertisers to bid on advertising space on websites and the search engine to display their creatives\footnote{\textit{Creatives} denote the visual appearance of ads, including but not limited to text, images, media and the respective styling.}~\autocite{Google.2019m}.
Furthermore, it hosts a tracking and analysis service that enables customers to gather information about web site visitors. To use \textit{Google Analytics}\footnote{https://www.analytics.google.com}, they only have to include a JavaScript snippet or \enquote{Google Tag}. Then they have access to a rich set of analysis tools and the opportunity to link insights and statistics to their respective advertising campaigns on the Google Ads.
Both, Search-Engine-Advertising (SEA) (listing ads as promotional results along on the SERPs of Google's search engine or its partners') and display ads (delivered over the AdSense program to a network of publishers, the \textit{Google Display Network}) can be purchased~\autocite{Google.2019t,Google.2019p}.
Google Ads offers both sponsored search and contextual ads in a generalized second price auction (GSP)\footnote{
	See \autocite{Edelman.2005} for an elaboration of the GSP}~\autocite{Google.2019t}.
It allows automation of ad delivery based on specified goals (clicks on the ad, conversions (some intended user action like a purchase or phone call) or impressions (mere display), e.g.) and automated bidding on advertisement slots~\autocite{Google.2019h,Google.2019l}.
If serving an ad via Google Ads, advertisers have multiple ways of targeting users. They can pick a specific audience (by demographics, affinity, purchase interests, specific behavior, similarity with another audience or by reconnaissance (remarketing)). Besides, they can address searchers by the topics and content of sites they search for or the keywords they type in. On top of that, users in a defined situation can be approached, for example at a distinct life event (marriage) or situation (time, place, mobile)~\autocite{Google.2020,Google.2020b}.
Herein, ads can be published automatically in an arbitrary fashion or deliberately on specific sites, apps or media~\autocite{Google.2019q}.
Ultimately, Google allows advertisers to address individual users by \enquote{Customer Match}, if it is compliant with privacy policies~\autocite{Google.2019s}.

Nevertheless, Google inhibits advertisers to imply knowledge of PII in their ads or market to a very narrow audience only. In fact, it also specifically prohibits promotion in sensitive categories such as clinical trials, personal hardships and health~\autocite{Google.2019r}.
Furthermore, there are numerous institutions that should guide advertisers in achieving ethical conduct of business\footnote{
	For example, the Interactive Advertising Bureau (IAB, \url{https://www.iab.com/}), the Network Advertising Initiative (NAI, \url{https://www.networkadvertising.org/}) or the European Digital Advertising Alliance (EDAA, \url{https://www.youronlinechoices.com/})}.

The auction process depicted above shows how an ad exchange acts as an intermediary between advertisers and user. Furthermore, we can conclude that based on the insights from \cref{sec:dataeco}, Google qualifies as both an advertising and data exchange. With its tracking services and analysis capabilities it leverages data collected about users and their online interactions to enable behavioral targeting. With this this technique, they are able to directly address specific users  that they assume to be in their target group. Unfortunately, this may include sensitive categories such as medical conditions. Even though they cannot be either immediately and the use thereof is prohibited, they can still be targeted through sophisticated combination and computation of user attributes. 

\subsection{The Integrated Search Engine's role as an intermediary}\label{ssec:iserole}
With their decisions on how to collect, index, rank and present results and advertisements, integrated search engines exercise great power. People turn towards them in search of all sorts of information. They confidently trust a search engine to objectively rank results of a query by true relevance \autocite{Pan.2007}. They shape searchers' perceptions of the web and intervene with their behavior online. This can have significant social and commercial implications as it assigns visibility and directs attention~\autocite{Goldman.2006}.

Grimmelmann enumerates three different views of scholars concerning Google's role in society \autocite{Grimmelmann.2013,Grimmelmann.2013b}. He contrasts the role as \textit{conduit}~\autocite{Chandler.2007} with those of an \textit{editor}~\autocite{Volokh.2011,Goldman.2006} and an \textit{advisor}~\autocite{Grimmelmann.2013b}.
These roles are described and discussed below.
As \emph{conduits}, ISEs appear as gatekeepers or bottlenecks that mediate between content providers, advertisers and consumers. Thus, a conduit can exercise power through blocking websites or neglecting certain advertising customers~\autocite{Grimmelmann.2010}. They could refuse to index content, manipulate auctions or introduce bias.
In \autocite{Chandler.2007}, Chandler juxtaposes \textit{speakers} and \textit{listeners} to stress how intermediaries can shape the communication between those parties. Comparing this communication as a form of verbal exchange relates it to the question of free speech as a foundation of fair use. She raises the question of how free speech can be guaranteed if gatekeepers like IREs have the opportunity to deliberately interfere with the interactions conducted on their platforms and automate their business with undisclosed algorithms. 
Chandler links this to \textit{net neutrality}, a principle by which selection intermediaries such as search engines and ad exchanges should not discriminate content and exercise bias~\autocite{Chandler.2007}. This idealistic approach means to maintain free speech online and is based on the idea of functional similarity between search engines and Internet service providers (ISP) and network providers. They all act as bottlenecks in data transmission, they argue, thus need to be treated accordingly. 
It is vigorously contested by Grimmelmann who argues that fulfilling all principles he derived from the idea of net-neutrality is just unrealistic and renders search useless\parencite[p.436f]{Grimmelmann.2010}\footnote{
	Grimmelmann enumerates eight principles that characterize neutral search in the way a radical conduit would perform it (additions for clarification added in parenthesis): equality (no differentiation among websites), objectivity (distinguishing between correct and incorrect results), (no) bias, (sufficient) traffic, relevance (maximize user satisfaction), (no) self-interest, transparency (full disclosure of algorithms), (no) manipulation.\autocite{Grimmelmann.2010}}.
Nonetheless, he adds that giving search engine operators free reign is not an option.
Granka agrees and elaborates in \autocite{Granka.2010} how following these principles would hamper quality of search results and competition through malicious manipulation and less market differentiation.
In addition, she notes that the most wide-spread components of search engine algorithms are already widely known and well researched.
Pasquale summarizes that net neutrality should be imposed on search engines only in regard to transparency concerning business relations, promotional content and paid results~\autocite{Pasquale.2008}.

The \emph{editor} describes another perspective on intermediaries. Selection lies in the very nature of ISE. All of their practices constitute a form of editorial judgment. Goldman points out how search engine providers decide upon what data to index, how to rank it and which part of it to eventually present. Even though most of these operations are performed automatically in a seemingly objective-computational rationale, the inner workings of these procedures, their weights and factors, parameters and input are clearly defined. These decisions generate an editorial act along with the manual adjustments that made in response to certain issues, he argues~\autocite{Goldman.2006}. Herein, the latter may reflect a company's values and willingness to self-regulate, though the actual criteria the algorithms ought to comply with usually remain unknown~\autocite{Diakopoulos.2013}\footnote{
	In a \textit{TechCrunch} article, an interviewee points out, how \enquote{[t]here are things Google has deemed relevant to the public interest that they’re willing to kind of intervene and guard against, but there really is not a great understanding of how they’re assessing that}~\autocite[Robyn Kaplan]{Dickey.2017}.}.
The misuse of editorial power though can mislead users~\autocite{Grimmelmann.2010}. Nonetheless, Grimmelmann demands platforms to take responsibility and moderate content, even manually, in order to cope with the \enquote{disturbing demand-driven dynamics}~\parencite[1]{Grimmelmann.2018} that scourge Internet platforms. He deems this measures necessary as algorithms cannot be conscious or self-aware about the entirety of consequences that entail their actions.

Grimmelmann notes how there is space left for another form of intermediary between the objective conduit and the subjective editor. While a conduit's job is to \enquote{deliver to each website the user traffic to which it is properly entitled}~\parencite[p.873]{Grimmelmann.2013b}, the editor only cares to satisfy the audience and keep it from switching to competitors.
In \autocite{Grimmelmann.2013b} users are introduced as the subject of interest, who are actively educating themselves on a certain topic. This underlines how the two approaches above are combined. Instead of being a passive audience, users formulate their goals and expect a specific mix of websites that cater to their needs. According to Grimmelmann, the advisory search engine answers to a user's query in a personalized way that is \enquote{uniquely relevant to the user’s unique interests}~\parencite[p.874]{Grimmelmann.2013b}.

The choices Google makes pertaining to ranking are relevant because research suggest that results higher up on the search result page receive more attention and generate higher click through rates \autocite{Granka.2004}~\autocite{Lorigo.2006}\footnote{
	Both studies were small scale eye-tracking experiments with only 26 and 23 validly reporting participants respectively but are widely cited and accepted.}.
Scholars assume that this observation can be attributed to two different factors. Firstly, search engines by design try to return the most relevant results on top of the list. This is perceived as an indication of quality which they call \textit{trust bias}. Users trust the algorithm to deliver the truly significant result at first.
Secondly, the relevance of an advertisement is assessed in comparison to other results on the page, leading to a \enquote{quality-of-context bias}~\autocite{Joachims.2007}.
This has ramifications for ad delivery as well.
If businesses in the stem cell tourism industry manage to get listed among approved clinics, governmental agencies and medical authorities in the health sector, they benefit from the quality-of-context bias. A slot on the SERP among those entities could be interpreted as a \textit{token of legitimacy}(see \cref{sec:digihealth}). 
They can also leverage the trust bias as ISEs seemingly convey objective importance. On top of that, keyword-based advertising campaigns might claim an association with a topic like emergent stem cell treatments or a form of therapy. They might try to \enquote{brand} a specific search term with their name and solidify their popularity among searchers in this field.

All of the above is equally relevant for advertising displayed on the SERP. Ads are located at the top and bottom of the result page and thus are perceived as being significant results deliberately chosen by an intermediary. However, no privately operated ISE can grant full disclosure of its workings. Nevertheless, its operators have to be aware of the ramifications that ensue their editorial choices.

\section{Application to Web Search}\label{sec:appl_sts}
From a constructionist perspective, one can model the socio-technical system of (sponsored) web search and affiliated online advertisement using the elaborations in \cref{sec:sts}. Below, this model will be constructed from the insights above.

The social system is represented by the fraction of society that is concerned with web search and online advertising. Herein, this subsystem is denoted the \enquote{Web Search Society} (WSS). In this analysis, the WSS consists of communications between four kinds of participants. The WSS is influenced by (1) consumers or users\footnote{
	This thesis refers to \textit{users} when it considers humans involved in a human-computer interaction (here: searching the web via a search engine). They conduct searches, review results and act based on the information they retrieved. In contrast, \textit{citizens} are concerned with their role and relationships within a society. Their perspective includes policies and governance issues and how the socio-technical system can be shaped.}
that search the web,
(2) the companies developing ISE\footnote{A platform that combines search engine and ad exchanges} and running ad exchanges and search engines and 
(3) advertisers promoting their products, services and ideas.
These influencers embody the WSS's environment. They can stimulate the communication within the social system.
Ultimately, content providers or publishers (website hosts) and governing institutions that regulate the WSS could be included as well. However, this thesis concentrates on the interactions of the first three and only covers the latter to a small extent.

Through the open-minded approach to information in \cref{sec:info} it is possible to identify communication processes induced by those agents. Below, the four-fold approach is reviewed with respect to web search.

\begin{itemize}
	\item \textbf{Representation of knowledge:} Website content, Knowledge Graph, algorithms
	\item \textbf{Data in an environment:} User data, implicit user feedback, WWW structural data, semantic ontologies
	\item \textbf{Part of process of communication:} Query semantics, advertisements, editorial selection, online behavior
	\item \textbf{Resource or commodity:} Ads, websites, user data, attention
\end{itemize}

\begin{figure}[htbp]
	\includegraphics[width=\linewidth]{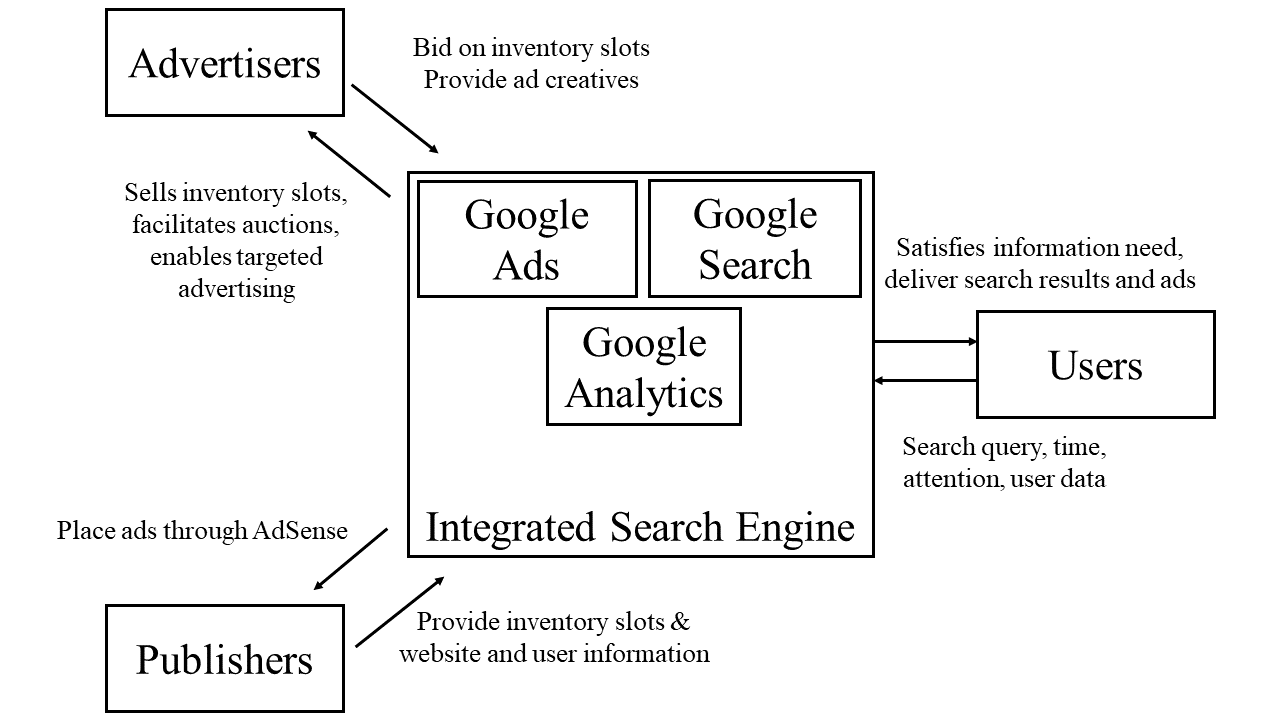}
	\caption[WSS communication processes]{Communication processes in the Web Search Society, illustration by author}
	\label{fig:comprocwss}
\end{figure}

Along these assignments, the communication processes in the WSS can be sketched. \cref{fig:comprocwss} shows them schematically, connecting the agents in the environment of the WSS through their mutual communication. The direction indicates sender and receiver, the arrows are labeled according to the information the respective communication carries.

The technical system (TS) manifests in an integrated search engine (ISE) which supports the above communication processes.
The ISE comprises algorithms that enable web search, collect and analyze data and organize online advertisement.
These algorithms constitute the entities or components of the technical system.
The communication processes it supports are evaluated and fed back into the system to re-calibrate its workings.
Herein, users seek to satisfy an information need and inquire about a subject. They want to find an informational resource on the WWW. 
They do so by inquiring the search engine providers via the search engine's web interface.
The company running the search engines executes algorithms to find relevant search results. First, it crawls the web and collects websites by publishers.
Then, it indexes the collection. Eventually, it displays a ranked list of findings on the search engine result page to answer the user.
Through selective presentation of publisher's content along with ads to users in a comprehensible way, Google Web Search supports the communication between publishers (producers), advertisers and users (consumers).
This allows users to satisfy their information need, advertisers to target consumers and publishers to reach their audience.
Concurrently, it facilitates negotiations and auctions about inventory slots on Google Ads' ad exchange. This also enables advertisers to place their promotional message on the SERPs or third-party sites which are eventually displayed to users through Google AdSense.
Furthermore, it collects and merges data from different sources. It computationally draws conclusions about the outer context of the web search ecosystem and the inner context of users with Google Analytics. This influences the capability of the communication partners to bridge the digital gap and base their interaction on more or less mutual context.

The technical system strongly influences the WSS. Through its editorial choices it determines what people perceive as relevant and shapes users' ways of formulating their questions. It furthermore dictates the code of conduct with respect to ad's and websites' quality in form and function. Through its dominant position in the market as a major search engine that accumulates a plethora of data and its actions have significant repercussions on the online experience of society. It also impacts individuals' sense of privacy since the dissemination of user data and targeted advertising are part of the technical mechanisms and social communications alike~\autocite{Hargittai.2016}\footnote{\autocite{Hargittai.2016} is from 2014 but remains relevant, with respect to the widespread use of social networks among young users and networked devices on one side and the evolution of tracking techniques on the other.}.

\begin{figure}[htbp]
	\centering
	\includegraphics[width=\linewidth]{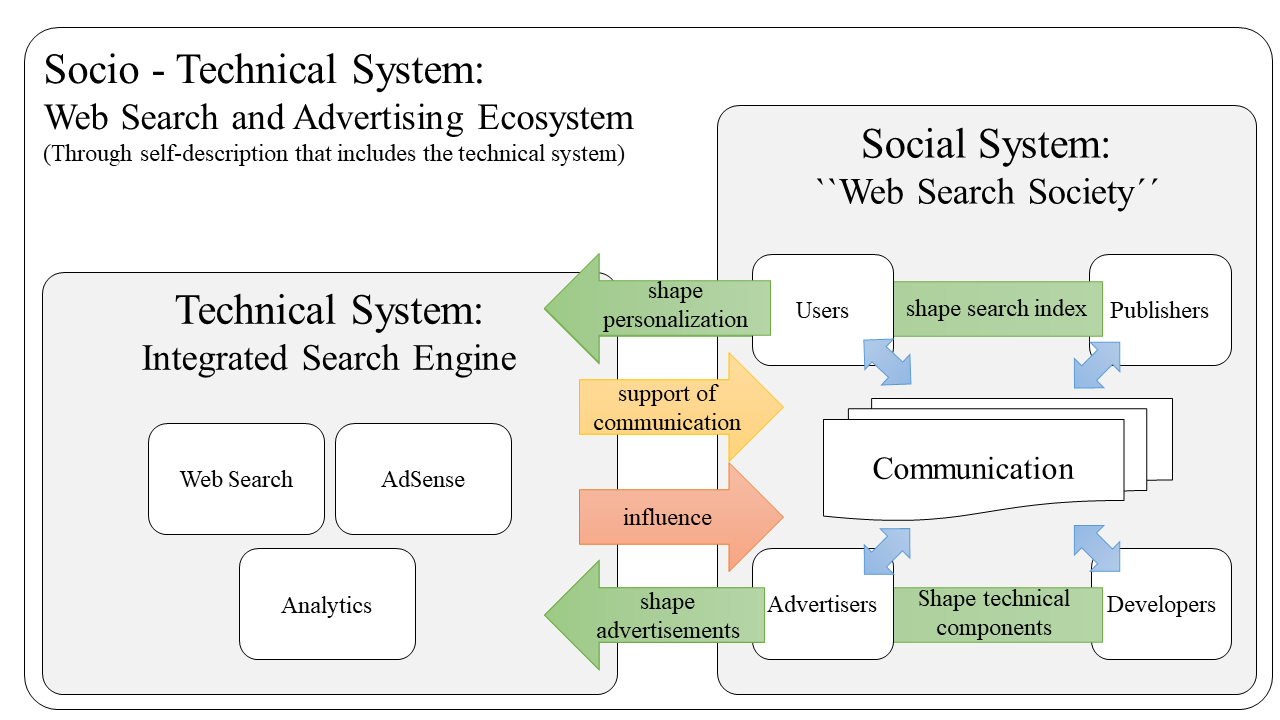}
	\caption[STS of web search and advertising]{A model of the socio technical system of web search and advertising, own illustration adapted from \autocite{Kienle.2014}}
	\label{fig:sts}
\end{figure}

As shown above, the WSS includes the technical system in various aspects of its communication as required by Kunau in~\autocite{Kunau.2006}.
Additionally, the WSS incorporates the mechanics of the ISE in its self-description. These include but are not limited to characteristics like instant answers, targeted advertising, realtime bidding on ads, as seen in \cref{fig:sts}.
Hence, without the traditional ISEs there would be no online search as we know it. The emergence of the term \enquote{to google}~\autocite{Duden.2020,MerriamWebster.2020} reflects this, as well as the rise of an online advertising industry~\autocite{Evans.2009} in the last decades and the comprehensive research in the field of search engine technology (see \cref{sec:se} and \cref{app:functsearch} and online behavior).

In this work, the focus lies on the communication between IRE and user, as it is the only immediately observable interaction. However, the analysis below tries to infer about the inner context of the intermediary and the motives of advertisers (the selection of ads).

\chapter{Related Work}\label{ch:relwork}
In academia, accusations of discriminatory and biased algorithms are not uncommon~\autocite{Sweeney.2013}.
Consequently, there have been numerous attempts by scholars of various disciplines to reverse engineer or scrutinize privately operated information systems that have social impact.
For example, researchers investigated;
\begin{itemize}
	\item Web search~\autocite{Willis.2012,Hannak.2013} and advertising~\autocite{Guha.2010,Sweeney.2013,Speicher.2018},
	\item E-commerce~\autocite{Mikians.2012,Valentino-DeVries.2012,Hannak.2014} and reviews~\autocite{Arjun.2013},
	\item Text completion~\autocite{Diakopoulos.2013}, correction~\autocite{Keller.2013} and detection~\autocite{Sap.2019},
	\item Perception of online environments~\autocite{Hannak.2014,Krafft.2017,Larson.2012},
	\item Finance~\autocite{Lazer.2014,Pulliam.2012}.
\end{itemize}

This chapter discusses literature related to my analysis of direct-to-customer marketing of stem cell-related services on integrated search engines.

First, in \cref{sec:digihealth} the realm of stem cell tourism is explored with its implications for patients and caretakers. This helps to understand how our investigations can contribute to the protection of vulnerable user groups like patients.
Then \cref{sec:gov} presents different approaches to regulation of Internet-based services. This is meant to emphasize the role that society plays in technological assessment.
Next, the concepts of algorithmic accountability with respect to transparency and responsibility are explored in \cref{sec:algacc}. This is important as it enables to designate moral agency of SRAs.
Lastly, Black Box analysis are described in \cref{sec:blackbox}. In this thesis, it is the method of choice to scrutinize opaque SREs.

\section{Digitalized Health in the Realm of Stem-Cell-Tourism}\label{sec:digihealth}

The health sector offers a growing number of opportunities to deploy digital technologies.
Health-related online research (specialized vertical search engines or websites for both novices and experts) is widely accessible to Internet users.
Social networking platforms host research communities, patient discussion forums, crowdfunding campaigns and lobby groups that strive for medical progress in one way or another.
Digitalization includes wearable devices that allow individuals to monitor or track the functions of their body~\autocite{Lupton.2012}\footnote{Lupton mainly discusses the implications of constant surveillance through mobile health trackers on the individual subject and society.}. It also encompasses gadgets and technical devices that record and analyze usage (e.g. \enquote{smart} toothbrushes).
Digital technologies allow stakeholders to actively engage in open discussion, lobby for progress, be involved in patient groups and steer public opinion as well as raise awareness or political attention~\autocite{Petersen.2019}.
The improved access to health related resources and social networks of people affected by a condition is especially important to individuals who are in any way \enquote{incapacitated, immobile and socially isolated through illness or disability}~\parencite[p.3]{Petersen.2019}.
It allows patients to engage actively in periods of near-hopelessness, when survival itself may be at stake~\autocite{Novas.2006}.

This active stance reflects patients' desire to take control and achieve subjectively significant improvements through SCTs, though most of them do not expect miraculous recoveries but merely slight improvements of their conditions.\autocite{Petersen.2014b}.
These technologies eventually provide a commodity to an increasingly large market that trades personal data and infers far reaching conclusions from it~\autocite{Petersen.2019,Tanner.2018}.
Unfortunately, this also includes an emerging black market with medical data being a casualty of data breaches~\autocite{Liu.2015,Tindera.2018}. Research suggests, health care providers were most often breached in the US from 2010 until 2017 (they accounted for 70\%). In this period, the frequency of incidents increased almost every year~\autocite{McCoy.2018}.
Recent regulations have pushed for commercial access to health records through questionable \enquote{empowerment} of patients which will further the dissemination of health-related data. This reinforces the imbalance between institutions with commercial interests and individuals concerned with their health~\autocite{Ebeling.2019}.
Concurrently, major Internet-based corporations tap into the market of health-related products to expand their portfolio. We see platforms like Amazon and Google acquire businesses that grant access to millions of people's health data~\autocite{Farr.2019,Scott.2019} or provide web services to health care institutions~\autocite{RoyalFreeLondon.2017}. Observers predict, operations like those will likely reshape the health care landscape\autocite{Tanner.2019}.
Ultimately, online marketing is an important aspect of the digitalization of the health sector. It allows offerors of health-care services to directly identify potential consumers and approach them in a personalized fashion through data amalgamated from different sources.

The development of health-related online activities is fueled by a new form of patient activism~\autocite{Petersen.2019b} and the \textit{right to try} that was already passed as a law in 36 US-states in 2017. It gave patients who suffered from \enquote{intractable or incurable conditions the opportunity to sample almost any last‐gasp therapy without interference from government regulators}~\parencite{Hiltzik.2017}.
It is assumed that patients turn towards the Internet in search of information and counseling about SCT due to the Internet-based nature of the stem cell tourism industry~\autocite{Master.2014}.
Patients may not be aware of the risks involved in the advertised treatments and ignorant of the information they need to gather~\autocite{Connolly.2014}.

Unfortunately, the Internet is the place, where the \enquote{politics of evidence} enfold, as Tanner puts it~\autocite{Tanner.2019}. He means, that it is hard to obtain reliable information and find credible advice among all the hype stories and anecdotal evidence (in crowdfunding abstracts, patient blogs e.g.\footnote{whose creation is encouraged by clinics themselves according to~\autocite{Ryan.2010}}).
A recent study found that there is a need for comprehensive information and active campaigning of medical authorities and professional organizations to meet the expectations that patients have when they conduct research on stem cell treatments~\autocite{Zarzeczny.2019}.
Some institutions issued advice on this topic to guide patients seeking to try experimental treatments~\autocite{Eurostemcell.2020b,InternationalSocietyofStemCellResearch.2019}
Observers see a rise in crowdfunding campaigns concerned with unproven stem cell treatments~\autocite{Petersen.2019,Tanner.2019} due to insurers refusal to cover expenses for experimental treatments~\autocite{Snyder.2019}. This way, they can circumvent scrutiny by professional medical institutions.

Stem cell clinics and affiliated businesses also list their treatments on popular platforms that register clinical trials\footnote{e.g. \url{https://www.clinicaltrials.gov}} to promote their unapproved therapies. Researchers found out that most of these studies are lacking scientific, ethical or regulatory review, charge patients for participation and are conducted with an unjustifiable risk~\autocite{Turner.2017}.
These phenomena amplify the narrative of stem cells treatments being a novel and universal cure and falsely grant them a scientific character.
On top of that, it enables marketers and providers of unproven treatments to advertise directly to consumers, circumventing regulation, expert review and professional oversight\autocite{Petersen.2019}.

For patients with severe conditions this poses a threat as they may be lured towards unproven treatments in the best case and fake medicine or dubious practices in the worst case. All of which come with possibly disastrous consequences such as physical harm, psychological distress and financial loss for the patients themselves or their caretakers~\autocite{Amariglio.2009,Nagy.2010,ODonnell.2016,Lysaght.2017}.
Furthermore, this leaves responsibility in the hands of a layman as patients must judge the validity of cutting-edge technology and emerging medical therapies with their limited understanding of the subject~\autocite{Petersen.2019}.

Providers of questionable SCT argue that freedom of choice and patient autonomy can be achieved through direct-to-customer marketing. However, they disregard the idea of informed-consent if they assume patients to make decisions based on unreliable and implausible claims or tokens~\autocite{Turner.2018}.
Online communication of SCTs mostly lacks medical information and truthful disclosure about a treatments details and efficacy~\autocite{Connolly.2014}.
Clinics and agencies concerned with either travel, advertisement, marketing, health or all of the aforesaid capitalize on tokens of legitimacy \autocite{Sipp.2017}.

The mostly private companies claim to be registered or certified in some way, assure the absence ethical or health concerns, refer to experts in charge and memberships in professional organizations and provide both testimonials and publications~\autocite{Munsie.2017,Lysaght.2018}.
On top of that, partisans often downplay risks, ignore warnings and do not emphasize patients' informed consent~\autocite{Master.2014,Enserink.2006,Ryan.2010}.
The advertised therapies themselves usually lack clinical trials, evidence of safety or efficacy, thorough patient information and follow-up and a clear process description\autocite{Enserink.2006,Ryan.2010,ODonnell.2016}.
They rarely publish actual data about their processes and the success rates \autocite{Gilbert.2018}.
The studies they do point to, are generally poorly conducted with respect participant structure and study design \autocite{Turner.2018}.
Since the questionable businesses involved in stem cell tourism use similar advertising techniques like legitimate medical authorities and facilities, it is hard for patients to distinguish malicious from lawful \autocite{Sipp.2017}.
The tokens of legitimacy listed above have a persuasive influence on patients that seek treatment \autocite{Snyder.2018}.
This probably explains why online direct-to-customer marketing is becoming the channel of choice when it comes to medical advertising \autocite{Mackey.2015}\footnote{\autocite{Mackey.2015} shows that marketing expenditure for Internet direct-to-customer marketing doubled from 2005 to 2009}.

The direct-to-customer marketing of SCT is seen as problematic as it leverages a narrative of hope, rides the hype of regenerative medicine and is mostly based on anecdotal success stories~\autocite{Enserink.2006}.
Medical travel meanwhile arrives at a new scale since it became a competitive online-based market with willing customers that seek health services abroad. Some of these services are experimental procedures in less regulated environments as well as treatments exclusive to only a group of patients~\autocite{Whittaker.2010,Hiltzik.2017}.
This industry has already flourished in the last years, with marketing and clinic networks spanning around the globe featuring hundreds of clinics worldwide~\autocite{Munsie.2017}.
The global market for stem cell therapies (SCTs) is expected to grow by almost 28\% in the next ten years making it a multi-billion dollar business~\autocite{BISResearch.2019}.
It comprises  institutions from travel, advertisement, marketing, health and government~\autocite{Turner.2007}

This \enquote{stem cell tourism} is described as an online, direct-to-consumer advertised 
Internet-based industry where patients and carers cross geographical or jurisdictional boundaries to receive stem cell treatments for which there exists little to no clinical evidence of safety or beneﬁt~\autocite{Master.2014,Petersen.2017}.
While the mainstream research community assumed the providers of SCTs to operate from Asia, Mexico and the Caribbean, there is evidence that the market is increasingly served by US firms and other middle men alike who strongly advertise their services online~\autocite{Turner.2016}\footnote{Turner identified 351 businesses engaged in direct-to-customer advertising and 570 clinics offering stem cell interventions in~\autocite{Turner.2016} and 432 businesses and 716 clinics in~\autocite{Turner.2018}, respectively.)}.
Many of those US companies advertise a plethora of unlicensed interventions for sundry conditions, some promote SCTs for more than 30 different diseases~\autocite{Turner.2016,Turner.2018,TaylorWeiner.2015}.
Unfortunately, local businesses involved with stem cell tourism are not yet subject to regulation concerning the therapies they promote the facilitating services they provide~\autocite{Turner.2015}.

Research suggests that people with poor health are not also newcomers to the web but also use it more frequently~\autocite{Houston.2002,Li.2016}. Over the years, this correlations remained stable but overall online information seeking decreased, possibly due to concerns about false information~\autocite{Li.2016}. 
Trust plays an important role in this activity, especially with elder users~\autocite{Miller.2012}.
This raises concern as researchers found that low digital literacy leads to online behavior that entails potential harm. In detail, Gangadharan worries that marginal users\footnote{By \textit{marginal users}, the author means members of historically marginalized or discriminated groups, like poor people, ethnic minorities or other groups at the fringe of society.} struggle with adoption of online activities. They could be discriminated and exploited because they are unable to identify malicious actors and distinguish promotional from organic content~\autocite{Gangadharan.2017}.

Already, we see how algorithms in health care endanger parts of the population. In New York, for instance, black patients were deterred from higher-quality health-care thanks to a biased algorithm that falsely inferred good health from low health-care spending. Contrary to that interpretation, it was bad access and distrust in institutions that made the discriminated groups to spend less on health care and treatments~\autocite{Akhtar.2019,Obermeyer.2019}.
With respect to SCT and medical travel, Turner criticizes in \autocite{Turner.2018} how neither government nor professional authorities (like the FDA) can oversee and regulate the market.
Sipp et al. conclude that stem cell tourism further grows even though scientific communities, media and governmental authorities issue warning~\autocite{Sipp.2017}.
Additionally, scholars point out that the financial and social implications are unpredictable, hence not included in today's discussions on how digital technologies should advance\autocite{Petersen.2019}.
The direct-to-customer marketing seems to be a crucial aspect of this industry. It leverages the insights from data collection and analysis that is enabled by digital technologies like web tracking and computational modeling. It allows SCT-providers to individually address potential candidates for stem cell treatments without publicly exposing their marketing efforts.
The increasingly personalized nature of these Internet services might undermine the notion of a \emph{public} opinion on the subject of SCT as users can be individually targeted with prseudo-informational content~\autocite{Tufekci.2014}.

Thus, some organizations demand to discuss algorithmic accountability (awareness of an algorithm's potential risks) and algorithmic justice (compensation for harm done by an algorithm)~\autocite{WorldWideWebFoundation.2017} before developing socially relevant algorithms in a sensitive field like health care. Below, in \cref{sec:algacc}, these ideas will be discussed more thoroughly.

\section{Governance}\label{sec:gov}
This section elaborates on different approaches to regulation of technologies like stem cell treatments of web advertisement. They are gathered from scholars of various domains. However, their universal applicability can support the analysis of the forming effects that actors in the socio-technical system express.
\subsection{The need for control}
A mix of laissez-faire attitude, unwillingness and wide-eyed astonishment has allowed tech companies to impose their algorithms, packaged in business models onto the world and its populations~\autocite{Mager.2012}. The history of search engine related cases shows that the interests of stakeholders are not aligned with policies and legislation, yet \autocite{Gasser.2006}.
Some authorities responded \enquote{perfunctory}~\autocite{Gasser.2006} to technological progress and deferred policies until the market has already created precedents.
Others embraced the technological advances and implement governance-supporting algorithms, better sooner than later \autocite{Kubota.2019}\footnote{
	\enquote{One significant highlight of these new rules is that the era of algorithm regulation is officially coming [$\cdots$] [A]lgorithms should have values, and they must have the right values. At the same time, algorithms should be rule- and law-abiding}, Zhu Wei, associate professor at China University of Political Science and Law, deputy director of the university’s Research Center of Communication Law~\parencite{Kubota.2019}}.

Some might argue that companies acting as intermediaries should not be held liable for content they republish or host. In the USA, this was integrated into legislation, so firms do not fear prosecution there \autocite{LLI.2018}\footnote{\textit{Communications Decency Act},47 U.S.C. § 230: Protection for private blocking and screening of offensive material \\
	\enquote{No provider or user of an interactive computer service shall be treated as the publisher or speaker of any information provided by another information content provider}~\autocite{LLI.2018}.}.
Supporters advocate this as being the sole way to protect free speech~\autocite{Ammori.2014}. The freedom of expression, they argue, is the foundation that allows platforms to operate on user-generated content, enable bloggers to communicate with their readers and to sustain life of communities that discuss controversial topics~\autocite{EFF.2019}.
In U.S. court rooms, cases involving the editorial characteristic of search engines were generally ruled in their favor, referencing the U.S. constitution's first Amendment and the right to free speech \autocite{Volokh.2011}.

In the meantime, the European Union has taken up a different stance. A voluntary agreement titled \enquote{code of practice on disinformation} was signed by big tech companies and the Union~\autocite{
Schulze.2019}. Along with these self-commitments, the EU wants to \enquote{upgrade liability and safety rules for digital platforms, services and products}~\parencite{Schulze.2019}.
They are willing to force regulation onto technology companies to protect citizens in their member nations \autocite{Ungku.2019}. Germany, for example, imposed significant fines of up to 50 million euros on misconduct or hosting of \enquote{criminal} content \autocite{Faiola.2017}.
Furthermore, the EU crafted the far-reaching General Data Protection Regulation (GDPR) to theoretically grant the right be informed about data collection to users~\autocite{EuropeanParliament.2016}.

Observers notice in the press how regional legislation (here: the aforementioned European advances) have an international influence on how services are provided in other countries. Thus, a reevaluation of an algorithm that was initiated due to local regulation often disseminates. Local adjustments leads to global adoptions.
This way, a public discussion about the suggestive nature of Google's autocomplete feature had ramifications for the global application of the algorithm, for example~\autocite{Dickey.2017}. The public was not informed of whether this was a matter of precaution or simply a measure to avoid multiple versions of code.
This shows that it is worth to scrutinize and question algorithms, as beneficial effects are contagious.

Scholars claim, it first takes a scandal pertaining to data privacy or discrimination for public to take notice, media to report or government to act ~\autocite{ONeil.2017b}.
But instead of precipitant legislation, scholars demand an open discourse and common understanding of values and policy objectives. Accordingly, this should steer discussions on regulatory strategies and yield sound policies that govern in agreement with all stakeholders~\autocite{Gasser.2006}.
In \cref{sec:digihealth}, it came clear that the realm of proprietary SRAs needs some sort of governance. The paragraph above showed how legislation can attempt to regulate Internet-based companies (in the case of Germany and the EU) or how they fail to do so due to conflicts of interest (free speech and content control). 

Goldman argues to let intermediaries fix the problems themselves as any regulatory interventions reduced their freedom to improve service quality and adapt to their environment~\autocite{Goldman.2006}. This might pose a problem as they are profit-driven companies that are possibly more concerned with customers than consumers.
Some algorithms disrupt and transform social systems and impose new rules of engagement~\autocite{Kitchin.2017}.
This prompts some scholars to propose that this kind of technological advancement is due to a \textit{technological determinism}~\autocite{Schelsky.1961,Habermas.1968,Mensch.1980}\footnote{
	The idea of \textit{technological determinism} was heavily criticized by scholars like Ropohl, who argued that technological progress can be controlled with appropriate methods. It requires a systemic approach though, as an individual cannot face the challenge alone. Furthermore, Ropohl rejects the idea of the \enquote{best solution} that technology supposedly strives to achieve. He argues that in the multiplicity of stakeholders, this is a naive simplifications~\autocite{Ropohl.1983,Ropohl.2013}.}
that imposes its reign on a society and shapes it accordingly to fit its functional requirements~\autocite{Grunwald.2002}. It infers that humans are doomed to \enquote{say certain words, click certain sequences, and move in predictable ways}~\parencite[104]{Ananny.2016} so an algorithm can anticipate their actions. Accordingly, advocates argue that in this sense, technological progress would strive to a single optimal solution in an almost Darwinian sense~\autocite{Ropohl.2013}. Ropohl immediately rejects the idea and points to the multiplicity of stakeholders and their variety of motivations and goals when it comes to technology~\autocite{Ropohl.2013}. This infers that there are diverse agents interested in shaping technological progress.

In consequence, society has to appoint agents to enforce governance if it does not want to surrender to technological progress that is both uncontrollable and unstoppable (or enforced and dictated by a single actor) as it is destined in the dystopia of technological determinism~\autocite{Grunwald.2002}.
Grunwald adds that society has to consistently reflect on its norms and regulations once it has a learning experience regarding emerging technologies. It must question the motives and intentions of stakeholders and the basis of their decisions. These reevaluations must be premised on the new insight that entail technological progress~\autocite{Grunwald.2002}. Black Box analysis are a possible tool to source these insights and fuel discussions on the subject of technology evaluation.

To sum up, herein this idea is rejected due to two reasons.
First, as described in \cref{sec:sts} and \cref{sec:appl_sts} a social system uses components of a technical system to facilitate its communication. It negotiates what technological advances it considers necessary to support its communication and freely decides what to include in its self-description. Thus, it has capability to shape the human-computer interaction that it integrates in its communication processes.
Second, as shown in \cref{sec:algacc}, the emergent behavior of algorithms can be accounted for by an agentic swarm. Its constituent actors make discrete design decisions concerning the technical system based among others on laws and norms.
Moreover, as pictured in this chapter, these decisions can be subject to a variety of governance forces that have a forming impact. In conclusion, society has the power to form technology in its respective socio-technical system by leveraging the various forces that are capable to shape an object to govern.

\subsection{Proposals of governance}
Due to an increasingly complex and interconnected world, scholars developed a new perspective on governance, that is no longer state-centered and monopolized by institutional authorities.
The \enquote{new} governance is concerned with the collective creation of rule through mechanisms that are not uniquely controlled by governmental agents by an autonomous network of interdependent actors~\autocite{Stoker.1995}.
Kooiman points out how it is more of a process than an entity. From now on, well-being, progress and security can no longer be achieved by one central agent alone. In contemporary societies, he argues, successful governance is a matter of interaction and cooperation between state, private, NGOs\footnote{Non-governmental organizations} and hybrid actors \autocite{Kooiman.2008}.
There is no longer one single authority that dictates and decides but a networked plurality of interdisciplinary stakeholders that engage in cooperation and confrontation and collectively come to a conclusion. The boundaries between traditional institutional actors, private sector companies, citizens and bystanders blur as they are more and more interconnected~\autocite{Introna.2016}.
Nevertheless, Grunwald notes that governance in a democracy has to be legitimized by state actors~\autocite{Grunwald.2000}. However, a government alone cannot achieve this. In \autocite{Grunwald.2000} he elaborates on four aspects that hamper state actors in meeting expectations as serious regulator. The factors are as follows:

\begin{description}
	\item[Knowledge:] In a decentralized and functional diversified society, a state actor cannot assemble all required knowledge to properly govern complex technology
	\item[Orientation:] The state itself cannot represent its citizens' concerns anymore. Instead of for the common good it acts on behalf of its own interests.
	\item[Implementation] In a differentiated society and political landscape, there is no central body of planning, implementing and controlling change that could consistently carry out the transformations.
	\item[Acceptance] Due to the first two problems, explicit and enforced measures will not be accepted by society
\end{description}

The bottom line is that governmental agents cannot solve this issue satisfactory due to the complex nature of the interconnected society and the multitude of stakeholders with contradicting interests. It needs some other sort of governance that is capable to act effectively, legitimately and extensive in both space and time in order to make claims relevant to society without the limits of national laws.

Ananny recommends a multivariate approach to algorithmic accountability. Code transparency, state regulation and user education on their own do not grasp the scope of a socio-technical system, he says~\autocite{Ananny.2016}.
Donzelot, who calls this emerging social tendency that arises in absence of conflict, oppression and poverty \enquote{mobilization of society}, suggests that problems must be solved by society in a bottom-up manner instead of the state implementing solutions top-down. He sees \textit{social partners} to self-manage and resolve issues in a decentralized manner. In this approach, he expects society to accept shared responsibility and find answers in the mutual fruitful conflict that used to be extinguished by states in the past~\autocite{Donzelot.1991}.
The actual government takes the role of a \textit{meta-government}, coordinating and stimulating discourse.
This perspective allows us to think of the entirety of society as an active body of citizens that engages in molding its future because it is aware of its own needs. It seeks confrontation with other agents and is willing to negotiate the processes that affect them.

In \autocite{Lessig.2006}, Lawrence Lessig labels these stakeholders and draws a framework of four \enquote{regulators} shaping governance of Internet-based agents.
Although they are distinct forces, they are highly interdependent. Not only can they shape the object of regulation, but they also affect how other forces behave through their interdependence. 
\begin{figure}[htbp]
	\centering
	\includegraphics[height=.5\textwidth]{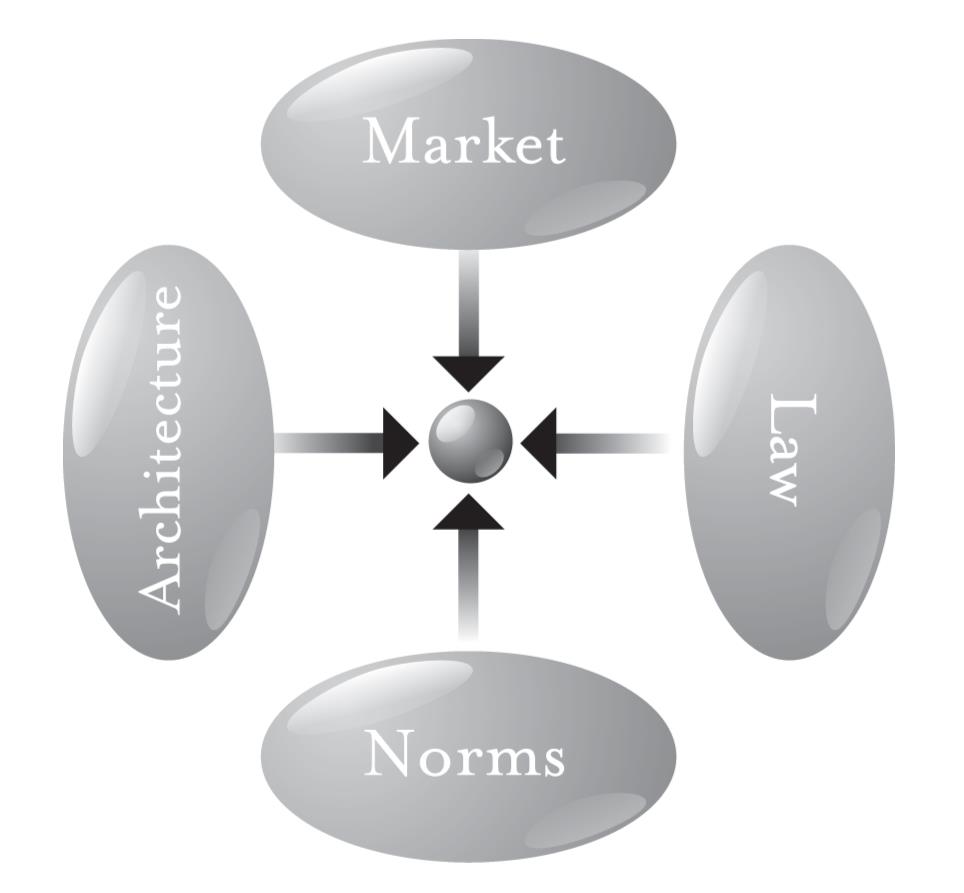}
	\caption[Lessig's four forces]{Lessig's four regulating forces, from~\autocite[123]{Lessig.2006}}
\end{figure}
\begin{description}
	\item[Law] is the state-driven regulator. It is equipped with the most immanent consequences. Misdemeanor entails prosecution and conviction might be severe (see the EU case above). Taxation and benefits incentivize decent behavior.
	\item[Norms] steer behavior through community-imposed punishment. Disregarding these rules (both explicit and implicit) might get an offender expelled from a social group or a company to fall into disgrace.
	\item[Markets] enact their force through supply and the nature of the services and products they provide. They steer through pricing, accessibility and marketing, for example. In doing so, they can restrict access, shape their supply and advertise their positions.
	\item[Architecture]  constitutes the last pillar. Technical infrastructure, protocols and code create a space for communication that is constrained by the limits of hardware and software. Behavior is limited to what is technically feasible and allowed by the programming\autocite[120ff]{Lessig.2006}.
\end{description}

Lessig adds that the regulators above can act indirectly and enact their power via another force.
For example, Google as a market agent, investing into academia~\autocite{GoogleTransparencyProject.2018,GoogleTransparencyProject.2017}~\autocite{HIIG.2020}\footnote{The Alexander von Humboldt Institute for Internet and Society (HIIG) was founded in 2012 by the Humboldt University of Berlin (HU), the Berlin University of the Arts (UdK) and the WZB Berlin Social Science Center, together with the Hans Bredow Institute for Media Research (HBI) in Hamburg as a partner through an initial donation from Google in the amount of \EUR{}4.5 million (until 2013). This funding was renewed again in 2014 (see below)~\autocite{HIIG.2020}.}~\autocite{Readie.2020}\footnote{Readie \enquote{promotes digital policies that benefit society and drive economic growth}~\autocite{Readie.2020} and sees itself as a network of organizations within the digital economy.},
in order to influence public discourse an thus norms or engaging with political organizations to shape law~\autocite{Vogel.2017,CorporateEuropeObservatory.2016}\footnote{
	The \textit{Google Transparency Project } and \textit{Corporate Europe Observatory} are two investigative transparency organizations concerned with the entanglement of Internet corporations and policy makers or  political authorities. They use publicly available data like meeting records or business reports to find and analyze the interlockings.}.
Google also gives incentives to agents who play by the rule and fear demotion for \enquote{gaming} the system~\autocite{Rashtchy.2007,Yuan.2012}. This allows the platform to shape the Norms and architecture of the web advertisement ecosystem and push customers or other affiliated agents to adapt or adopt a  practice~\autocite{Edelman.2011}.
Ultimately, platforms might also engage in politics, specifically concerning regulation of the world wide web and Internet-based services~\autocite{GoogleOfficialBlog.2012}.
Journalists as well as scholars raise awareness of big platforms' capability to influence offline behavior of citizens. In \autocite{Bond.2012} more than 60 million Facebook users were mobilized to vote\footnote{The researchers found only marginal effects on increased voting willingness but conclude that the experiment only consisted of one message displayed to each user, so it might be extensible.}
This urges scholars to speak against what they call \enquote{digital gerrymandering}~\parencite[p.335]{Zittrain.2014}.
Lessig warns that this heavily undermines credibility and acceptance if done non-transparent.
Analogously, if markets enact their power through opaque code and infrastructure, it creates an imbalance that is perceived as unfair~\autocite{Lessig.2006}.

Drawing from this fourfold forcefield of regulation allows us to put a label on some of the entities in the \textit{agentic swarm} influencing the socio-technical system of web-advertisement.
Law is enforced by the body of government. In legislation, politicians determine fair conduct in the online advertising business by a set of commands and threats. By this, they sketch the values of the respective community and impose punishment on those who disregard them by a centralized authority.
Social norms on the other hand are enacted in a decentralized manner through entities of a social systems like professional associations, advertisers, net activists, citizens, users or cultural distinct parts of the population. They are enforced through societal sanctions following violations.
The market forces are shaped by Internet-based companies and ISEs like Google and their business partners, in this case advertisers.
However, most importantly, the regulation imposed by architecture (code and technical infrastructure) can be attributed to the \enquote{architects of our society}~\autocite{Glaser.2009}, namely informaticians and computer scientists. They shape cyberspace with the values and norms they embed in code. Their structural perspective is molded into technical infrastructure whose performativity or emergence defines the means of communication in a socio-technical system. Therefore, they play an important part in governing Internet-based services, as explained above in \cref{sec:algacc}.

Due to the interdependent nature of these forces, it is hard to assess a net impact of single measures or one regulator as a whole. Nonetheless, it is sufficient to show that computer scientist play a significant role in establishing governance. They are required to contribute their expertise, both domain-specific but also interdisciplinary, as Glaser argued in ~\autocite{Glaser.2009}. Only through their participation, a balance of power can be established and maintained~\autocite{Lessig.2006}.

According to \autocite{Grunwald.2000}, the social partners need to determine five aspects, in order to jointly shape technology:
\begin{enumerate}
	\item An object to shape
	\item Involved actors that are willing to design
	\item Goals and intentions (non-discrimination, privacy boundaries)
	\item Means to influence the formation
	\item Reasonable expectation of success
\end{enumerate}

All requirements can be satisfied to a certain extent with respect to Black Box testing of SRAs as it has been presented in related work and this thesis.
The object to shape is either an algorithm (albeit unknown in its specifics) or the whole web-advertising ecosystem.
The actors willing to do so are researchers that lay their finger on unwanted side effects of those objects, citizens that demand change and politicians who invite the collaborative efforts of all parties to craft a socially acceptable system through legislation.
Herein, members of the society could be integrated as auditors, enacting governance via an auditing platform or participating in a\enquote{bug bounty}~\autocite{Eslami.2019}.
Removing discrimination, harmful bias or ensuring safe conduct on Internet platforms are the common goals of the actors.
The tools and measures used to scrutinize the technical systems and justify change include software and methods like the ones mentioned at the beginning of \cref{ch:relwork} and the outcome of this thesis as presented in \cref{ch:datadonation}.
For the last item on the list, one can only hope to make a valid and convincing case to persuade all involved actors to accept a regulatory measure. The past shows, that this is possible. Apparently, Google is generally willing to make their services safe and sane, as seen in the examples above.

Lessigs model enables us to understand how technical systems can be regulated. It supports to find actors that engage in governance and opens new perspectives on the challenge of algorithm accountability. With this in mind, the academic body, media and citizens can scrutinize SRAs and punish misbehavior and ignorance of common norms accordingly.

\subsection{Challenges}
Scholars saw a rise in attempts to govern search engines over the years.
According to \cite{Gasser.2006} future debates will have to consider a wide array of subjects. Discussions will include
\begin{itemize}
	\item infrastructure (physical and logical characteristics of search engines),
	\item content (free speech and limitations on it, cultural bias),
	\item ownership (proprietary code, indexed content),
	\item security (fraud, safe conduct),
	\item identity and privacy (governmental access, commercial exploitation),
	\item participation (impact on political and cultural processes),
	\item ethics (tension between localized laws and morality of conduct).
\end{itemize}

Furthermore, in \cite{Gasser.2006} Gasser highlights how the high variety of topics poses a challenge for regulators and identifies some key aspects. Social partners must prioritize issues, reconcile policy goals, find appropriate strategies and most importantly find timely solutions that are internationally and interculturally acceptable.
To meet these challenges, Gasser derives three democratic key principles that are generally consistent with ethical concept like human right and agreed upon across cultural boundaries. He suggests guiding policies with informational autonomy, diversity and information quality.
The first comprises free speech, freedom of choice and possibility to participate. Diversity is concerned with variety of information and source thereof. An environment with high-quality information encompasses functional and cognitive as well as aesthetic and ethical dimensions~\autocite{Gasser.2006}. All of these aspects support sound decision-making, for example with health-related issues (see \cref{sec:digihealth}) and should guide a technology assessment like the Black Box analysis.
We can benefit from the insights in this chapter to develop methods of collaborative examination later in this work.

\section{Algorithm Accountability}\label{sec:algacc}
\begin{quote}
	\enquote{\textit{Explainability is a social agreement. We decided in the past it mattered. We’ve decided now it doesn’t matter.}}~\autocite[p.35]{Heaven.2013}\footnote{Nello Cristianini in \autocite{Heaven.2013}. He is with the University of Bristol, UK and writes about the evolution of AI research}
\end{quote}

The purpose of Algorithm Accountability is to assess \enquote{power structures, biases, and influences that computational artifacts play in society}~\parencite[p.3]{Diakopoulos.2015}.
In recent years the field has developed in an interdisciplinary discussion spanning the domains of law, tech, business, sociology and psychology.
In 2017, the \textit{ACM US Public Policy council} came up with the following seven principles to foster algorithmic accountability~\autocite{USACM.2017}.
\begin{enumerate}
	\item Awareness
	\item Access
	\item Accountability
	\item Explanation
	\item Data Provenance
	\item Auditability
	\item Validation and Testing~\autocite{USACM.2017}
\end{enumerate}
The council aimed to encourage algorithm designers to act responsibly, knowing how their choices in algorithmic design can introduce bias and entail harm. They demand them to provide interfaces for public scrutiny, explanations of algorithmic decisions and documentation of data and procedures used in testing and training~\autocite{USACM.2017}.
In this context, an explanation is a \enquote{comprehensible representation of a decision model associated with a black box, acting as an interface between the model and the human}~\parencite[6]{Pedreschi.2018}.

Thus, the goal is to communicate an algorithm's functionality and purpose so that humans can understand it. The explanation needs to be interpretable by stakeholders at their respective level of domain-specific literacy\footnote{Interpretability is defined as the \enquote{ability to explain or to present in understandable terms to a human}~\parencite[2]{DoshiVelez.2017}. The perspective on interpretability from researchers in the field of Machine Learning (\textit{Explainable AI}) is used here because it is equally complex with similarly far-reaching consequences for society.}.
Hence, Algorithmic Accountability strives to establish transparency of algorithmic decisions for the sake of public scrutiny and responsible development that is aware of potential bias.

\subsection{Transparency}
Lessig argues that \enquote{in at least some critical contexts, the kind of code that regulates is critically important}~\parencite[p.139]{Lessig.2006}. By \enquote{kind of code} he distinguishes between open and closed code. Herein, he is concerned with the transparency of its functionality. Transparency, he argues, depends on the kind of architecture and code a computer scientist choses. It creates credibility and legitimacy because users are aware of how the architecture component regulates them~\autocite{Lessig.2006}.
Moreover, transparency enables informed decisions~\autocite{Diakopoulos.2014}.
Hence, critical scholars see the urge to reestablish transparency in domains that require consumers to exercise \textit{information literacy}. This ability allows consumers to \enquote{recognize when information is needed and have the ability to locate, evaluate, and use [it]}~\parencite{ACRL.1989}. Opaque technologies, they argue, hamper this ability and thus harm the credibility and trust that organizations rely on to provide their services~\autocite{Albright.2017}\footnote{
	Though this was meant to apply to journalism, reporting and the dissemination of news, we can clearly see how this can be generalized to search engines and SRAs alike.}.
Naturally, there are limits to open code, especially with respect to proprietary code of private companies. It usually constitutes a trade secret and loss thereof would diminish competitive advantage and put the company at risk~\autocite{Diakopoulos.2014}. On top of that, disclosure would open the gates to malicious actors who arbitrarily manipulate or \enquote{game} an algorithm which would degrade the quality of search or advertising ~\autocite{Bracha.2008,Granka.2010}.

People might turn against algorithms that do not perform correctly in their eyes~\autocite{Dietvorst.2015}.
However, research suggests that users defend or challenge an opaque algorithm, even if they perceive it as biased, depending on whether they benefit from it~\autocite{Eslami.2019}.
On another platform of similar dominance (Facebook), researchers found that ad explanations can be incomplete and misleading. Moreover, they allow malicious advertisers to obfuscate their intention to target sensitive attributes~\autocite{Andreou.2018}.
Dietvorst also showed that if participants observed forecasting algorithms perform, they showed less confidence in its performance. This could have implications about advertising algorithms as well. Irrelevant ads after targeting could disappoint users but transparency about choice of inputs might churn trust in a SRA.

Of course, the more complex an algorithm, the more complicated an informational description gets. To bridge the gap between complexity and explainability scholars suggest a standardized label like the \enquote{Nutrition Label for privacy}~\autocite{Kelley.2009} that allows quick and easy understanding of an algorithm's \enquote{ingredients}.

On this basis, some scholars demand a standardized disclosure of an algorithm's basic aspect.
Diakopoulos, for example, suggests the following in \autocite{Diakopoulos.2014}:
\begin{enumerate}
	\item Criteria of prioritizations, classifications, rankings and associations including their definitions, implementations, thresholds and possible alternatives.
	\item Input and other relevant parameters
	\item False positives and false negatives as well as the method of balancing those two
	\item Training data, potential bias and the ensuing evolution
\end{enumerate}

Nonetheless, there is more to an algorithm's performativity than code. In the case of Google, company values, hiring procedures, hidden labor of quality raters and culture play an important role~\autocite{Bilic.2016}.

\subsection{Responsibility}\label{ssec:resp}
The question of responsibility concerning SRAs in complex socio-technical systems is not trivial.
Letour comes up with the notion of an \textit{actant} describing an artificial actor (like an algorithm or any arbitrary technical entity) that requires a human actor to enact agency~\autocite{Latour.2005}. But once they collectively act, they can only be held accountable together\footnote{Like a human firing a gun. In his sense, both are to be held accountable. The human for pulling the trigger, the gun for shooting the bullet.}. Consequently, this perspective holds all entities accountable that fall in line with the algorithm's purpose. Design decisions, emergent effects as well as interpretation of outputs and ensuing actions are all interdependent and rely on each other.

Introna points out that only through their execution, algorithms have the ability to \enquote{enact objects of knowledge and subjects of practice in more or less significant ways}~\parencite[p.27]{Introna.2016}. Introna uses Law's idea of \enquote{empirical practice with ontological contours}~\autocite{Law.2013} to stress how algorithms perform in the real-world, and have the capability to create entities, rules, norms and social measures.
What they call \textit{performativity}\footnote{
	Introna describes performativity as an \enquote{ontology of becoming}~\autocite{Introna.2016}. In this sense, an algorithm does not exist solely for the sake of execution of its step-wise instructions but for to be \enquote{enacted as such by a heterogeneous assemblage of actors, imparting to it the very action we assume it to be doing}~\parencite[p.23]{Introna.2016}}
stresses how the code is not an end in itself, but it exerts agency through empirical, ontological and normative artifacts that emerge from its execution. These artifacts may have a significant impact on society.
This effect is concerning in a sense that the inscrutable instructions and how they produce their outputs often remain obscure Black Boxes to those who are affected~\autocite{Heaven.2013}.

From the definition of algorithms in \cref{ssec:algo}, it can be inferred that all computational steps as well as inputs and outputs are well-defined. Introna argues how algorithms express a nature of flow, inheriting from prior and imparting to subsequent actions~\autocite{Introna.2016}.
Thus, the specifics of all actions are significant regarding its following practices. Because the operations are interrelated, an algorithm's outcome can never be accounted for or associated with a single act or actor alone. All involved actors partake in design, development, execution and interpretation of the algorithm.
Especially in large and complex SRAs, a heterogeneous \enquote{agentic swarm}~\parencite[p.32]{Bennett.2010} collaborates to creatively construct distributed, sophisticated algorithms.
This collective authorship is motivated by various goals at different times~\autocite{Seaver.2014}.
This creates a complicated and ever-changing structure. Seaver concludes: \enquote{once these systems reach a certain level of complexity, their outputs can be difficult to predict precisely, even for those with technical know-how}~\parencite[418]{Seaver.2014}.

As a consequence, they cannot be judged separately from their development or deployment~\autocite{Geiger.2014}.
However, design of code is not self-sufficient, but it is deliberately determined by programmers.
Observers assume that algorithms of large software systems incorporate values and attitudes of their creators and users through criteria choices, training data, semantics, interpretation and possibly feedback ~\autocite{Diakopoulos.2014,Grimmelmann.2017}.
Even the notion of relevance with respect to search results and personalized advertising is highly subjective~\autocite{vanCouvering.2007}.
Seaver understands these properties as intrinsic parts of culture that will find their representation as technical details in an algorithm's code~\autocite{Seaver.2017}. He further points out that one should especially pay attention to the logic that guides the decisions on algorithmic workings, data structures and methods. Seaver expects them to be more persistent than the technical details.
Thus, assessment of algorithms has to consider their respective \enquote{relational, contingent [and] contextual}~\parencite[18]{Kitchin.2017} features and the socio-technical system they perform in~\autocite{Kitchin.2017}.
This suggests, that no one involved can fully grasp the multitude of purposes, intentions and motivations that a piece of software was built on.

Conclusively, we need to understand algorithms in their respective context and how they are embedded in the social system. Thus, one should not assign agency to the algorithmic actor or the developer of single instructions alone, but rather to the entirety of participants in the flow of actions along its development, deployment and usage~\autocite{Introna.2016}.
In this sense, Datta notes that online advertising is a result of complicated mechanisms and interactions between data collection, user profiling, keyword bidding and inventory auctions. Thus they admit that it is unrealistic to assign blame for a specific ad delivery to a single actor only from external observation~\autocite{Datta.2015}.
Ananny even holds the users accountable since they contribute to the algorithms output through their interaction\footnote{
	He asks :\enquote{[W]ho is the maker and who is its target when algorithms dynamically adapt to the users they encounter? Should users be held partly accountable for an algorithm’s output if they knowingly provided it with data?}~\parencite[108f.]{Ananny.2016}}.
Bilic also notes how their \enquote{free labor} and commodified transactions are an integral part of the STS of web search~\autocite{Bilic.2016}.

In \cref{sec:gov} different approaches of governance to shape SRA were discussed. Lessig's proposal described four forces, one of which was concerned with architecture. This perspective is concerned with algorithms than sustain a socio-technical system. Above, this thesis argues that the collective of creators has to ensure the correct behavior of algorithms. 
Glaser points out how informaticians partly carry responsibility for the radical changes that transform our society today. They encode laws and norms into software and provide infrastructure for society to operate on. Thus, he concludes, they are indeed architects of tomorrow's society and are therefore accountable for the repercussions of information technology on society~\autocite{Glaser.2009}.
Though in this thesis, \textit{informaticians} is used equivalently with \textit{computer scientists}, the latter suggests that professionals and academics in this field are merely concerned with the design and development of hard and software and the networking of computers alone. This reduces the role of informaticians, computer scientists and all IT-professionals to that of technical suppliers.
Unfortunately, this resembles the public opinion, argues Glaser in \autocite{Glaser.2009}. He points out how the portrayal of computer scientists as only being occupied with technical aspects of computation deprives them of their qualification or authorization to evaluate the social or systemic ramifications of their actions due to their supposedly techo-centric world view.
Glaser's insists on repositioning the discipline as a science concerned with structure and communication of technical system. He claims that informaticians' have the ability to identify and analyze structures and mechanisms of technical and non-technical systems (organizational and social) and transform them into computational processes. This competence can be applied interdisciplinary to evaluate and improve socio-technical systems.

\begin{figure}
	\centering
	\includegraphics[width=\linewidth]{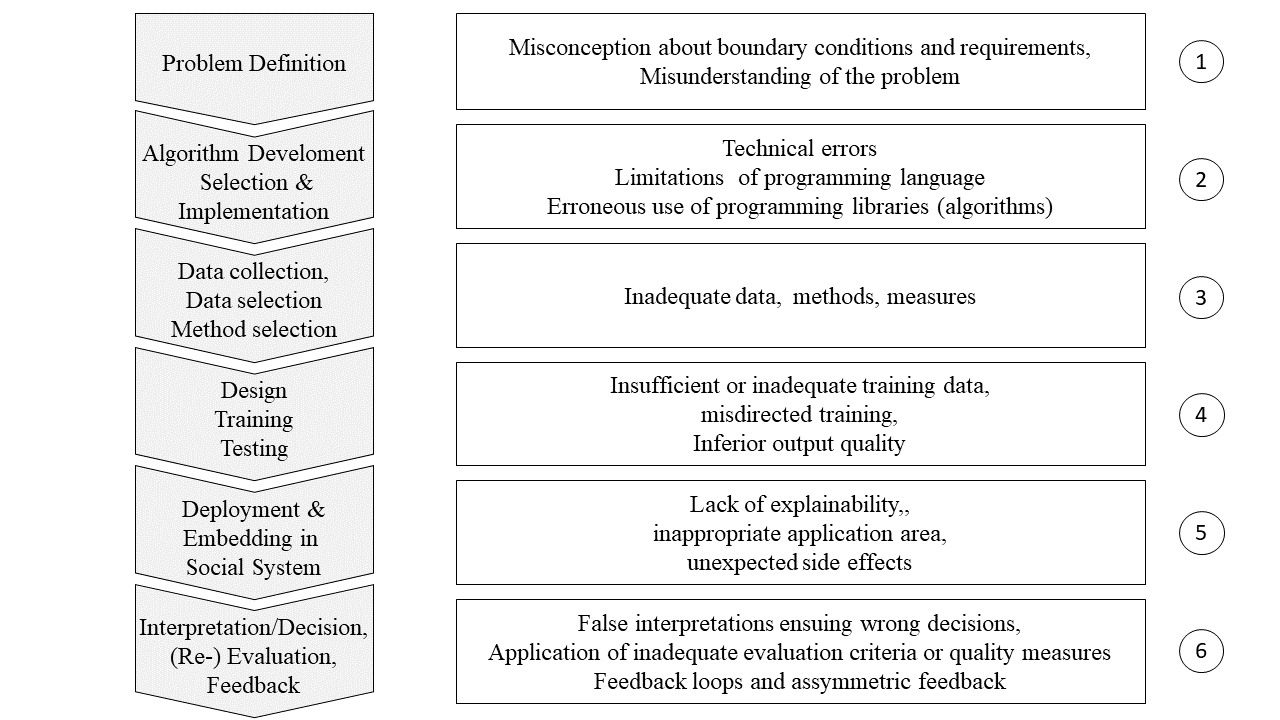}
	\caption[Zweig's Chain of Responsibility]{Chain of Responsibility on the left, possible pitfalls in the development and deployment process in the respective phases on the right, adapted from~\autocite{Zweig.2018} and altered with respect to orientation of the graphic and wording of the pitfalls}
	\label{fig:chainofresponsibility}
\end{figure}

In the view of this, the \textit{Chain of Responsibilities} is introduced to describe pitfalls throughout the lifecycle of an algorithm from development to deployment including evaluation. The concept is drawn from ~\autocite{Zweig.2018b,Zweig.2018,Zweig.2016}. It is adapted to shift the focus from Automated Decision Making towards SRAs in general because both domains face similar challenges, such as a high degree of complexity, an unknown array of (confounding) variables and high significance for those affected by its outcomes.
The metaphor of a chain underlines, how an algorithm can only live up to expectations if all links hold (or can only be as reliable as its weakest link). The similarity to the \textit{waterfall} model of software development is not a coincidence. Errors early in the process are propagated throughout the progress of the development, as subsequent steps are based on their predecessors.  It emphasizes how every actor involved in the development and deployment process is responsible for the algorithm as a whole due its interrelated creation.
On the left side of \cref{fig:chainofresponsibility}, the responsibilities of the respective phases in software development and deployment of SRAs are listed. On the right hand challenges and risks are enumerated. These pitfalls need special attentions in the process of creating SRAs and releasing them into the wild. Below I elaborate on the distinct phases' most important tasks that are introduced in \autocite{Zweig.2018b}, \autocite{Zweig.2018} and \autocite{Zweig.2016}.
\renewcommand{\labelenumii}{\roman{enumii}}
\begin{enumerate}
	\item \textbf{Problem definition:} First, the problem to be solved has to be clearly defined. Here, misinterpretation of requirements or wrong assumptions can lead to misconceptions about the purpose of an algorithm. Especially in multi-causal and interdisciplinary problem spaces, this is a great challenge that requires the cooperation of domain experts from different fields.
	
	\item \textbf{Algorithm development:}
	\begin{enumerate}
		\item \textbf{Algorithm selection:} Failures in problem analysis can lead to misinformed choices of methods and algorithms. Some algorithms may be more suitable to solve the problem than others. Detecting these problems is facilitated by access to code, concise specification with respect to purpose and function and a large user base. For example, what existing code to reuse or which class of algorithms might be appropriate for a certain problem?
		\item \textbf{Algorithm implementation:} The transformation of algorithms into machine-readable code bears the risk of wrong translation or usage of programming language with limited or inappropriate applicability.
	\end{enumerate}
	
	\item\textbf{Data and method selection:}
	\begin{enumerate}
		\item \textbf{Data collection:} Availability, purpose and origin of data can have an impact on data quality, bias and relevance. Data needs to be accessible and usable. On top of that, the method might require a certain sample size to work properly.
		\item \textbf{Data selection:} Developers must determine which subset of data they assume to be a meaningful input to the algorithm. Here, noise or irrelevant data might hamper an algorithm. The choice has to be made regarding the specific problems nature.
		\item \textbf{Operationalization:} The translation of data into informational measures (like relevance) can lead to errors due to misconceptions about certain causations and interpretations.
		\item \textbf{Method selection:} Developers have to come up with an idea of how to solve the problem. Here, misconceptions about a model, its structure and construal, can lead to errors. This includes parameter space, fidelity criteria and intended scope of a method.
	\end{enumerate}
	\item \textbf{Design, training, testing:} Intelligent software systems must be trained on training data that can include biases. Developers ought to determine adequate training parameters and decide whether the data sufficient in quality and quantity to find patterns and draw conclusions. When to end training and testing and how to define success or correctness is another important decision. In this phase it is vital to explora all possible usage scenarios.
	
	\item \textbf{Deployment:} Deployment to a social context entails a learning experience for all its user. It requires them to interact with the system as intended. This requires the system to be explainable. Naturally, some cannot or want not to comply with these demands. Moreover, unwanted effects can emerge from the human-computer interaction. Furthermore, the system could be used in an improper manner or its results could be misinterpreted.
	
	\item \textbf{Re-Evaluation:} The behavior of a software systems and the quality of its output are compared with the expectations it has to satisfy. This feedback can be used to improve the system or detect issues. Feedback loops that reinforce negative effects due to asymmetric feedback~\autocite{ONeil.2017} are a threat to this endeavor. Here, finding an appropriate quality measure is a challenge. Methods to evaluate and analyze software systems are described below, in \cref{sec:blackbox}.
\end{enumerate}

Zweig et al. recommend several measures to solve the aforementioned issues in \cite{Zweig.2018}. They suggest institutionalizing a watchdog authority for algorithms to guide and assess their development and demand professional ethics for data scientists\footnote{
	\textit{Data Scientists} extract knowledge from data. They \enquote{[require] an integrated skill set spanning mathematics, machine learning, artificial intelligence, statistics, databases, and optimization, along with a deep understanding of the craft of problem formulation to engineer effective solutions}~\parencite[1]{Dhar.2013}.}.
Furthermore, input monitoring and Black Box experiments should ensure unbiased foundations and correct performance of an algorithmic system.

\section{Black Box Analysis}\label{sec:blackbox}
One method to establish Algorithmic Accountability is to conduct a \textit{Black Box analysis}. Black Box analysis is a form of reverse engineering\footnote{
	Diakopoulus denotes \textit{Reverse Engineering} as \enquote{the process of articulating the specifications of a system through a rigorous examination drawing on domain knowledge, observation, and deduction to unearth a model of how that system works}~\parencite[16]{Diakopoulos.2014}}
where an opaque system is scrutinized by analyzing observable in- and outputs, deducing the inner mechanics that transform the former into the latter and approximating the inner workings with models. This can be achieved by manipulation and observation of the box~\autocite{Ashby.1956}. The insights are usually juxtaposed to expectations with respect to certain statistics, norms or standards of stakeholders about how the system is intended to work~\autocite{Diakopoulos.2014}. This kind of analysis tries to produce a model (computational or mathematical) of an algorithm.

To analyze SRAs, scholars have made up different approaches. In \cite{Mikians.2012}, for example, crowdsourced user requests were rerouted over the researcher's proxy and captured in a man-in-the middle fashion to examine price and search discrimination. Other work has spawned various software solutions to run Black Box experiments on web search and targeted advertising. Below, some programs to scrutinize the workings of ISE are listed without intention to be exhaustive.
\begin{description}
	\item[XRay] leverages differential correlation to examine targeted advertising and educate users how their input (email, web search, shopping behavior) translates into certain outputs (personalized ads, prices, product recommendations)~\autocite{Lecuyer.2014},
	\item[AdScape] examined user interest based personalization on 175k display ads from 180 websites~\autocite{Barford.2014},
	\item[AdReveal] analyzes targeting mechanisms for ad delivery~\autocite{Liu.2013},
	\item[AdFisher] examines the the relationship between behavioral tracking and Google~Ads and the impact of Google's Ad Settings~\autocite{Datta.2015}.
	\item[AdAnalyst] reviews Facebook's ad explanations and collects data on ads and explanations to give users an understanding of the advertising algorithms and data sources~\autocite{Andreou.2018}. It also examines the advertisers behind promotions and \enquote{measures} the ad ecosystem~\autocite{Andreou.2019}.
	\item[Datenspende Project] crowdsourced data collection for an analysis of SERP personalization during the last German election (Bundestagswahl) with a Browser extension~\autocite{Krafft.2017}.
\end{description}

Most models in this field can be classified in either reverse engineering (Black Box explanation) or design (transparent box design) approaches. While the first is concerned with the general logic of mechanics within the Black Box and an explanation thereof and how outputs correlate with inputs, the latter tries to re-create the outputs of an algorithm with a given set of training data~\autocite{Diakopoulos.2014}.

This work elaborates on the Black Box explanation problem, specifically outcome explanation~\autocite{Guidotti.2018}~\autocite{Pedreschi.2018}. It attempts to reconstruct an explanation of an algorithm from only the output.
In our case, just a fraction of the input was available. We could only collect the information that participants submitted via the surveys they filled out when they downloaded the plugin. Contrary to this, a fully observable In-Out-Relationship would require an API that serves as single source of input~\autocite{Diakopoulos.2014}. But even then, an opaque system may use more than that input. Thus, in our study, the variety of input variables that the algorithms takes into account remain mostly unknown and uncontrollable.

It has no be noted that reverse engineering SRA is a highly complex endeavor, as there is constant feedback from the social system and the workings of the technical system usually are in an ever-changing state. Thus, Seaver argues that in analyzing them, the Black Box algorithm is more of a social construction created by outsiders that differs for each observer as it is influenced by cultural background~\autocite[413 \& 419]{Seaver.2014}.
Herein, Seaver's notion is adopted as he accepts a variety of interpretations to exist.	The attempts to analyze the technical system in this thesis are part of the social system's communication processes and thus can yield different descriptions of the same algorithm depending on which communication processes the Black Box analysis observes\footnote{Theoretically, the best we could do is create an isomorph representation of the algorithm~\autocite{Ashby.1956}.}.
The detailed specifics of an algorithm cannot be determined by observers outside of the Black Box. Eventually, they do not need to be known in their completeness to infer about an algorithm's workings and effects in practice~\autocite{Diakopoulos.2014}. It is sufficient to \enquote{develop a critical understanding of the mechanisms and operational logic}~\autocite[86]{Bucher.2016}.
Rather, the examination should be conducted with focus on relevant aspects only and consider those conditions that are required to understand a  phenomenon~\autocite{Grunwald.2002}. Hence, the Black Box analysis of the web-advertisement algorithms of Google conducted in \cref{ch:datadonation} can be restricted to the question of whether there still are questionable advertisements delivered via Google Ads after the announced policy changed that are harmful to patients. In this sense, it is irrelevant to examine the technical systems of Google's ad exchange and search engine as an integrated Internet-service. Rather, the implications for the distinct social system of patients of Parkinson's Disease, Multiple Sclerosis and Diabetes are of interest.

Nevertheless, the results and interpretations of the analysis can have consequences for the socio-technical system. Ideally, it facilitates understanding of the technical system. It might influence the use and perception thereof among the entities of the social system. This can spark new motivations and communication and a changed behavior of interactions with the algorithm. For a responsible society, methods of algorithm accountability like the Black Box analysis are integrated in their respective self-description, thus into the STS. Hence, this thesis claims that the Black Box analysis itself is an SRA.

In this thesis, we strive for empirical quantifiable evidence of the phenomenon and do not try to create an accurate representation of the algorithm.

\subsection{Methodology}\label{sec:bb_meth}

Ashby \autocite{Ashby.1956} points to the three central questions below that researchers have to consider in a Black Box analysis.
\begin{enumerate}
	\item What is the analysis process?
	\item Which properties can be uncovered, which remain disclosed?
	\item What methods should be used?
\end{enumerate}
In the paragraphs below, these questions will be discussed in more detail.

\subsubsection{Analysis process}\label{sec:analysis-process}

\begin{figure}[bhtp]
	\centering
	\includegraphics[width=\linewidth]{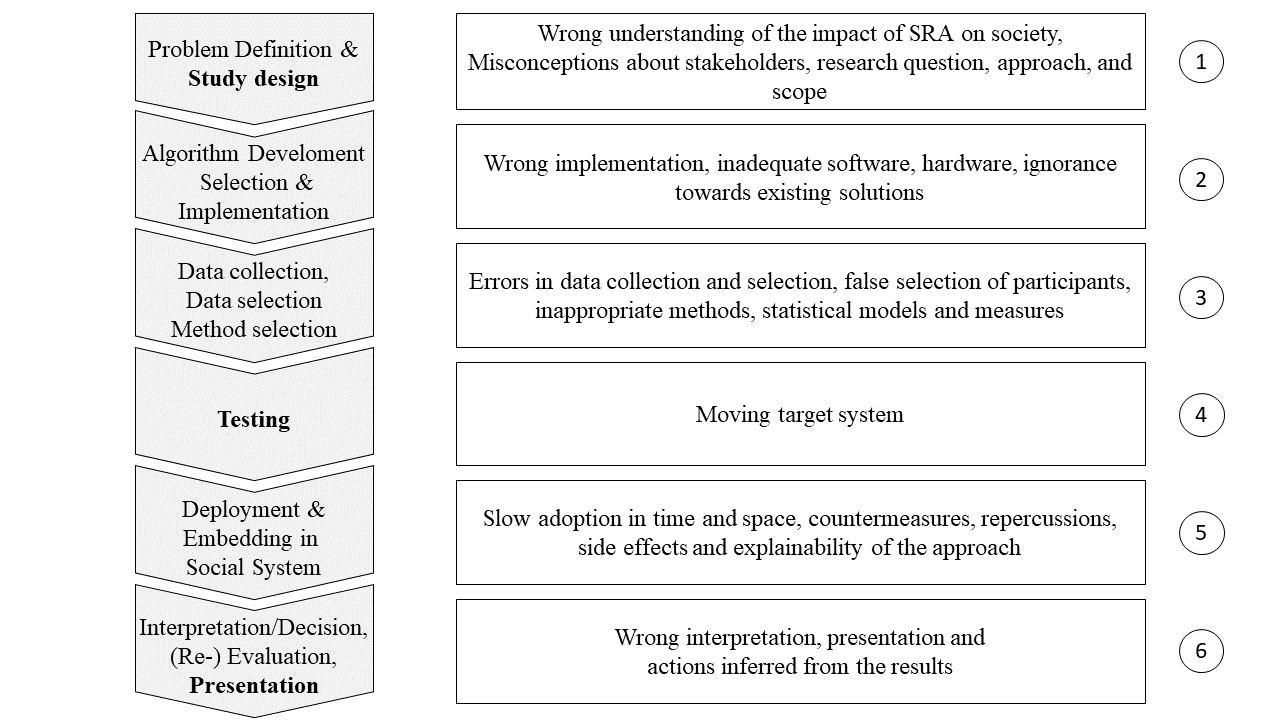}
	\caption[Black Box Chain of Responsibility]{Translation of the Chain of Responsibilities to the Black Box analysis process, own illustration, adapted and altered from \autocite{Zweig.2018}}
	\label{fig:chainofresponsibilitybb}
\end{figure}

As argued at the beginning of \cref{sec:blackbox}, a Black Box analysis can be denoted a socially relevant algorithm as it has repercussions on both the technical and social system. It furthermore facilitates the discourse about the adequacy of algorithmic decisions.
Thus, the Chain of Responsibility from \cref{ssec:resp} can be used to design the analysis process along its axis. Again, \cref{fig:chainofresponsibilitybb} illustrates how each phase of the development process should receive attention according to its specific concern. To the right of each phase, the Black Box-specific pitfalls are listed.
Below, a list of exemplary questions was compiled that can support the execution of a Black Box analysis. They guide the design, development and deployment of a Black Box analysis study and assist in the post-analysis process as well. They are mostly based on lessons learned in the process of conducting the EuroStemCell Data Donation.

\begin{enumerate}
	\item \begin{itemize}
		\item What phenomena emerge from the SRA's deployment?
		\item How are they interrelated and what dependencies exist?
		\item How can the scope of interest be determined and limited?
		\item In this scenario, what is the real impact of the SRA in question?
		\item How can this translate into a testable hypothesis?
		\item Who are the stakeholders that need to be considered?
		\item How should the study be sized in time and space?
		\item How are they affected by the SRA, the Black Box analysis and its outcome?
		\item What are their motives and attitudes towards the analysis?
		\item How can they contribute?
		\item Which resources can be used (crowdsource labor and hardware)?
		\item How to design the study to analyze the Black Box?
		\item What is the ideal scientific approach in terms of efficacy, effectiveness, efficiency and validity?
	\end{itemize}	 
	\item \begin{itemize}
		\item How can the phenomenon be analyzed?
		\item Are there reliable (software-) solutions available?
		\item Are there accessible APIs?
		\item Which hardware is required to conduct the analysis?
		\item What programming approach is adequate in functionality and sustainability?
	\end{itemize}
	\item \begin{itemize}
		\item Which variables are of interest?
		\item Which inputs to the Black Box are observable?
		\item What are the limitations of data collection?
		\item How to clean the data and remove noise?
		\item Which methods are most suitable to approach the problem with respect to data collection and analysis?
		\item How are participants recruited?
	\end{itemize}    
	\item \begin{itemize}
		\item Can the application be tested in a realistic environment?
	\end{itemize}
	\item \begin{itemize}
		\item How can the change of the target system be controlled for?
		\item Could there be countermeasures by the target system?
		\item Does the study need to be adapted?
		\item How to evaluate the quality of the results?
		\item What repercussions and side effects can the analysis produce?
		\item Is the approach explainable and reliable?
		\item How can the analytic process can be guaranteed to be consistent across time and space?
	\end{itemize} 
	\item \begin{itemize}
		\item How to interpret the results?
		\item What implication do they have?
		\item How can the results be presented in a comprehensive and unbiased way?
		\item Are the results actionable?
	\end{itemize}  
\end{enumerate} 

\begin{figure}[htbp]
	\centering
	\includegraphics[width=\linewidth]{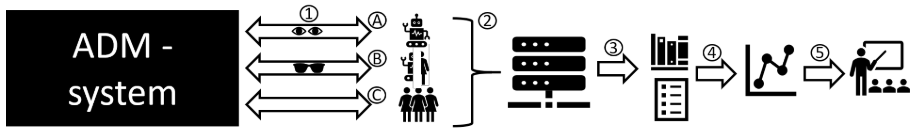}
	\caption[Krafft's Black Box Concept]{Conceptualized process of a black box analysis. The numbers represent the different steps in which errors can occur, from~\autocite[forthcoming]{Krafft.2020}}
	\label{fig:krafftbots1}
\end{figure}

To scrutinize the crucial steps of an analysis, Krafft introduces a conceptual pipeline of generic Black Box analyses in \autocite[forthcoming]{Krafft.2020}. He emphasizes the crucial steps in the process and points out possible sources of errors and misconceptions.
His ideas will be used to assess the EDD along the pipeline (seen in \cref{fig:krafftbots1}) in \cref{ssec:technical}.

According to this, errors can by introduced in probing the system (1) with either a Scraping Audit (1A), a Sock Puppet Audit (1B) or a Crowdsourced Audit (1C). Then, central data collection (2) can fail and data cleaning (3) can degrade quality. The choice of data analysis methods (4) is crucial as well. Eventually, the presentation of the results also needs careful attention (5)~\autocite[forthcoming]{Krafft.2020}.

\subsubsection{Properties}
The nature of the discoverable properties is mainly dependent on the applied method and how the challenges reviewed in \cref{ssec:challenges} can be met. Most times when dealing with proprietary systems, researchers can just assume the inputs and manipulate only a fraction of these variables. Consequently, inferences from the output are mainly based on informed statistics and subject to noise and methodological limits.

\subsubsection{Methods}
Kitchin proposes six different methods of algorithm examination in \autocite{Kitchin.2017}. They are reviewed below to show alternative approaches and why they were not applied\footnote{He notes, however, that \enquote{[e]ach approach has its strengths and drawbacks and their use is not mutually exclusive}~\parencite[22]{Kitchin.2017}.}. Below they are assessed with respect to their applicability in the EDD.

\begin{itemize}
	\item	Examining pseudo-code / source code
	\item	Reflexively producing code from task formulation and design ideas
	\item 	Interviewing designers or conducting an ethnography of a coding team
	\item 	Unpacking the full socio-technical assemblage of algorithms
	\item 	Examining how algorithms do work in the world
	\item 	Reverse Engineering
\end{itemize}
Approach (1) fails at the \textit{Access} challenge as well as the second method (2), which is also impracticable due to the networked nature of the algorithm. Interviewing designer could possibly yield interesting insight in design decisions, constraints and implementation details, but again it breaks down due to access. Reviewing the entire social impact poses a complex problem due to the algorithm being \enquote{performative} (see above) having emergent effects.~\autocite{Diakopoulos.2014}.
As we have not had immediate contact with Google's algorithm designers and unpacking the full socio-technical system of web advertising would exceed the scope of this work, we dropped the first four alternatives. Nevertheless, the real-world effects of algorithms (5) were examined in \cref{sec:digihealth}, possible implementations (1,2 and partly 3) were derived from academic literature and patents in \cref{app:functsearch} and conducted a small-scale study (6) in \cref{ch:datadonation}.
To expand the approaches to reverse engineering, five different \textit{Algorithmic Audits} of opaque Internet-platforms are proposed in \autocite{Sandvig.2014}. All come with distinct advantages and drawbacks. 

\begin{description}
	\item[Code Audit] Code review of proprietary code by expert third parties~\autocite{Pasquale.2010}
	\item[Noninvasive User Audit] Self-reported measures of users' normal interactions
	\item[Scraping Audit] Observing the results of repeated scripted queries to a platform or requests to an API
	\item[Sock Puppet Audit] Programmatically impersonate specific user behavior or traffic
	\item[Crowsdsourced or Collaborative Audit] Recruit real users to collect data
\end{description}

From the approaches above, we merged \textit{Scraping Audit} with \textit{Crowdsourced Audit}. This way, we were not forced to find affected individuals in person to observe for a noninvasive user audit or construct reliable and authentic but artificial user profiles. As Google does not provide an API or discloses code or data for this cause, we had to discard these approaches, too.
The benefits of the methods we applied are natural interaction with the web service by participants with real profiles and the opportunity to get a broad selection of input configurations through a variety of users. The disadvantages of procedurally collecting data from a platform are the risk of detection (and subsequently blocking requests or adapting outputs to them), the possibility of violation of the service's terms of service\footnote{
	Sandvig presumed that under the US Computer Fraud and Abuse Act (CFAA)~\autocite{18USC1030.2008}, any unauthorized access to any computer is prohibited. He argues, that this unnecessarily broad definition would penalize any access to a website unwanted by the provider.\autocite{Sandvig.2014}}
and the lack of fully controlled real-user data as regular Internet user would produce it\footnote{For privacy reasons, we only collected self-reported demographic and statistical data at registration and the eventual submissions}. 
It was the most cost- and time-efficient approach that allowed us to quickly distribute our software and gather data via data donations.

Data donations are an emerging topic in the scientific community and spark interdisciplinary discussions. Scholars make a case for donations as an act of sovereignty that can \enquote{generate social bonds, convey recognition and open up new options in social space}~\autocite[48]{Hummel.2019}. This way, patients can be involved in scientific progress and be invited to take an active stand enacting their autonomy on behalf of solidarity~\autocite{Prainsack.2019}.
We also faced the challenges of data donations with respect to trust, future use, invasiveness, affected people and voluntariness pointed out in~\autocite{Hummel.2019}. To do so, we collaborated with a trustworthy institution (EuroStemCell), declared the possibility of future accessibility of the data \autocite{Couturier.2019}. We further minimized invasiveness through reduced data collection and a non-obtrusive software implementation. As to voluntariness, the study was proposed to affected and non-affected individuals alike. In contrast to donations in the purely medical field, the participants were not subject to moral pressure or an alluring expectation of direct reciprocity. After all, this study was not concerned with researching curative therapies but misconduct in online advertising. Concerning the affected people, we did not check whether only the people who decided to contribute donated but also other users of the respective browser. As consent was given at installation, anyone using the browser took part in the study.

The author further admits and accepts the dissonance between the notion of transparency and trust that we established through publication of the collected data and the possibility of uncertain and possibly problematic future use.

\subsection{Challenges}\label{ssec:challenges}
There are numerous challenges to the Black Box analysis of an algorithms though. They range from the most trivial pitfalls to sophisticated technical restrictions and from adversarial efforts to systematic complications.

As scholars have noted in \cref{ssec:resp}, the algorithm cannot be divorced from the conditions it was developed under or the contexts it is applied in. Thus, a wholesome analysis of an algorithm and the effects thereof require an interdisciplinary team of examiners~\autocite{Zweig.2018}. They need to understand not only the technical aspects, the mathematical models or methods but also the domain-specific preconditions and ramifications.
Furthermore, interviews with designers and programmers of an algorithm can be helpful, as Sandvig suggested. After all, their motivations, beliefs, ideas, visions and corporate culture may be weaved into the code.

The inability to analyze clear code is due to the following challenges described in \autocite{Kitchin.2017}. They were enriched with examples and related problems below:
\begin{description}
	\item[Access] Proprietary algorithms of large Internet-based companies are simply not meant to be analyzed from the outside~\autocite{Obermeyer.2019}. It often is a trade secret and disclosure would allow \textit{gaming} the algorithm~\autocite{Diakopoulos.2014}. \enquote{[The algorithms] are designed to work without human intervention, they are deliberately obfuscated, and they work with information on a scale that is hard to comprehend}~\parencite[26]{Gillespie.2014}, concludes Gillespie.
	Eventually, there might be inputs that the algorithm considers but that are not observable, thus not measurable~\autocite{Pedreschi.2018}.
	This relates to the problem of correlation vs. causation, because statistical significance cannot guarantee a causal relation or design intention~\autocite{Diakopoulos.2014}. The origin of an output can remain undetected and an effect might be misattributed to a non-causal source.

	\item[Heterogeneous and embedded] The algorithms of complex software systems are highly interdependent networked algorithmic systems\footnote{\enquote{In fact, what we might refer to as an algorithm
			is often not one algorithm but many}~\parencite[12f.]{Gillespie.2014}, says Gillespie. This reflects the capability of an algorithm to be flexible and adapt to context due to its nature as intelligent agent (see \cref{ssec:algo})}.
	They are embedded in socio-technical assemblages of various types of entities that all may feedback into the system. Their constituent parts are the work of collective authorship, created \enquote{with different goals at different times}~\parencite[418]{Seaver.2014}.
	A plethora of distinguished configurations of an Internet-based service can be A/B-tested and the variety of actors engaging in networked systems make it hard to determine the reason behind marginally different outputs\footnote{
		A/B-Testing is widely applied in web development to assess the efficacy of design changes in either processes or presentation on the feedback of uninformed users by providing slightly different versions of a service ~\autocite{Christian.2012}(Journalistic source, but gives a concise and comprehensible description of the technique).
		}~\autocite{Diakopoulos.2014}. Furthermore, Internet-based services are delivered over a network of multiple middle-men. Routing and load-balancing of traffic make the route of requests and the actual source of an answer opaque~\autocite{Guha.2010}. Thus, it is hard to establish a truly identical experimental setup for two experiments.

	\item[Ontogenetic, performative and contingent] Algorithms of large-scale software systems are constantly changing, either being updated or adapting to context. They are fluid in their manifestations in code and need to be assessed with respect to their \enquote{contextual, contingent unfolding across situation, time and space}~\parencite[21]{Kitchin.2017}. Moreover, they are highly adaptive to the user as they are personalizing their outputs~\autocite{Bucher.2016}. It was early acknowledged that in a Black Box experiment, the examiner and the subject of interest form a system with feedback. Thus, the process of examination may affect the Black Box and thus alter its inner workings, making it harder to reproduce experiments~\autocite{Ashby.1956}.
	Gillespie concludes that the entanglement of algorithms with its audience creates a moving target meaning the relationships are constantly changing~\autocite{Gillespie.2014}. On top of that, the emergent effects of an algorithm can only be assessed with respect to the context it performs in~\autocite{Introna.2016}.
	Drawing from the fact that inputs are unknown, countless and arbitrary and outputs are fluid, contextual and subject to personalization, the real challenge is to find a stable representation of a system and its environment to analyze\footnote{
		Bucher puts it in a nutshell: \enquote{If the Black Box by definition is a device of which only the inputs and outputs are known, what remains of the metaphor when we can no longer even be certain about the inputs or outputs?}~\parencite[94]{Bucher.2016}}.
	
\end{description}

After all, it seems like it is impossible to fully \enquote{unbox} complex Black Box systems. Nevertheless, scholars like Hilgers argue that even with the epistemological limits of the method and the sheer impossibility to deduce all specifics, the analysis still yields insights and allows knowledge acquisition. Even if all we learn is that we need new methods and practices to analyze Black Boxes~\autocite{Hilgers.2011}.

\chapter{EuroStemCell Data Donation 2019 / 2020 (EDD)}\label{ch:datadonation}
As introduced before, EuroStemCell is concerned with educating the public and patients especially about stem cells. They work closely with patient groups, educators, policy makers and regulators to develop material that caters to their respective needs~\autocite{Eurostemcell.2019}. They produce material on forms of treatments, specific therapies, scientific works and commercial aspects. One of their major concerns is to inform the public about questionable applications of stem cells pertaining to Parkinson's disease, Multiple Sclerosis and Diabetes and the respective clinics or providers. Anna Couturier, Digital Manager at EuroStemCell and PhD candidate in Science, Technology and Innovation Studies at the University of Edinburgh  found, along with many other scholars that these agents make use of online advertising, possibly in a targeted manner (behavioral advertising) to market directly to affected individuals.
She contacted us with the intention to scrutinize these practices with respect to the underlying algorithms.

She initiated the EuroStemCell Data Donation project (EDD) to examine online advertising pertaining to unapproved stem cell treatments. As a part of that project this thesis intends to answer whether;
\begin{enumerate}
	\item There is no more evidence of questionable advertising on Google's search engine result page concerning unproven stem cell treatments of Parkinson's Disease, Multiple Sclerosis or Diabetes (I and II), e.g..
	\item Users affected by any of the diseases (Parkinson's disease, Multiple Sclerosis, Diabetes) receive more critical advertisement than members of a control group.
\end{enumerate}

The cooperation between Couturier and the AALAB started in summer 2019. In a two-day workshop, the project's keystones were discussed. Following these agreements, a plugin\footnote{Herein, the terms plugin, extension and addon are used interchangebly} for both Firefox\footnote{\url{https://addons.mozilla.org/en-US/firefox/addon/EuroStemCell-data-collection}} and Chrome browser\footnote{\url{https://chrome.google.com/webstore/detail/EuroStemCell-data-collect/mdlalccnlkekigohghfbifkibgphaick}} was developed along with a Django server that received and stored data. While the server and backend were constructed by a fellow student on AALAB's payroll, the plugin was part of this thesis.

The server counterpart was developed by a fellow student (Roman Krafft\footnote{\href{mailto:r\_krafft14@cs.uni-kl.de}{r\_krafft14@cs.uni-kl.de}})and is not part of this thesis.
The data collection and study design were administered by researchers of EuroStemCell, Anna Couturier\footnote{\href{mailto:Anna.Couturier@ed.ac.uk}{Anna.Couturier@ed.ac.uk}} and AALAB, Tobias Krafft\footnote{\href{mailto:krafft@cs.uni-kl.de}{krafft@cs.uni-kl.de}}.

After development, just before go-live, Google announced in a new healthcare and medicines policy to \enquote{prohibit advertising for unproven or experimental medical techniques such as most stem cell therapy}~\autocite{Biddings.2019} in a  blog post which lets assume that the company was aware of the issue~\autocite{Biddings.2019}.
We expected this change to degrade the quality and quantity of the collected material with respect to our research question (see below). However, data collection ran for about 3 months. Despite the high attention the subject received and the wide reach of EuroStemCell's partnership network, installation numbers stagnated at a low two-digit range. As a result, this thesis' focus was shifted from a quantitative to a qualitative analysis.

\section{The Donation Plugin}\label{sec:pl}

Because of the small scale of the development project, the manageable amount of expected requirements, the time constraint imposed by Google's policy change and the proof-of-concept nature of the plugin, we omitted an extensive documentation of requirements and project planning and in turn used a SCRUM-like approach to development. Below, explanations of the plugins workings are enhanced with screenshots and UML diagrams\footnote{The \textit{Unified Modelling Language} is a popular language to model software systems. Among others, it comprises graphic notations to express structure, activity and flow of software.}.

\subsection{Requirements}\label{ssec:pl_req}

Couturier acted as product owner, the initial product backlog was compiled during aforementioned  workshop (see \cref{app:backlog}).
The software should regularly search Google for keywords, collect content from the SERP and send this to a server dedicated to storing the results.
Its goal was to imitate a user who repeatedly queries Google for specific search terms (see \cref{app:userstory} for the \textit{User Story} of a typical user).
User should experience easy installation and on-boarding and only little disruption in their browsing experience.
Upon registration, a survey should provide statistical background information about participants.
The infrastructure should be scalable and maintainable with respect to updates.

\subsection{Conceptual Design}\label{ssec:pl_concept}
To allow a crowdsourced audit (see \cref{sec:bb_meth}), we decided to collect data via a browser plugin. This way, the study gets easily scalable on the client side. Moreover we could capitalize on the realistic nature of participant's requests as they would engage with Google using their natural browsing profile and behavior.

\begin{figure}[t]
	\centering
	\includegraphics[width=\linewidth]{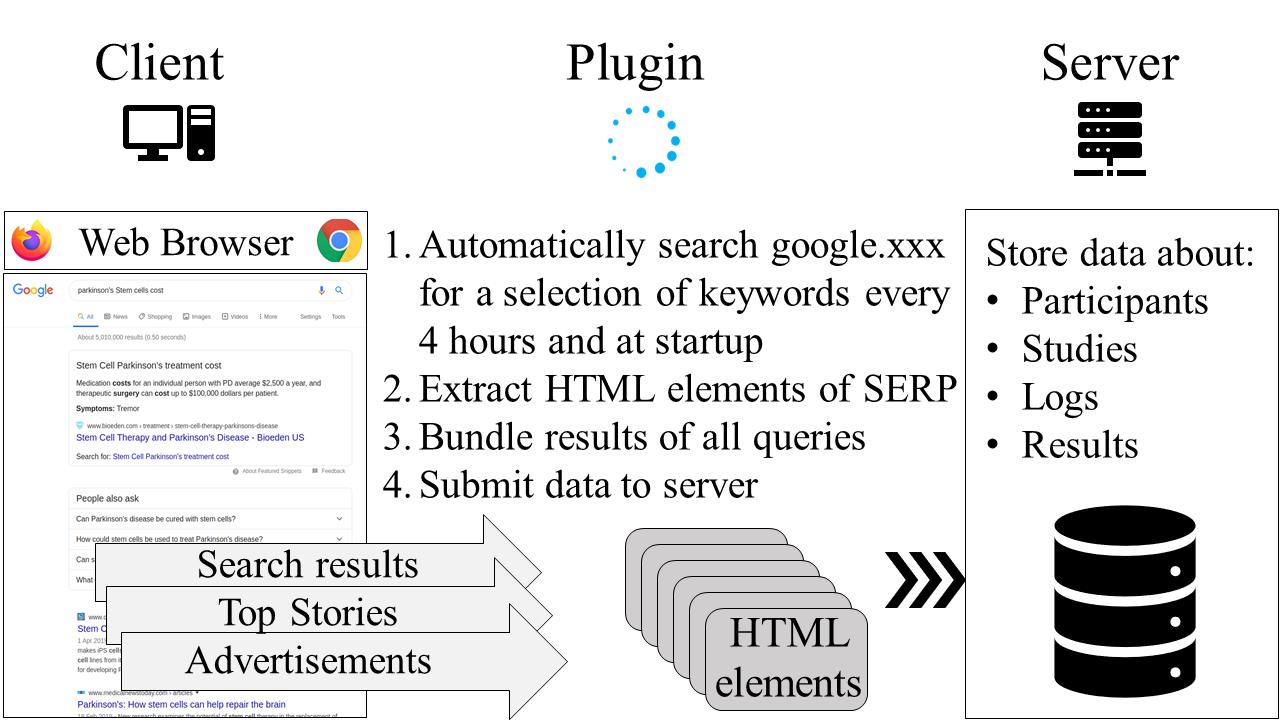}
	\caption[Plugin-Server-Communication]{Sketch of the plugin-server-communication of the EuroStemCell Data Donation, by author}
	\label{fig:plugin-server-communication}
\end{figure}

The Plugin was designed to operate on the current versions of Mozilla Firefox and Google Chrome. They were picked because they cover a majority of users as they are among the most popular web browsers \autocite{statcounter.2019b}.
By using two major platforms, we could benefit from their infrastructure that allowed easy distribution, download and install, uncomplicated updates and possibly gave us an air of legitimacy as being hosted from the official store site.
The usage process was derived from the requirements compiled in the product backlog (see \cref{app:backlog}). It is illustrated in \cref{fig:plugin-server-communication}.

As most participants / donors\footnote{Herein, the notions of \textit{participants} and \textit{donors} are distinct. Participants describe users that only downloaded, installed the plugin and registered, whereas donors are active contributors who submitted their respective collected data.} were assumed to be patients of the aforementioned diseases, thus elderly people with limited technological literacy and willingness to cope with complicated software, we needed to provide an unsophisticated piece of software.
It required a seamless onboarding process and automatic execution with minimal user involvement. Hence, we minimized the number of steps in the registration process and provided FAQs.
Additionally, it should not interfere with everyday browsing and operate in an unobtrusive manner. That is why the collection runs in non-active tabs in the current browser window.
Nevertheless we provided transparency through a utility that showed the recent submissions to give users an idea of how their contribution looked like.

Upon downloading, participants should be walked through a gapless onboarding process.
First, they were to accept a privacy statement\footnote{see \url{https://www.Eurostemcell.org/datadonation\#paragraph-1576}}, then they should be redirected to a survey. Here, we wanted to request information about participants for statistical reasons and to assign them to a study group.
We furthermore planned to gather information to control for frequency of use, domain-specific results (in the case of academic researchers)
Groups should be allocated server-side in a country-by-disease manner plus an additional control group each.
Users impacted by a disease were to be allotted to the respective group, unaffected people were to be used as control. Controls should successively fill the control groups. This would ensure that the users were not scattered among the groups and we could guarantee to provide at least one comparative study.
Their donations should be assigned to a participant and group identifier.

From then on, the plugin should automatically crawl the SERP of Google at browser startup and every 4 hours (starting at midnight). Upon completion, it submitted the collection to the server along with participant- and plugin-related administrative and statistical data (IDs, version, time, language). The plugin queried the Google search engine with terms according to the study group a participant was associated with. We denote the results of the individual queries (searches for keywords) \textit{donations}. The wrapped up collections that were sent to the server were called \textit{submissions}.
Every four hours, the terms were subsequently sent to Google in a randomized order. The plugin requested the website \texttt{https://www.google.\textit{[top level]}/search?q=\textit{[term]}}, where [top level] corresponds to the respective top level domain of a participant group's region and [term] relates to the search terms or query. 
The queries were composed of either a \texttt{[disease]} prefix (\enquote{parksinson's}, \enquote{multiple sclerosis}, \enquote{diabetes}) followed by clinical terms or \enquote{stem cells} in a more general wording (see \cref{app:comp} for details).

\subsection{Development}\label{ssec:pl_dev}
Sprints lasted about two weeks and were loaded with about three work packages each. The software then evolved in a planned manner, as prioritized by the product owner.
Each sprint concluded with a working prototype of the plugin that was critically reviewed by the product owner.
Versioning was ensured on an university-based github repository\footnote{\url{https://git.cs.uni-kl.de/m_reber16/EuroStemCell}}.

Development was guided by Mozilla's online documentation of browser extensions~\autocite{MDNcontributors.2019}.
According to Mozilla's documentation, a browser extension consists of a \textit{manifest file}, a \textit{background page}, \textit{content scripts}, an \textit{options page}, \textit{browser actions} and others.

Additionally, to ensure browser interoperability, the \textit{webextension-polyfill} library was included in my code\footnote{\url{https://github.com/mozilla/webextension-polyfill}, licensed under Mozilla Public License 2.0}. This allowed me to development the Firefox version only. If ported, the library checks the environment it runs in and adapts the code to use \textit{callbacks} on Chrome and \textit{promise}-based APIs on Firefox. This pertains to all functions of the \textit{chrome} and \textit{browser} namespaces, respectively.

Furthermore, the uploaded package included HTML files for on-boarding, off-boarding and overview over submitted results, a privacy statement, a configurations file, their respective CSS and JavaScript files as well as icons for the addon's button.

On- and off-boarding sites comprised informational content whereas the options page contained the mandatory survey. They were plain HTML pages styled with CSS. We used a design similar to the EuroStemCell corporate design to create a feeling of familiarity and leverage the legitimacy of said organization. After all, trust is deemed an important success factor in data donations~\autocite{Prainsack.2019}.

The manifest file declared version number, extension name and other specifics that are required for upload to the browser addon stores in a JSON\footnote{Java Script Object Notation} file format. Moreover, the file details the scripts to run and the required permissions.
We minimized the amount of permissions to increase the acceptance rate for privacy-sensitive users through explicitly stating the domains we intended to crawl.

The \textit{background page} incorporated the main script, the \textit{background script}, that runs once the browser starts if the addon is active. It administers registration, manages communication with the server, keeps track of the study schedule and initiates the crawls. In addition, it checks for updates and loads the configuration file that comprises all parameters for data extraction and server communication. Eventually it contains handlers to process browser-actions that are triggered after a click on the addon's button on the browser interface.

First, the \textit{page-crawl.js} script was developed. It extracts information from HTML elements on the Google SERP according to the respective parameters (see \cref{app:crawl} for detailed descriptions). It receives them upon invocation through the parameters passed by the background script. Finally, it returns the donation to the background-script.

\begin{figure}
	\centering
	\includegraphics[width=\linewidth]{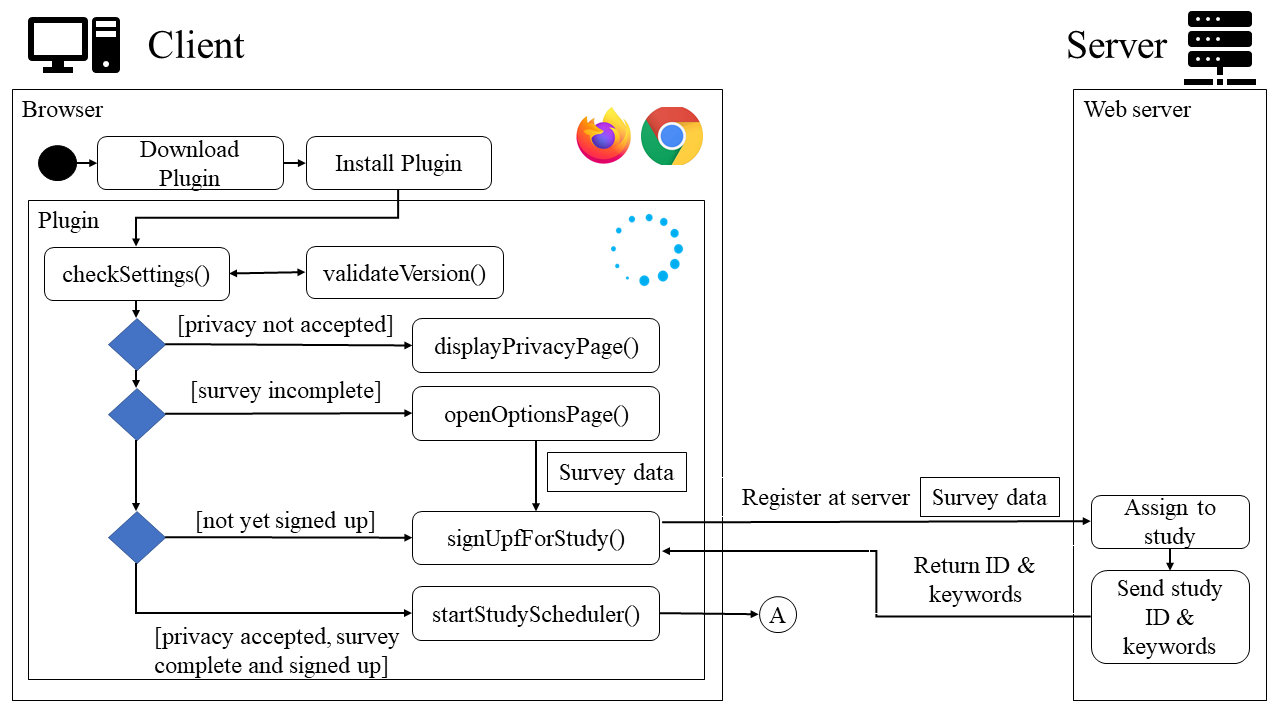}
	\caption[EDD Registration]{UML activity diagram of the registration process, by author}
	\label{fig:umlregistration-process}
\end{figure}

Then the registration process was implemented as illustrated in \cref{fig:umlregistration-process}. At installation users were prompted to read, understand and accept a privacy statement, see~(\cref{fig:addonprivacy}). Users were directed to an options page, where they filled out a survey, see~\cref{fig:addonsurevey}. They were interrogated with respect to health condition, demographics and stem cell-related experiences (see \cref{app:survey} for details)\footnote{Here, we included a question concerned with the participants being contacted by direct-to-marking practitioners of stem cell therapies. We hoped that this would encourage contribution, underline the high topicality of the issue and give patients the feeling that they are seen and their problems are acknowledged. Research suggests that this can assist reconciliation from harm~\autocite{Prainsack.2019}.}.
The client submits this information to the server and registers as a new user there. The server answers with a participant ID, a study identifier and a list of keywords associated with the study (for a detailed list of query compositions, see \cref{app:comp}).

After registration, scheduling was initiated by the \textit{background-script}. The scheduler is also started at each browser startup. First, it executes a crawl, then it uses the \texttt{browser.alarms} API to fire every 4 hours (or more specifically at 0, 4, 8, 12, 16 and 20 o'clock).

To start a crawl, the background page opens a new tab for each search term in the background and injected the \textit{page-crawl} script. After each keyword-crawl, the respective tab was closed, the results were collected and returned. Then, the next result page was requested according to the randomized keyword list. We decided to run the collection in the background to provide a less intrusive experience. The crawl code was injected directly into the newly opened tabs to circumvent the implementation of \textit{content scripts} which would have applied to all requests to Google. That could have been seen as privacy invasion by participants, thus we reduced the scope of the crawl to only those tabs that the extension itself opened.

\begin{figure}
	\centering
	\includegraphics[width=\linewidth]{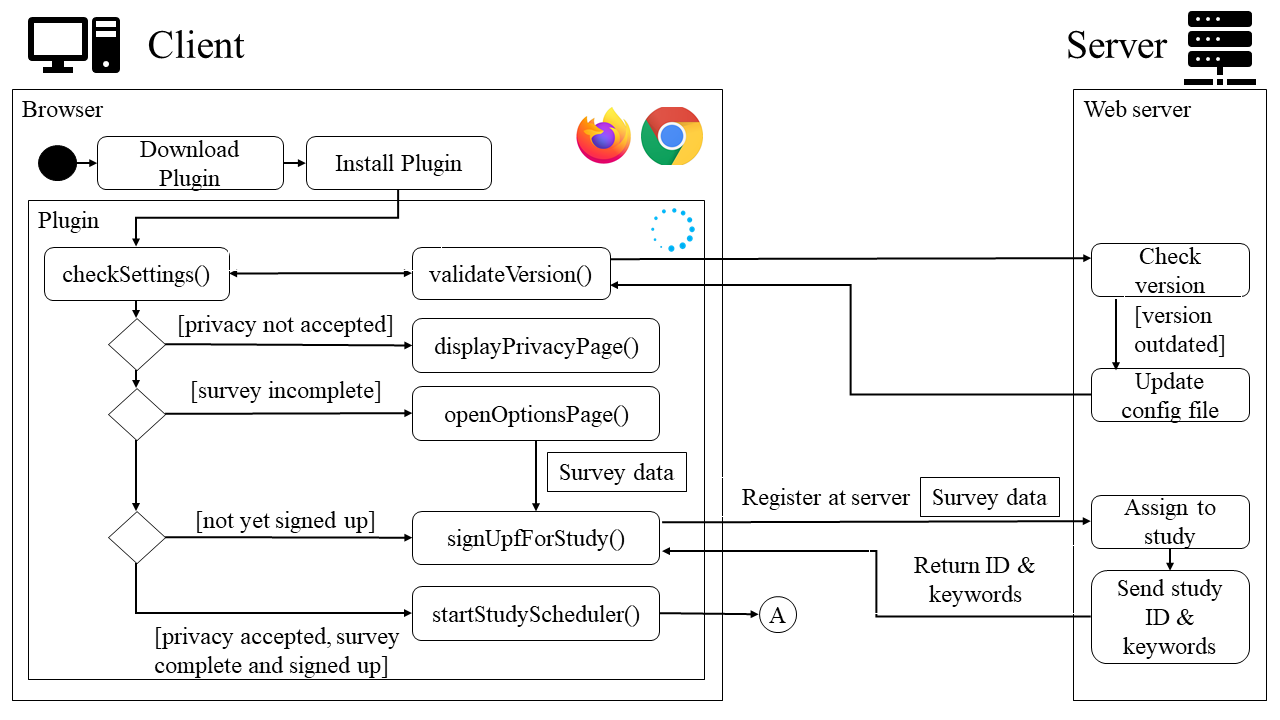}
	\caption[EDD collection process]{UML activity diagram of the data collection process, by author}
	\label{fig:umlcrawl-process}
\end{figure}

The \textit{page-crawl} script returned the results as listed in \cref{app:crawl} to the background script, which added context information like user and study identifiers, packaged them and submitted them to the server.
As we intended to deliver the plugin to different time zones, we decided to include a time zone offset identifier with the submissions. For a sketch of the collection process, see \cref{fig:umlcrawl-process}

The \textit{browser action} button was added to increase both engagement of users and transparency of the addon. While users where not actively participating, the button was styled with an attention-grabbing exclamation mark. Upon click or after the first automated crawl (about 30 seconds after browser startup), the button would resolve to a clean EuroStemCell symbol, if a user was ready to participate. If not, the registration procedure was imitated. If a user was signed up and actively donating, the click on the browser button revealed a page showing recent donations. This was implemented for transparency reasons and to consider the \textit{relationality} of the donated data. This way, we could honor the participants' work through a display of their contributions~\autocite{Prainsack.2019}. 

\subsection{Deployment}\label{ssec:pl_depl}
The first release candidate was uploaded to the addon / extension stores of the respective browsers after testing. 
The upload consisted of packaged code, privacy statements, explanatory screenshots and a \textit{Readme} file.
The raw code was also published and updated on a public repository under GNU GPL v3 for the sake of transparency and reproducibility\footnote{Repository: \url{https://github.com/AALAB-TUKL/EuroStemCell-data-donation}}. 
Following feedback from stakeholders and to resolve issues that came up during production, the plugin was continuously improved in terms of usability, recognizability and stability.
Subsequent updates versions were distributed via the stores update mechanisms.

Then, 13 Virtual Private Servers (VPS) were used to provide region-specific baseline data. Three servers were set up in each of the regions in scope\footnote{Australia, Canada, United Kingdom, United States of America. In the course of the study, one VPS was added in Florida as our partners noted a large density of firms practicing stem cell therapy there.}.
The machines operated on a clean-slate Ubuntu 18.04 LTS and ran Firefox and Google Chrome browsers which would only access Google search websites of various domains (.com,. ca, .co.uk, respectively). The VPS providers were each based in and offered services from one of the respective countries, so we could accommodate for regional effects.
The machines were regularly monitored and updated to assure duly operations.
A server-side logging process was established to give a rough overview of VPS performance. IP logging of only virtual clients was rejected by project partners due to privacy concerns.
Though running on the same specifications, some servers suffered from unexplainable loss of performance while others operated flawlessly.
Hence, operation of the Firefox browsers was switched to headless mode to decrease the processing load.
As the reiteration of the plugin process strained the working memory of the servers which caused the browsers to crash, \texttt{cron-jobs} were isntalled to schedule reboots for the machines and restarts for the browsers. The automatized behavior of our plugin allowed us to initiate the donations computationally.

The overall study period lasted from September, 30th 2019 until March 2020.
After that, an offboarding prompt was delivered via an update of the respective plugins informing the participants of the end of the study and inviting them to fill out an offboarding survey. Finally, they were asked to uninstall the plugin.

\section{Findings}
The donation data\footnote{Data is among the work's uploaded files for the reader's examination.} was downloaded at the beginning of February. Thus, the study period in scope ranges from September, 30th 2019 until February, 11th 2020.
The data was compiled to a CSV file\footnote{
	A Comma-separated value (CSV) is a textfile containing data that is delimited by a distinct separator}
and analyzed with Python after a first evaluation in Microsoft Excel. The illustrations were created using \textit{Jupyter Notebook} in combination with \textit{pandas} for data cleaning and formatting and \textit{matplotlib} as well as \textit{bokeh} for visualization.

\subsection{Data Analysis}
In summary, 162 \textit{participants} registered their plugins on the server. 102 of them were actively contributing. They are denoted \textit{donors}. Of those, 24 were VPS servers automatically submitting data as described above (the VPS represented 23.5.\% of contributing participants). The VPS accounts are addressed by \textit{VPS} or \textit{VPS donor} and the supposedly human donors as \textit{\enquote{real} donors}. \cref{fig:statisticsffdownloadsusersovertime,fig:statisticschromeweekly,fig:statisticschromeregs} in \cref{app:download_stats} show the download statistics of both versions of the extension\footnote{The figures are drawn from the addon stores' statistics dashboard.}.
The store statistics in \cref{app:download_stats} showed that download figures plateaued after the mid of November (a third of the study period).

\subsubsection{Participants}

\begin{figure}[htbp]
	\centering
	\includegraphics[width=0.7\linewidth]{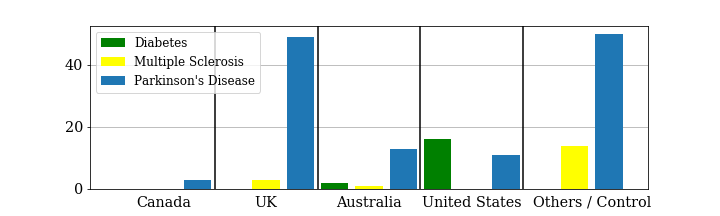}
	\caption[Overview all studies]{Cardinality of all study groups, grouped by region, color-coded by condition}
	\label{fig:ptpsbyallstudies}
\end{figure}

The scope of this thesis was limited to the Parkinson's Disease (PD) study groups because the numbers were too low in the groups concerned with the Diabetes and Multiple Sclerosis conditions, as seen in \cref{fig:ptpsbyallstudies}. The chart visualizes the respective group sizes and shows that there are as low as zero participants in some groups. The study groups were encoded by numbers. Groups 3, 6, 9, 12 and 15 thus accommodated the users affected by PD from Canada, the UK, Australia, the US and the global control respectively. Users who indicated the absence of a relevant medical condition in the survey were assigned to the control. In the following, this thesis will refer to them as \textit{control} or \textit{control group}. These participants were assigned in a fashion that ensured a certain control group size that would allow comparability. The control \enquote{buckets} for each condition were subsequently filled. First, all unaffected participants were assigned to the PD control bucket. After this reached a size of 50 participants, another condition's bucket was going to be filled. We chose to firstly fill the PD bucket as it was Couturier's primary concern to investigate the situation in the realm SCT with respect to PD. 

\begin{figure}[htbp]
	\centering
	\includegraphics[width=\linewidth]{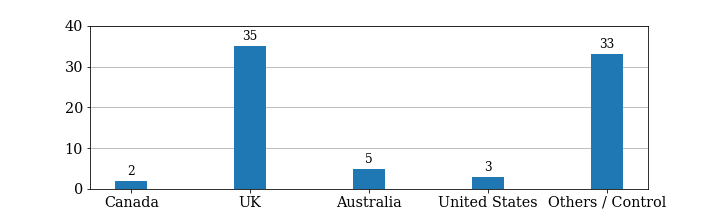}
	\caption[Overview of PD studies]{Numbers of \enquote{real} participants (donors) in the Parkinson's studies (without VPS participants)}
	\label{fig:ptpsbystudy}
\end{figure}

\cref{fig:ptpsbystudy} illustrates the cardinality of the PD study groups. It shows how many real participants were assigned to the respective groups. From this analysis we could have inferred the efficacy of our communications strategy. Because we partnered with medical institutions to promote our cause in the different regions, the numbers would possibly reflect the success of the respective communication strategy. Nonetheless, their numbers were too low to draw statistically significant conclusions.

\subsubsection{Donations}
\begin{figure}[htbp]
	\centering
	\includegraphics[width=\linewidth]{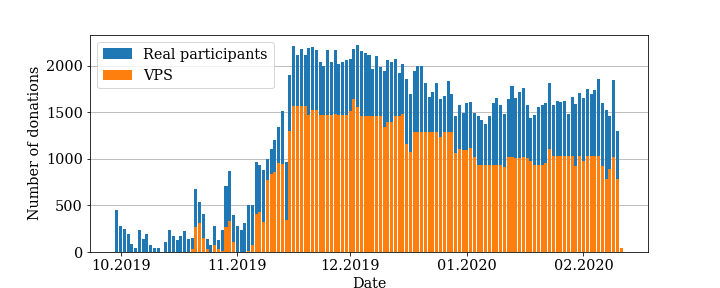}
	\caption[Donations over time]{Total donations of real and VPS donors per day (encoded with blue and orange bars, respectively), from September, 30th 2019 until February 2nd, 2020}
	\label{fig:donationsovertime}
\end{figure}

In the study period, 177,756 donations\footnote{The terms \enquote{donations}, \enquote{submissions}, \enquote{entries} are used interchangeably to refer to the data from an individual crawl that was submitted by actively contributing participants. A crawl denotes one request-scrape-collect cycle of the plugin with one of the study's respective keywords.} were submitted to the collection server. The contributing participants averaged at 21,747 submissions with a median of 105. This measure and the 80th percentile of 3270 donations show how the distribution of donations among donors fits a long tail distribution, thus is heavily skewed. \cref{fig:donationsovertime} shows that the collection server received regular daily donations on a stable level from mid-November on. Although the VPS' submission frequencies may vary slightly as we see in \cref{fig:donationsovertime}, they were a reliable source of donations as they continuously submitted data as planned.

\begin{figure}[htbp]
	\centering
	\includegraphics[width=\linewidth]{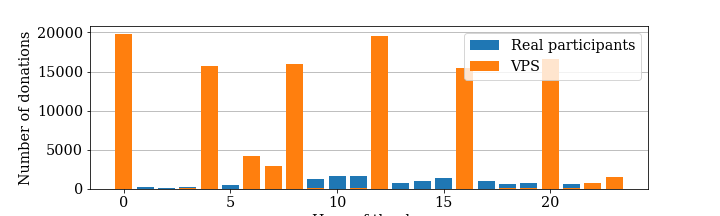}
	\caption[Daily donation distribution]{Distribution of donations over hours of a day}
	\label{fig:donationsoverhours}
\end{figure}

The VPS donors worked as expected, submitting in a recurring manner, as depicted by the regular four-hour pattern of the orange bars in \cref{fig:donationsoverhours}. The contributions in \cref{fig:donationsoverhours} between the scheduled donations show the submissions at browser startup (by real participants, encoded with blue bars) or reboot / restart (by VPS donors, indicated by orange bars). As most \enquote{real} donations were submitted between the four hour spikes (the blue bars in \cref{fig:donationsoverhours}), triggering the initial donation at startup was a vital function for our data collection. This allowed us to capture data even when users were just briefly browsing the web. 

\begin{figure}[htpb]
	\centering
	\includegraphics[width=\linewidth]{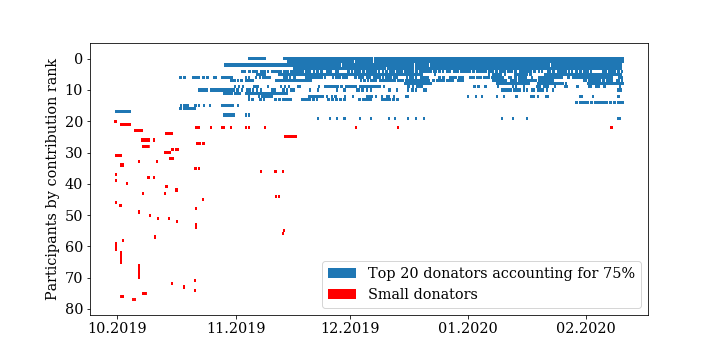}
	\caption[Donation events]{Submission events of real donors over the course of the study. The blue markers indicate the top-20 donators.}
	\label{fig:donevents}
\end{figure}

Many real donors collected only little data, but there are some users that consistently submitted, as seen in \cref{fig:donevents}. The figure shows the individual submissions of each real donor. Each donors contributions are depicted by data points along the x-axis which measures time. The donors are sorted top down by contribution rank. The lower part of the illustration shows that there were about 140 users that only occasionally donated (red data points). The illustration reflects the rise in donation numbers in mid-November, as visualized in \cref{fig:donationsovertime}. Also, we can derive usage patterns from this data that would allow us to validate the self-declaration of users concerning computer- / internet usage. \cref{fig:donevents} further visualizes that there are about 20 donors who account for about 75\% of the donations. 

\begin{figure}[htbp]
	\centering
	\includegraphics[width=\linewidth]{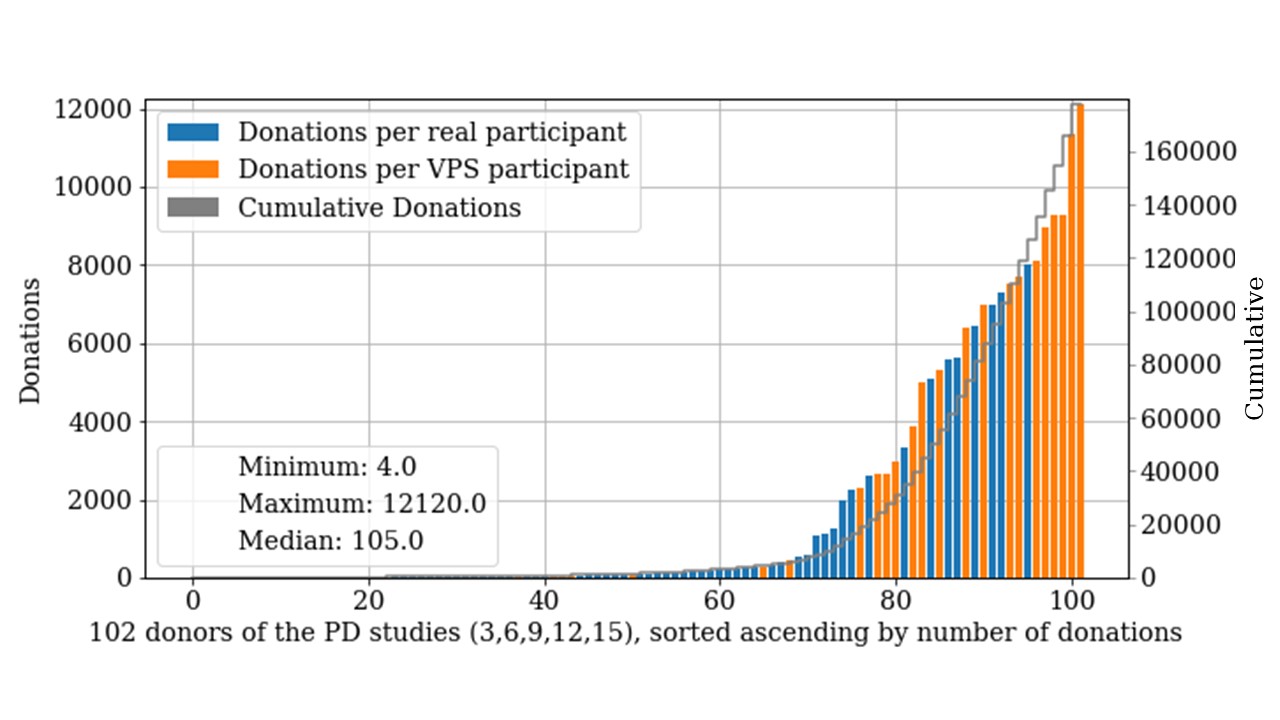}
	\caption{Donations by individual participant and cumulative submissions}
	\label{fig:donbypart}
\end{figure}

\begin{figure}[htbp]
	\centering
	\includegraphics[width=\linewidth]{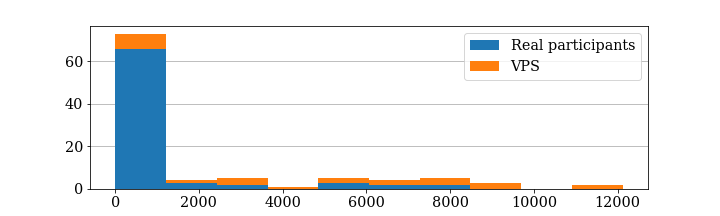}
	\caption{Histogram of donor's donation distributions}
	\label{fig:histsubbydonor}
\end{figure}

\cref{fig:donbypart} supports the lead from above concerning the 20 most active donors. On top of that it visualized the large contribution of VPS donors which amounted to 63.8\% of all entries. However, if we would narrow the research down to the top 20 donors, we would lose many of the real donors. As shown in \cref{fig:histsubbydonor}, the majority of real donors only submitted low quantities, most of them for a very short period of time (as low as a single day, see in \cref{fig:donevents}. 

\subsubsection{Advertisements}
\begin{figure}[htbp]
	\centering
	\includegraphics[width=\linewidth]{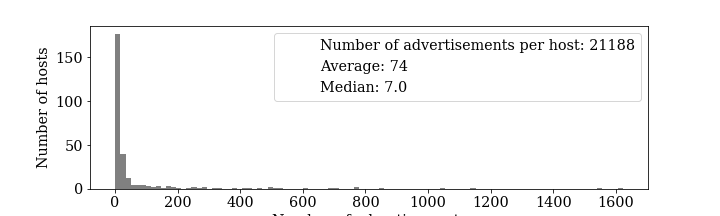}
	\caption[Advertiser histogram]{Histogram of advertisement host distribution by ad count}
	\label{fig:adhosthist}
\end{figure}

Among the 177,756 donations stored at the server, only 5.7\% contained ads. This number is derived by selecting only those entries that contain values in the \texttt{ads} field. As some submissions included more than one advertisement per page, they were extracted, which lead to 21,188 single advertisements. According to the domain of the landing pages\footnote{
	A click on the advertisement directs a user to a landing page. This can happen immediately or via a proxy which enables an ad exchange platform to monitor click-through rates. In the latter case we could seldom capture the destination of a link due to obfuscation (see \cref{sec:lessons}}
about 285 hosts accounted for the paid slots on the SERPs.
\cref{fig:adhosthist} shows that the advertisements on the SERP originate from many small-time advertisers and only few large companies. This is reflected by an average ad-count per host of $74$ and a median of only $7$. $80\%$ of advertisers appeared less than 50 times in the data. This leads to the conclusion that there are many minor players in the field who compete with very strong actors that have significant impact as their ads are regularly delivered and thus dominate the field of advertisements on the SERPs of SCT-related searches.

Because the intricate nature of the stem cell therapy ecosystem and my limited knowledge thereof, I consulted with Anna Couturier, PhD candidate in Science, Technology and Innovation Studies of the University of Edinburgh to assess the background and validity of the ads and their respective promotional messages. Due to her proficiency and experience in the field of science communication and stem cell-related research, she undertook the task of labeling the hosts\footnote{
	\enquote{This masters thesis is contributing to on-going work at the University of Edinburgh through the PhD work of Anna Couturier, PhD candidate in Science, Technology and Innovation Studies. As such, the qualitative analysis of the sources of advertisements is still on-going and will include a detailed coding of the advertising sources according to a number of factors, including relationship to stem cell tourism, potential risk to patients, scientific credibility, and financial impact. The coding included here is a rough "first pass" finding for the purpose of this master's thesis and has been designed to mark potentially problematic advertising sources. These sources have been marked as problematic according to a number of factors derived from an overview of the landing pages. These factors include explicit referencing of scientifically unproven treatments, vague claims about medical outcomes, promotion of stem cell tourism, and targeting of vulnerable patient communities with financial impact.}(Anna Couturier)}. This being said, it has to be noted that the categorization and labeling as well as the distinction of problematic ads do not reflect my educated decision.
 The labels were selected by Couturier to reflect the background of advertisers as it could be inferred from the contents on their website (the advertisement's landing page).\footnote{
 	The complete list of hosts with their respective labels and risk score can be reviewed in \texttt{Ads\_Data\_labeled\_by\_Couturier.csv} and \texttt{problematic\_mapping\_by\_Couturier.csv}. The files are among the uploaded files. Both documents were filled out by Anna Couturier.}

\begin{table}[htbp]
	\centering
	\begin{tabular}{lr}
		\rowcolor[HTML]{FE0000} 
		Most Problematic                                                  & commercial clinic                 \\
		\rowcolor[HTML]{ff8000} 
		\cellcolor[HTML]{ff8000}                                          & clinical trials - private                        \\
		\rowcolor[HTML]{ff8000} 
		\cellcolor[HTML]{ff8000}                                          & clinical trials - commercial                        \\
		\rowcolor[HTML]{ff8000} 
		\cellcolor[HTML]{ff8000}                                          & complementary treatment - commercial                    \\
		\rowcolor[HTML]{ff8000} 
		\cellcolor[HTML]{ff8000}                                          & blood banking - commercial                            \\
		\rowcolor[HTML]{ff8000} 
		\multirow{-4}{*}{\cellcolor[HTML]{ff8000}Quite Problematic}       & health news - commercial                           \\
		\rowcolor[HTML]{ffbf00} 
		\cellcolor[HTML]{ffbf00}                                          & political lobby organization      \\
		\rowcolor[HTML]{ffbf00} 
		\cellcolor[HTML]{ffbf00}                                          & pharmaceutical company            \\
		\rowcolor[HTML]{ffbf00} 
		\cellcolor[HTML]{ffbf00}                                          & commercial non-health specific            \\
		\rowcolor[HTML]{ffbf00} 
		\cellcolor[HTML]{ffbf00}                                          & conference - commercial           \\
		\rowcolor[HTML]{ffbf00} 
		\multirow{-4}{*}{\cellcolor[HTML]{ffbf00}Potentially Problematic} & biopharma supplies                \\
		\rowcolor[HTML]{006600} 
		\cellcolor[HTML]{006600}                                          & \textcolor{white}{health news - public}              \\
		\rowcolor[HTML]{006600} 
		\cellcolor[HTML]{006600}                                          & \textcolor{white}{research institute}            \\
		\rowcolor[HTML]{006600} 
		\cellcolor[HTML]{006600}                                          & \textcolor{white}{blood banking - public}            \\
		\rowcolor[HTML]{006600} 
		\cellcolor[HTML]{006600}                                          & \textcolor{white}{clinical trials - public}          \\
		\rowcolor[HTML]{006600} 
		\cellcolor[HTML]{006600}                                          & \textcolor{white}{conference - public}               \\
		\rowcolor[HTML]{006600} 
		\cellcolor[HTML]{006600}                                          & \textcolor{white}{governmental}                      \\
		\rowcolor[HTML]{006600} 
		\cellcolor[HTML]{006600}                                          & \textcolor{white}{healthcare provider - institution} \\
		\rowcolor[HTML]{006600} 
		\cellcolor[HTML]{006600}                                          & \textcolor{white}{non-profit health organization}    \\
		\rowcolor[HTML]{006600} 
		\cellcolor[HTML]{006600}										  & \textcolor{white}{patient groups}                    \\
		\rowcolor[HTML]{006600} 
		\cellcolor[HTML]{006600}                                          & \textcolor{white}{social}                            \\
		\rowcolor[HTML]{006600} 
		\cellcolor[HTML]{006600}                                          & \textcolor{white}{crowdfunding}                      \\
		\rowcolor[HTML]{006600} 
		\cellcolor[HTML]{006600}                                          & \textcolor{white}{other}                             \\
		\rowcolor[HTML]{006600} 
		\multirow{-12}{*}{\cellcolor[HTML]{006600}\textcolor{white}{Neutral}}                & \textcolor{white}{news}                            		\\
		\rowcolor[HTML]{CCCCCC} 
		Not to determine                                                  & unknown                                \\                                        
		\rowcolor[HTML]{808080} 
		Possibly drugs                                                  & Needs review  
	\end{tabular}
\caption[Host categories]{Advertisement host labels and categorization proposed by Anna Couturier}
\label{tab:host_labels}
\end{table}

Commercial clinics were deemed to be the \textit{Most Problematic} actors, aggressively advertising questionable SCT as it was described in \cref{sec:digihealth}.
The \textit{Quite Problematic} category contains mostly commercial actors that capitalize on patients' conditions through complementary services or are interested in involving them in clinical trials. As described in \cref{sec:digihealth}, private and commercialized clinical trials are that charge for participation are a threat for affected people as they might exploit their dire need for a cure.
\textit{Potentially Problematic} institutions need to be evaluated in a more detailed way. Their influence do not have immediate impact on patients, but pharmaceutical companies and lobbying groups might have an interest in branding keywords or \enquote{framing}~\autocite{Kahnemann.1984} the search domain around stem cell research and treatments (as described in \cref{ssec:bizmodels}). This can be interpreted as the Market-Force of Lessigs regulation framework presented in \cref{sec:gov}.
The entities in the \textit{Neutral} category were deemed unbiased by Couturier in a sense that they would not actively engage in advertising questionable therapies.
\begin{figure}[htbp]
	\centering
	\includegraphics[width=\linewidth]{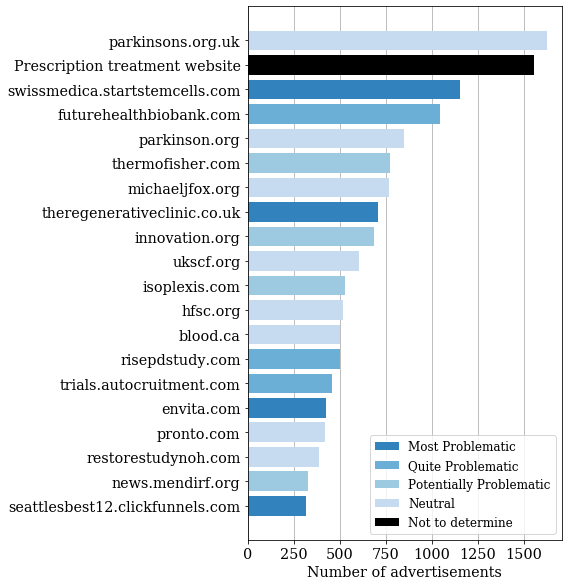}
	\caption[Top 20 advertisers]{Top 20 advertising domains with respective ad count and labeling by Couturier}
	\label{fig:top20hosts}
\end{figure}

\cref{fig:top20hosts} shows that the top-20 of advertisers by number of ads in the data sample comprise many different categories. A multitude of advertisers with varying motives compete for users' attention on the SERP.
This is especially interesting in the field of emerging technologies like SCT where persuasion by commercial actors and lobby interests clash with educational efforts by NGOs and legitimate medical authorities.
Among the top-20 there are 6 foundations dedicated with educating about PD and fostering scientific research\footnote{
	The hosts \texttt{parkinsons.org.uk}, \texttt{parkinsons.org}, \texttt{michaeljfox.org}, \texttt{ukscf.org}, \texttt{hfsc.org}, \texttt{blood.ca} all non-profit NGOs concerned with funding independent research in the field of medical application of stem cell and providing educational material about conditions, clinics, treatments and other health-related procedures.}
The second largest source of advertisements titled \textit{Prescription Treatment Website} accounts for promotion of PD drugs related to Carbidopa / Levodopa\footnote{Carbidopa / Levodopa are two medications used to treat symptoms of PD. They do not alter the progression of the disease.} that are direct-marketed to both patients and practitioners\footnote{This can be inferred from the creatives of the ads that specifically address practitioners}. This shows that not only affected people are addressed but also health-care professionals. Nine providers of drugs were identified in the obfuscated links that direct users to their respective landing pages via an ad network.
Furthermore, there are four clinical trials being advertised among the top-20, three of which are deemed problematic. Apparently, there is also recruitment for clinical trials via online advertising, which may be a hint to the marketing strategies of providers of unproven SCT to acquire customers through ostensible research.

\begin{figure}
	\centering
	\includegraphics[width=\linewidth]{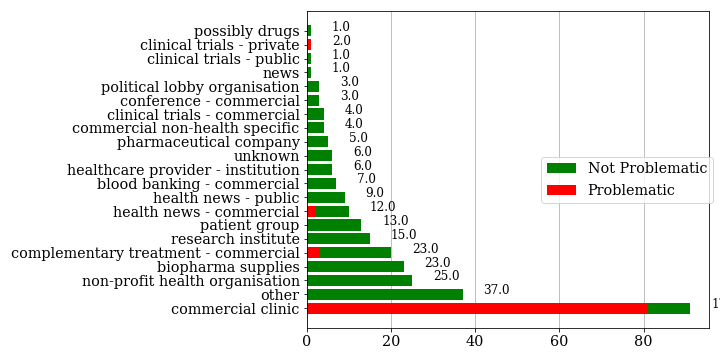}
	\caption[Critical actors in host categories]{Proportion of critical actors in each of the host categories as proposed by Couturier}
	\label{fig:adcatcrit}
\end{figure}

This categorization was further boiled down to a binary classification of \textit{critical} / \textit{noncritical} hosts.
Although the majority of critical actors were in fact commercial clinics, there were also entities labeled as commercial clinics that were not deemed critical, as seen in \cref{fig:adcatcrit}. Additionally, some providers of health news, private clinical trials and complementary treatments qualified to be critical. False claims with respect to treatment efficacy, open promotion of stem cell tourism and claims of applicability of SCT for sports injuries, hair transplants and cosmetic treatments accounted for the categorization as a problematic actor.

\begin{figure}[htbp]
	\begin{subfigure}[l]{0.4\linewidth}
		\centering
		\includegraphics[width=\linewidth]{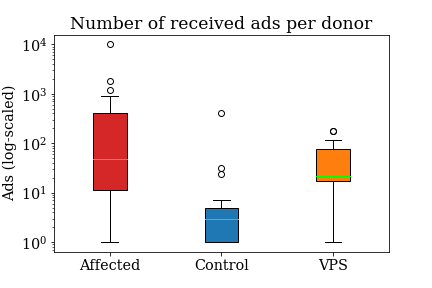}
		\caption[Ads received]{Total number of ads received by participants of each group, y-axis is log-scaled}
		\label{fig:adcountpergroup}
	\end{subfigure}
	\begin{subfigure}[r]{.4\textwidth}
		\includegraphics[width=\textwidth]{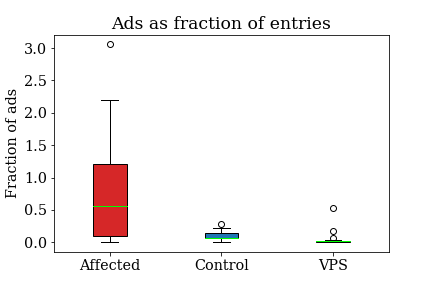}
		\caption[Proportion of ads received]{Advertisements received by participants of each group as a fraction of total entries, y-axis as a proportion}
		\label{fig:adfracper-group}.
	\end{subfigure}
	\begin{subfigure}[l]{.4\linewidth}
		\centering
		\includegraphics[width=\linewidth]{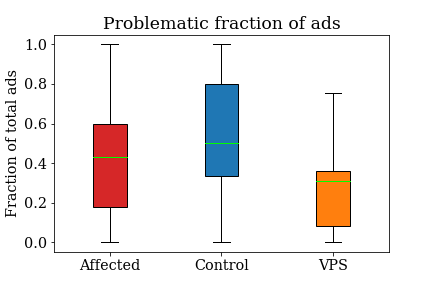}
		\caption[Critical Ad fraction]{Critical advertisements per group as a fraction of total ads received, y-axis as a proportion}
		\label{fig:probfraccountper-group}
	\end{subfigure}
	\caption[Overview of donation statistics]{Overview of donation statistics by group}	
\end{figure}

To investigate the differences between the three groups (affected, control and VPS) with respect to entries, advertisements and critical ads, the donors' entries were grouped by study ID for further analysis. \cref{fig:adfracper-group} showed that users from the affected study groups (study groups 3, 6, 9, 12) received more advertisements in proportion to the number of requests than participants assigned to control or VPS groups)\footnote{
	Note that one entry can yield multiple advertisements. This explains the values higher than 1. Those participants received more than 1 ad on average each time they queried Google.}.
The fraction of critical ads was surprisingly low in the affected groups, as seen in \cref{fig:probfraccountper-group}.
\cref{fig:adfracper-group} raises the suspicion the there is some sort of fitting of the ad delivery algorithm to repeated probing through research as described in~\autocite{Guha.2010}. This can be inferred from our VPS servers in~\cref{fig:adfracper-group} as they see a particularly high number of ads.

\begin{figure}	
\begin{subfigure}{\linewidth}
	\centering
	\includegraphics[width=\linewidth]{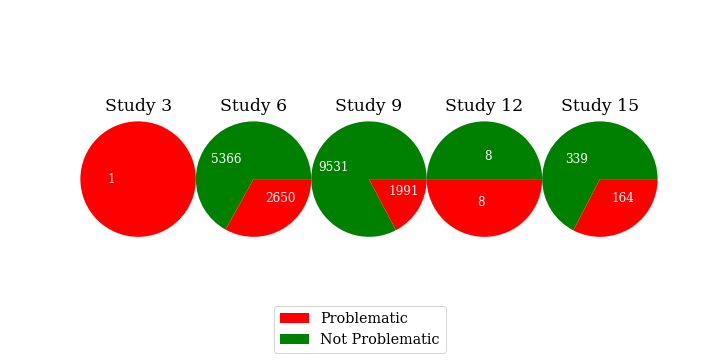}
	\subcaption{Real donors}
	\label{fig:pistudycritreal}
\end{subfigure}
\begin{subfigure}{\linewidth}
	\centering
	\includegraphics[width=\linewidth]{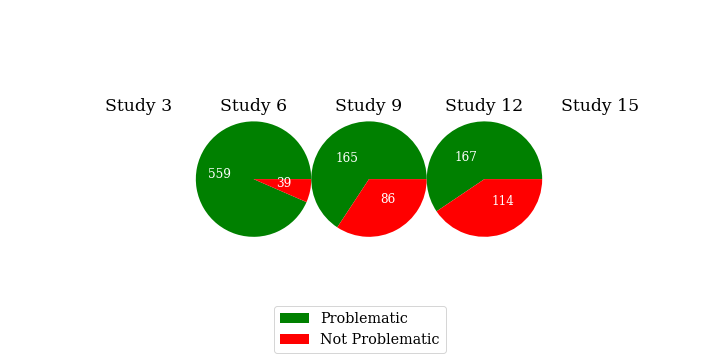}
	\subcaption{VPS donors}
	\label{fig:pistudycritvps}
\end{subfigure}
\caption[Proportion of critical ads]{Proportion of critical ads delivered to real donors, grouped by study}
\label{fig:crit-pies}
\end{figure}

\cref{fig:crit-pies} contrasts the proportion of ads between VPS and real donors. Again, there is no clear sign of targeted advertising between groups or due to real / VPS distinction. This is probably due to the small number of participants and the skewed distribution of contributions. Interestingly, the VPS servers that operated from the UK received the ads and the smallest share of critical ads.

\begin{figure}
	\centering
	\includegraphics[width=0.7\linewidth]{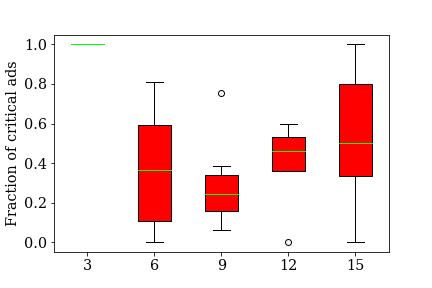}
	\caption[Critical ads by group]{Proportion of critical ads received among all ads, by study group}
	\label{fig:critpergroup}
\end{figure}

A further analysis could confirm this, as seen in \cref{fig:critpergroup}. Study group 3 from Canada can be excluded from this examination as there were not enough donors.
There was no significant difference between the proportion of critical ads among the study groups as \cref{fig:critpergroup} shows and a Kruskal-Wallis Test confirmed.

\begin{figure}
	\centering
	\includegraphics[width=0.7\linewidth]{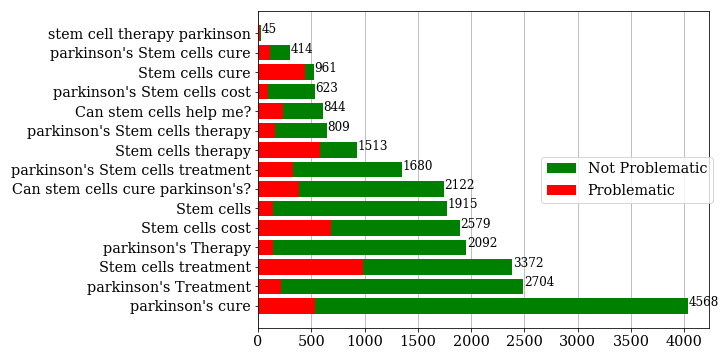}
	\caption[Critical ads by keyword]{keywords}
	\label{fig:pikeywordcrit}
\end{figure}

\cref{fig:pikeywordcrit} shows that keywords associated with PD were not necessarily more targeted as other. Critical advertisers seemed to concentrate on advertising stem cell treatments, therapies and cures in general.

\subsubsection{Detailed inspection of exemplary SCT advertisers}

The host named \texttt{swissmedica.startstemcells.com} was selected by Couturier to be a typical source of problematic advertisements\footnote{
	It might be of interest that their ads were the only ones in our collection whose URL features smileys and emoticons. Admittedly, this is eye-catching.}.
\cref{fig:screenshotresultswissmedica} proved an insightful example of its ads.
It was consistently placing ads in the  course of the study and was the third-largest source of advertisements in this study

\begin{figure}[htbp]
	\begin{quote}
		\textit{Cure with the new technology. Proven results. Higher success rate. The latest treatment. Save \& effective. No side effects. In details! Revitalization. Diagnostic. Post treatment. Stem cells treatment. Accommodation. Treatment for 60 diseases. Higher success rate.} -- swissmedica ad content on Septmeber 30th, 2019
	\end{quote}
	\begin{quote}
		\textit{The latest treatment. Proven results. No side effects. Cure with the new technology. High success rate. In details! Post treatment. High success rate. Dementia. Diabetes 2. Diagnostic. Arthritis. Autism. Multiple sclerosis. Innovative treatment. Treatment for 60 diseases.} -- swissmedica ad content on February 8th, 2020
	\end{quote}
	\caption{Typical examples for swissmedica advertisement creatives}
\end{figure}

Typical example for problematic ads were the ones hosted by \textit{swissmedica.startstemcells.com}
were composed of a certain number of keywords in alternating arrangements. The terms included but were not limited to
\begin{itemize}
	\item \texttt{Proven results}
	\item \texttt{Cure with the new technology}
	\item \texttt{Higher success rate}
	\item \texttt{No side effects}
	\item \texttt{Treatment for 60 diseases}
	\item \texttt{Higher success rate}
	\item \texttt{Save \& effective}
	\item \texttt{Destinations: Switzerland, Slovenia, Serbia, Russia, Austria}
\end{itemize}
These keywords relate to the narratives of the stem cell tourism industry. They usually advertise their treatments as safe, successful, advanced and approved. They furthermore offer international travel and claim applicability for a wide array of conditions.

\subsection{Limitations}
As further discussed below, we could not determine the mechanics behind the targeted advertising of questionable SCT. This is due to the limited number of actively contributing participants and the nature of the data collection. Our approach refrained from large-scale data collection for the benefit of privacy and data security. We did not want to stimulate privacy concerns among potential participants or endanger them through uncertain future use of the published data. Furthermore, we cannot guarantee, that our research had not repercussions on the target system because our intent was discovered.

Since selection processes are at the heart of an ad exchange, the entirety of advertisements in the online advertising ecosystem are subject to rigorous selection. From the plethora of available ads, only few make it to the bidding process due to quality or policy reasons. Then, they are subject to an intricate and opaque auction. In the end, only those ads that have optimal value with respect to user personalization, bidding price, quality, inventory slot and competing content are delivered to a searcher. As a consequence, we can never grasp the entirety of ads related with a subject. We can only assess a subset thereof, in a specific environmental configuration regarding user profile, time and space of a request. In conclusions, there might be ads out there that are highly significant to a research question but there is no way to guarantee that they are eventually being delivered to participants.

The VPS services were locally sourced from Australia, Canada, the United Kingdom and the United States of America. Even though the providers were located there, we could not guarantee that the virtual servers really operated in the respective ZIP codes. We found that some location declarations from the provider deviated from our contractual agreements or details retrieved from a third party localization service. Nevertheless, we cannot expect server farms to located in an average neighborhood, so the IP location probably reveals the artificial nature of a web request, anyway.

Due to privacy concerns we were not tracking Google login status, cookies or fingerprints. To better understand targeting, the insights with respect to user tracking would have enabled an analysis through Google's lenses and control for tracking protection measures possibly employed by users. Also, we could have examined whether users receive different ads and results depending on wehther they are logged in on Google.

\section{Lessons learned}\label{sec:lessons}

In the course of the EuroStemCell Data Donation, the remarks of \cref{sec:analysis-process} were implemented if feasible. Nevertheless there some learning experiences that are described in the following paragraphs. They originate from the review of literature, discussions with peers, the deployment of the plugin, data collection and analysis and the interpretation thereof. Some were of conceptional nature, while others just took time to review and fix. They are presented so future research can built on top of them.
First, learnings concerning the study design are listed.
Then, the individual learnings pertaining to technical aspects are assigned to the crucial phases of Krafft et al.'s Black Box Analysis Process~\autocite[forthcoming]{Krafft.2020}. He describes the critical steps of a Black Box analysis and illustrates how practitioners can fail to conduct a sound analysis. However, his model is only concerned with the actual execution and evaluation of an analysis. Thus, the study design aspects are not including in these assignments.

\subsection{Study Design}
The following thoughts were compiled after the data collection, when some shortcoming of the approach became evident. Herein, the learnings with respect to study design are described in chronological order pertaining to the analysis process depicted in \cref{sec:blackbox}. They describe the two initial steps in the development process as depicted in \cref{fig:chainofresponsibilitybb} and precede the actual analysis, that will be covered in \ref{ssec:technical}.

\textbf{Pre-Study:} A pre-study on existing solutions in the field of Black Box analysis could have facilitated the development of the plugin. \cref{sec:blackbox} gives a brief overview of developments made so far. Most of the Black Box sofware solutions are open source, though some of them operate on outdated browser versions. They provide insights with respect to technologies of browser automation (like \textit{Selenium}) or other libraries concerned with web crawling. However, as the EDD project was forced to quickly deliver a working plugin after the surprising announcement of the policy change, those alternatives could not be reviewed in-depth.
	
\textbf{Target audience:} In the projects introductory workshop, the plugin, the usage scenario (including the search terms) and the typical users were modeled. Usage scenarios, search term formulation and search strategies were discussed among young and tech-savvy academics. However, research suggestes that these properties vary by demographic and motivation~\autocite{Weber.2011,Lorigo.2006}. Investigating real usage scenarios, personal backgrounds of potential users and Internet and technology literacy distributions among them could have supported a more refined understanding of the plugin's target audience.
		
Supposedly, it would have been advisable to consult with representatives of the target user audience which were assumed to be elderly due to the nature of the diseases we covered. The opportunity to connect with them through patient groups associated with Eurostemcell was left untouched due to the constrained time and the geographical distance.
\begin{figure}[ht]
	\centering
	\includegraphics[width=0.7\linewidth]{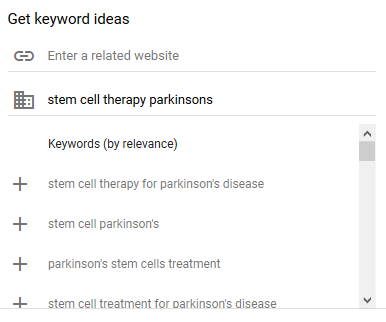}
	\caption[Keyword suggestions]{Keyword suggestions in the Google Ads campaign setup process, screenshot of the Google web interface of Google, by author}
	\label{fig:keyword-suggestions}
\end{figure}
To learn more about popular search terms in a certain field, Google's suggestions can be examined (see \ref{fig:keyword-suggestions}. They are presented in the process of setting up a new advertising campaign via Google Ads. Researchers could infer popular keyword combinations or search queries from these suggestions, as they are probably compiled for advertisers (who aim to maximize reach or efficacy of their ads).	
In conclusion, these approaches could strongly facilitate a more customized development. 
	
\textbf{Organization:} VPS services were ordered and managed abroad from a German location and payed with Scottish credentials, it was a common view to see services suspended due to measures of automatic fraud detection. It probably streamlines organizational processes with respect to payment and management if resources and agency were allocated more closely.

\textbf{Reach:} Distribution and promotion of the plugin and EDD's mission were only conducted via EuroStemCell's network of affiliated researchers and patient groups. We did not try to advertise our cause to other groups that may have been enthusiastic to join. After all, \href{https://twitter.com/search?q=datadonation}{\#DataDonation}\footnote{https://twitter.com/search?q=datadonation} is a thing on social media, and a broadly discussed topic in medicine, sociology, law and, of course, business. There are various NGOs, interests groups and individuals that are engaged with medical data donations and its personal and societal implications. For example, the Hasso Plattner institute recently introduced a \textit{Data Donation Pass}~\autocite{HassoPlattnerInstitut.2020,Schapranow.2017}. It might give research endeavors like this an uplift to connect with like-minded projects and leverage their respective networks or advances in the field of societal data donations. Additionally, this gives the chance to take a participatory role in the development of data donation concepts and infrastructures.
	
\textbf{Crowdsourcing Recruitment:} Since a crowdsourced audit approach was selected, this was the most critical step for the EDD project. Recruiting real-world participants leverages the opportunity to probe a Black Box with real-world user profiles. However, the target audience we meant to address is hard to mobilize, apparently. Though there are reportedly strong ties between EuroStemCell and patient groups, we failed to get affected people onboard. \enquote{A number of patient recruitment events were held including three events with Parkinson's UK, two events with the Anne Rowling Clinic and a number of internal recruitment drives (using mailing lists and direct mailings) with the Australian Stem Cell Network, the University of Texas in Austin, Yale-New Haven Hospital, and the Edinburgh Parkinson's Research Interest Group. However, these events produced more one-to-one structured interview opportunities rather than translation to study recruitment. This may have been due to the demographic targeted as well as the difficulty in translating in-person engagement into digital engagement}(Anna Couturier).
	
\textbf{Education:} Presumably, the target audience consisted mostly out of senior citizens. This can be deduced from the fact that the project is directed at people suffering from Parkinson's disease, Multiple Sclerosis and Diabetes. It also reflected Couturier's experience in this field. She further assumed that these users have low technology-literacy. Thus, educational material or on-boarding guides regarding the plugin and the EDD might have supported the cause. Hands-on trainings or explanatory videos could have boosted adoption. However, these measures must be specifically designed to address the target group and convincingly engage them to join. Unfortunately, this was not in the scope of this thesis as it requires comprehensive analysis of demands, expectations and motivations of the target group as well as an investigation of available methods and their respective efficacy.
	
\textbf{Survey:} The survey was composed for statistical purposes. In fact, the correctness of the submitted data was never controlled. We trusted users to truthfully fill out the survey and not falsify information. Also, this kind of information gathering is a balancing act between invading the privacy of sensitive groups and detailing a user's characteristics which facilitates analysis. On another note, the survey could have been expanded by questions like \enquote{\textit{How did you learn about the study?}}

This would enable researchers us to evaluate the success of your recruitment efforts between regions and among partner institutions and communication channels. In consequences, this would have allowed us to strengthen some bonds and emphasize our efforts to push the EDD to some regions.
	
\textbf{Time period:} It remained unclear whether the time period allocated for the study had any impact on the results. We see that some academics allotted as little time as a week to their study \autocite{Yan.2009,Guha.2010}, while others processed data from longer intervals.
As described in \cref{sec:se}, some web services update on a daily basis, which infers highly volatile algorithms. However, the changes might be so marginal that for narrowed-down research questions it may be unlikely to see an impact. Nevertheless, the longer the study interval, the greater the effect of aggregated changes. This being said, for a snapshot-like investigation of a specific question (like in this case) it could suffice to reduce time and broaden the search effort in this time in exchange (e.g. expand queries, create sophisticated profiles, create more variety among participants). If there is no major advancement in the field of stem-cell related research or a major shift in the web advertising ecosystem, the structure of results supposedly remains stable. Nonetheless, these are interesting effects that should definitely be accounted for.

\subsection{Technical}\label{ssec:technical}
\begin{figure}[htbp]
	\centering
	\includegraphics[width=\linewidth]{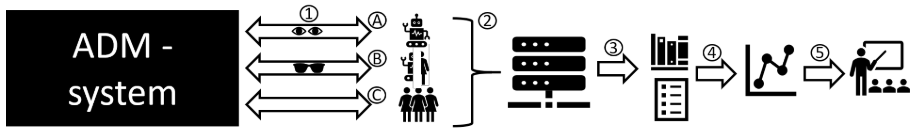}
	\caption[Krafft's Black Box Concept]{Conceptualized process of a black box analysis. The numbers represent the different steps in which errors can occur, from~\autocite[forthcoming]{Krafft.2020}}
	\label{fig:krafftbots2}
\end{figure}

The lessons learned with respect to technical aspects are structured along Krafft's concept of Black Box analyses \autocite[forthcoming]{Krafft.2020}. \cref{fig:krafftbots2} schematically displays the analysis process in the last four steps of the Chain of Responsibility described in \cref{sec:blackbox} and \cref{fig:chainofresponsibilitybb}, especially.

\textbf{(1) \enquote{Fluid} Internet:} As seen in \cref{app:functsearch}, ISEs like Google and other modern Internet-based platforms are in a constant flow. They dynamically adapt their websites to follow trends, update their algorithms daily and improve their services through A/B-Testing. This makes web crawling strenuous, as website structure can change any time.
Thus, it is hard to identify different instances of the same ad. Incomplete information due to real-time auction among several other advertisements, load balancing and network routing may affect delivery~\autocite{Guha.2010}
Thus, if relying on HTML tags, one has to closely monitor the online documents to register changes and appropriately tweak the respective software. A slight change in website (DOM-) structure or naming conventions (element IDs) would have rendered our crawl useless, as no data would have been extracted. This can be countered with storing the whole website.

\textbf{(1 A/B) Infrastructural Limitations:} By using VPS hosts that accommodate numerous virtual systems, there is the risk that an IP range will be blocked by Internet services. This happened at our US-based VPS server location in Dallas, where requests to Google's web search were consistently blocked. Some other locations required us to solve \textit{captchas} to prove the truthful intentions and non-robotic nature of the user. As the servers were meant to automatically deliver baseline results, this turned out to be impracticable as it required constant manual interaction.

\begin{figure}
	\begin{subfigure}[b]{\textwidth}
		\centering
		\includegraphics[width=0.8\linewidth]{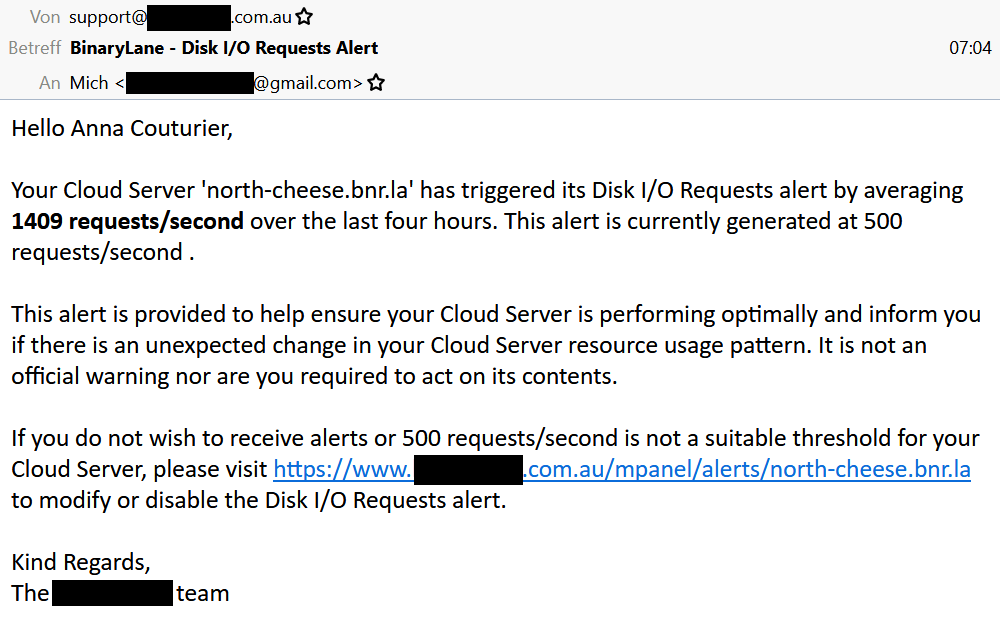}
		\caption{VPS provider alert after a spike of 1409 request per second, screenshot by author}
		\label{fig:ddos}
	\end{subfigure}
	\begin{subfigure}[b]{\textwidth}
		\centering
		\includegraphics[width=0.8\linewidth]{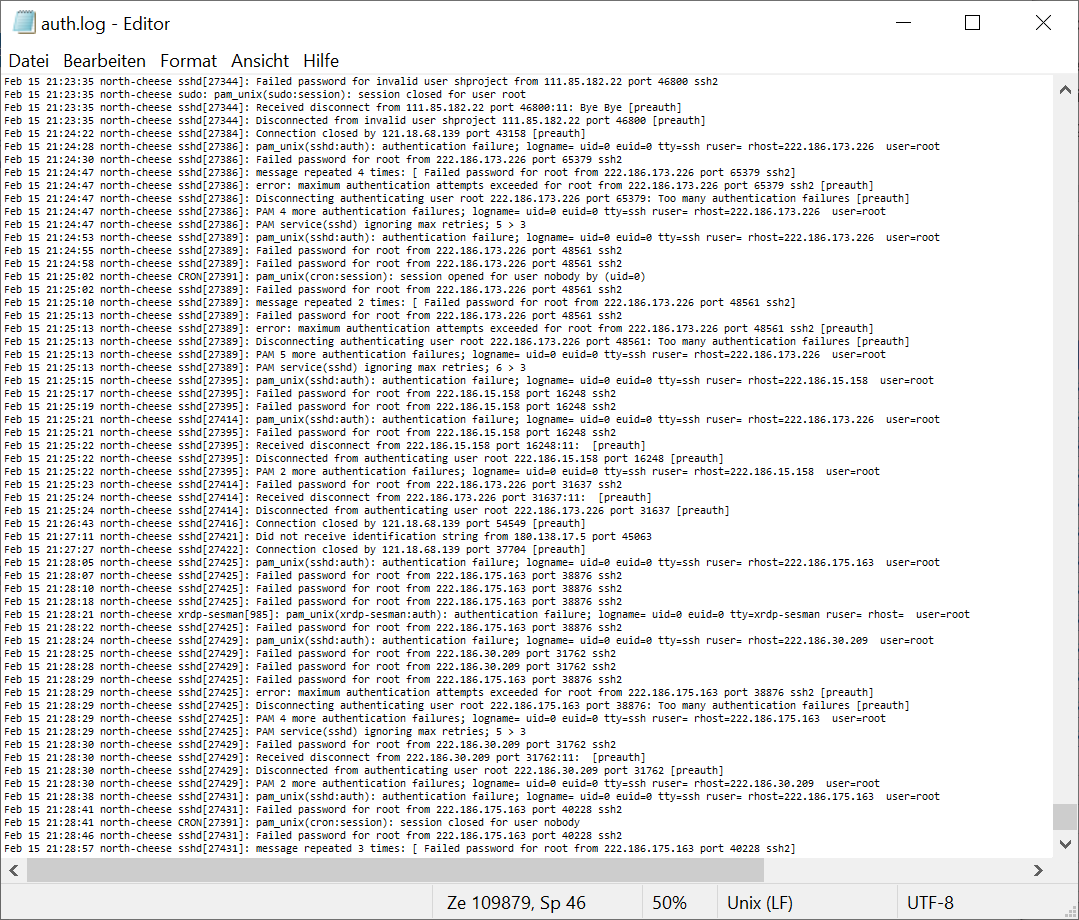}
		\caption{\texttt{auth.log} of the attacked server, screenshot by author}
		\label{fig:ddos2}
	\end{subfigure}
	\caption[Security incidents]{Brute-force attack on one of the VPS in the Australia control group on 15.02.2020}
\end{figure}

\textbf{(1 A/B) VPS Security} Although the VPS' operating systems were regularly updated to the latest version, we received alerts of increased Disk I/O requests during the study on one of the Australian VPS (see \cref{fig:ddos}). As we did not perform recurring high-load operations on these machines, they possibly received malicious attention from the outside. The server logs showed the patterns of a distributed brute-force authentication attack over SSH on almost all of the servers (see \cref{fig:ddos2} for an example log). The server becoming a target of coordinated attacks disqualifies it as reliable control for the study. However, due to the structure of the attacks I assumed that we were dealing with an arbitrary non-targeted online attack with either leaked or widely used \enquote{standard} credentials. As our servers were protected with strong passphrases, they were not shut down. As a countermeasure, we could have used a SSH port different from the standard Port 22, blocked all access from IPs other than the ones on a whitelist, entirely prohibit SSH remote logins or only allowed SSH login via public/private RSA keys. These approaches were rejected, because the problem occurred at the end of the machines' lifecycle. For future studies that use a similar setup, it would be advisable to use a whitelisted VPN server to connect to the VPS, so all stakeholders have access through a protected tunnel. Other than that, enabling authentication via fingerprint is also effective but requires the stakeholders to collect their respective keys first and add them to every server.

\textbf{(1 A/B) Human-Computer Differences:} As discussed in \autocite{Diakopoulos.2014}, a SRA might behave differently if queried by an automatic agent. Diakopoulus therein experienced this phenomenon as results of human-computer interaction did not line up with bare API requests. He argues that in order to conduct a truly reliable study, one has to closely imitate users and simulate the usage scenario as close as possible. This has to be adapted to the respective target audience of a Crwodsourced Audit as well since demographics might have an effect on Internet and media literacy.

\textbf{(1 B/C) Research detection:} Internet-based service providers of Google's scale might have the capability to detect automatized audits. There is no such evidence but some of our VPS were blocked because of increased traffic from the respective Internet node. This proves that there are at least some mechanisms to deal with suspicious traffic. Scholars already noted the possibility of this to happen in \autocite{Datta.2015}.\\
This being said, organized computational approaches like Sock Puppet or Crowdsourced Audits that operate in very predictable patterns are easy to be identified and might see countermeasures such as captchas, traffic thresholds, IP-range blocking and adapted responses. They were all hosted by a single provider, thus it seemed like other virtual machines on the respective server already produced too much traffic. This being said, relying on third-party hardware, especially virtual machines can impede a research endeavor. Actions by other clients of the respective virtual machine can arouse suspicion. This might entail punitive measures by the researched Black Box system against the whole IP range allocated to the virtual machines of a server. 

\textbf{(1 B/C) Bot-Control:} A scraping audit like the one used in this thesis must be easily manageable. This requires centralized roll-out, administration and controlling of VPS as well as real-time information about every machine's  performance. It took an unnecessarily long time to set up the VPS due to the multitude of providers, procedures, requirements. Although the rented VPS had equal specifications, the runtime behavior of the machines differed greatly from optimal to unstable to unusable. Some would perform flawlessly, others crashed at low loads. It would have saved a lot of time and effort to order VPS services from only one provider that operates globally and serves with scalability both in size and reach. This could have greatly reduced setup times, administrative overhead and configuration efforts. Also, it would have greatly simplified logging of VPS performance.

\textbf{(1 C) Timing Intervals:} Data analysis showed that the majority of submissions by real users occurred in between the 4-hour intervals. Thus, including the data donation in the startup process of the browsers was vital for the collection. Of course, the subject of interest (stem cell treatments) may show some topical advancement over time, but it is not as time-sensitive as for example news-related political data shortly before a major election (as in \autocite{Krafft.2017}).

\textbf{(2) Collected Data:} \enquote{Raw}\footnote{meaning unfiltered} data is superior. By pre-selecting the attributes to store, the chance to re-analyze the results is missed. Thus, an evaluation from a different perspective or with an alternative research question at a later point of time is basically impossible. Also, the snapshot of the real result page is lost. On top of that, future research is hampered by this limitation. After all, we decided to publish the collected data after the study concludes.

\textbf{(3) Timeliness and obfuscation of ads:} Many ads were delivered over ad networks like Google's \textit{doubleclick} or \textit{googleadservices}. To enable performance tracking and billing, these referrer links contain uniquely identifying sequences and are often obfuscated with respect to their actual destination. Moreover, those links are only valid for a limited time. Therefore, the destinations of the links collected during the study period were not accessible for further examination at the time of the analysis.\\
The source of an advertisement was inferred only from the data that was available on the SERP, namely the respective \texttt{name} of each ad as it was denoted in the crawled HTML element. In the process of creating ads though, one is not hindered to put any arbitrary URL as a redirect destination. Theoretically, an entity other than the promoted one may have created the ad.\\
Consequently, neither their origin, nor their destination could be retrieved. Further research should consider capturing the eventual landing pages (possibly after a user interaction like a click) as well to allow a reliable association of ads and websites. Nevertheless, some links included clear text destination URLs that could be extracted and scrutinized.
	
\textbf{(3) Data Format:} The data was made available as a csv-file download. The collected data was very heterogenous with respect to symbols (some even included smileys) and special characters like commas were not escaped in the first place. Thus, the delimiter (we used a semi-colon, \enquote{;}) has to be carefully picked to correctly structure the downloaded data. Moreover, the server download function initially changed all double-quotes to single-quotes making parsing the string data to JSON impracticable.

\chapter{Conclusion}\label{ch:conclusion} 

\begin{quote}
	\enquote{I think our findings suggest that there are parts of the ad ecosystem where kinds of discrimination are beginning to emerge and there is a lack of transparency,[t]his is concerning from a societal standpoint.}~\parencite[Anupam Datta, one of the developers of AdFisher]{Simonite.2015}
\end{quote}

In this thesis, I examined the socio-technical system of web-advertising using the example of Google's integrated search engine. I developed a browser plugin to crowdsource data that was used to conduct a Black Box analysis of said system. I wanted to scrutinize whether a change in Google's advertising policy had any effect on problematic health-related ads.

The data from our collection shows that Google's policy change did not eradicate questionable stem cell advertisements on its online platform. Thus, patients of severe diseases are still being targeted by providers of unproven stem cell treatments and other questionably actors. This poses a societal risk because a vulnerable user group is being discriminated.
The second research question cannot be fully answered. Although there were no significant effects, this might be due to our small and possibly biased sample.

Besides, we learned that there are several competing actors that advertise in the realm of stem cell treatments. Those actors have distinct motivations with respect to either commercial or educational intentions. There is a constant struggle for attention between cautionary medical associations and questionable actors. The narrative of stem cell tourism as described in \cref{sec:dataeco} could be confirmed as there were multiple agents that openly advertised unapproved treatments.

Discrimination can occur through an advertiser's questionable motivation, the targeting process or the targeted audience (the eventual outcome)~\autocite{Speicher.2018}.
Due to the high complexity and interdependency of the platform, we cannot determine which of the three causes ultimately lead to this condition.
In order to sustain the web search ecosystem, it is vital to guarantee users safe interaction with advertisers' content~\autocite{Donnell.2015}.
Society and especially advertisers and intermediaries in the online advertising ecoystem need to consider users' perception of ads, including potential confusion as well as concerns regarding personalization and abuse. 

To summarize, it should be possible to scrutinize socially relevant algorithms as they have significant impacts on society. Because society decides which parts of a technical system to adopt, all involved parties have to assess technical components collaboratively to establish fair and safe communication processes. Either providers of SRE should enable examination or society should strive to analyze, evaluate and correct these systems.

\chapter{Future Work}
On a last note, we found that socially relevant algorithms like the ones deployed in Google's ISE are impossible to scrutinize from the outside. Any conventional small-scale study fails because of unobservable variables, timeliness of algorithms, interdependence of actors, Personalization and A/B-Testing of online services make it make it hard to retrieve a comparable snapshot of a system. 
Due to the opaque nature of these SRE, researchers are compelled to use Black Box analysis and demand \enquote{infrastructure and tools to study these systems at much larger scale}~\parencite[Roxana Geambasu]{Simonite.2015}. This would allow for a \enquote{widely applicable, systematic approach with a real impact}~\parencite[5]{Pedreschi.2018}.
This being said, academics concerned with the field propose two main approaches.
Along with an (possibly selectively) accessible API to test SRAs (possibly by a watchdog authority~\autocite{Zweig.2018}) it would be helpful to establish methods and infrastructures that allow for crowdsourced and publicly available data donations.

The first approach intends to probe SRAs or socio-technical systems via an interface that enables researchers to gather receive output for a specified. In our case, outputs are usually heavily personalized, so this would require computing input configurations based on the variables that the algorithm uses. As these remain undisclosed, this option falls short.
Further research may come up with software to facilitate crowdsourced data collection and standardized Black Box frameworks to scrutinize online platforms. Regulatory efforts should encourage developers of algorithms to comply with principles of algorithm accountability and foster public scrutiny~\autocite{USACM.2017}.

However, SRAs must be evaluated in the respective contexts or environments they are applied in to account for emergent effects. Thus, involving the affected social system is crucial for a sound analysis.
Thus, it has been suggested to establish trustworthy and honest \textit{Donation Brokers}~\autocite{Vaught.2012}. These could act as an intermediary between data donors and researchers. They could enable donors to determine the terms of usage with respect to time, research subject or involved parties\footnote{
	However, this would undermine the notion of a donations as gifts in the sense of \enquote{conscious, deliberate, uncoerced acts of giving, informed by beliefs about a need that is being addressed through the donation}~\parencite{Hummel.2019}. Nonetheless, it has to be made clear to users that digital data donations are subject to an uncertain future use and an unknown degree of comprehensibility through emerging methods of gathering and analysis.}.
In turn, the broker would ensure proper use and conduct as well as fair licensing~\autocite{Hummel.2019}.
Crowdsourcing data collection in a privacy preserving manner would enable society to take part in the process of algorithm accountability and support the scrutinizing of algorithmic systems that affect them.
Herein, future research could develop frameworks of transparent and reliable donation platforms were society can contribute to public scrutiny of private technical systems.

In addition, comprehensible information pertaining to data sources and algorithmic decisions can be a field of future research. Similar to the \enquote{Nutrition Label for privacy}~\autocite{Kelley.2009,Kelley.2010}, this may improve users' understanding of underlying mechanics and risks and improve technological literacy. This might increase user acceptance and reduce perceived discrimination, questionable advertisements and data privacy scandals.

Finally, interdisciplinary research might yield interesting insights in how systems can be governed. Kooiman draws a framework that helps to characterize interactions and mutual influences of interdependent systems. His model of governance could in the future be applied to STS to understand the  \textit{governability} of the systems and their respective interactions~\autocite{Kooiman.2008,Kooiman.2013}.

With respect to the EDD, there is still a lot to uncover. This thesis only provided a glimpse at the workings of the web-based SCT industry. The multivariate data that was collected, provides new perspectives on the web advertising ecosystem that evolves around stem cell treatments.
Including questions like;
\begin{itemize}
	\item Who is your go-to information source pertaining to stem cell treatments?
	\item Do you search for health-related information online?
	\item What are your concerns with respect to stem cell treatments?
	\item What are the first 3 terms that come to your mind when you think of stem cell treatments?
	\item Are you willing to try experimental therapies?
\end{itemize}
on future surveys might shed a light on the motivations of patients to search for health related information online and who they trust. It would also be interesting to learn whether some advertisers succeeded in \enquote{branding} a search term. If donors are willing to submit more data about themselves, researchers are able to deduce targeting mechanisms. They could investigate the relation between types of advertisement and medical condition or sensitive attributes like religious beliefs, risk affinity and Internet literacy. Because the evaluation of online offers of SCT is probably highly dependent on familiarity with the Internet and the health sector in general, correlations between those factors could be subject of future research.

The majority of creatives was composed from a collection of terms that are used in alternating order and combinations. It would be interesting to analyze these compositions in the future to search for patterns with respect to personalization.

Another interesting field is the to be found at the second largest host of ads in our study. The subset of data concerned with drugs can be analyzed with respect to the targeting behavior of advertisers. The peek into some of the advertisements revealed that they are equally addressing patients and practitioners. Future research could be concerned with the degree to with this targeting occurs.

As Couturier proposed above, the classification of advertisement hosts is ongoing work and needs some more scrutiny by medical professionals and people who are familiar with the field of SCT. The comparison of advertisement creatives and landing page content could reveal whether misleading lures are used to capture users' attention.

To further explore the international targeting of providers of SCT, it would be interesting to collect more detailed user information with respect to their residence and examine the regional scope of the various advertiser categories.

\shorthandoff{"}
\printbibliography
\shorthandon{"}


\appendix
\chapter{EuroStemCell Data Donation: Development}
\section{My Code}\label{app:Code}
The full code of the plugin is obtainable from \url{https://github.com/AALAB-TUKL/EuroStemCell-data-donation}.

\section{User Story}\label{app:userstory}
A patient of Parkinson's disease, Multiple Sclerosis or Diabetes perceives an information need. She wants to inform herself about the condition and the respective medical perspectives, especially in the field of stem cell-related medical applications. She decides to consult the Internet. She uses a search engine to find the most relevant website that answer her questions. Then she reviews advertisements, search results and top stories on the website to gather information and educate herself as a basis of future decisions with respect to clinical treatments and therapies.
\newpage
\section{Product Backlog}\label{app:backlog}
Below, the requirements to the plugin are listed, ordered by priority and thus, order of implementation;

\begin{enumerate}
	\item	Client-Server architecture with browser plugin dedicated to data collection and a web-server concerned with storing the data
	\item	Based on popular browsers (Firefox / Chrome), allow cross-browser implementation
	\item	Capable of crawling websites
	\item	Enable straightforward installation
	\item	Register users on the server
	\item	Receive a unique identifier from the server and attach this to submissions
	\item	Submit data to server
	\item	Enable uncomplicated on-boarding process
	\item 	Display privacy statement and obtain obligatory consent
	\item 	Include an options page to capture demographics and participant's details
	\item	Request demographics (age, gender, residence, impact of Parkinson's disease, Multiple Sclerosis and Diabetis on participant, researcher status, frequency of computer or search engine usage, experience with paid stem cell therapy, next largest city)
	\item 	Receive a study group identifier and attach this to submission
	\item 	Automate queries
	\item 	Make automated queries unobtrusive to browsing
	\item 	Enable updates of crawl specifications
	\item 	Display recent submission and informational content

\end{enumerate}

\section{Participant survey}\label{app:survey}

\begin{figure}[H]
	\centering
	\begin{subfigure}[t]{\textwidth}
		\centering
		\includegraphics[width=\linewidth]{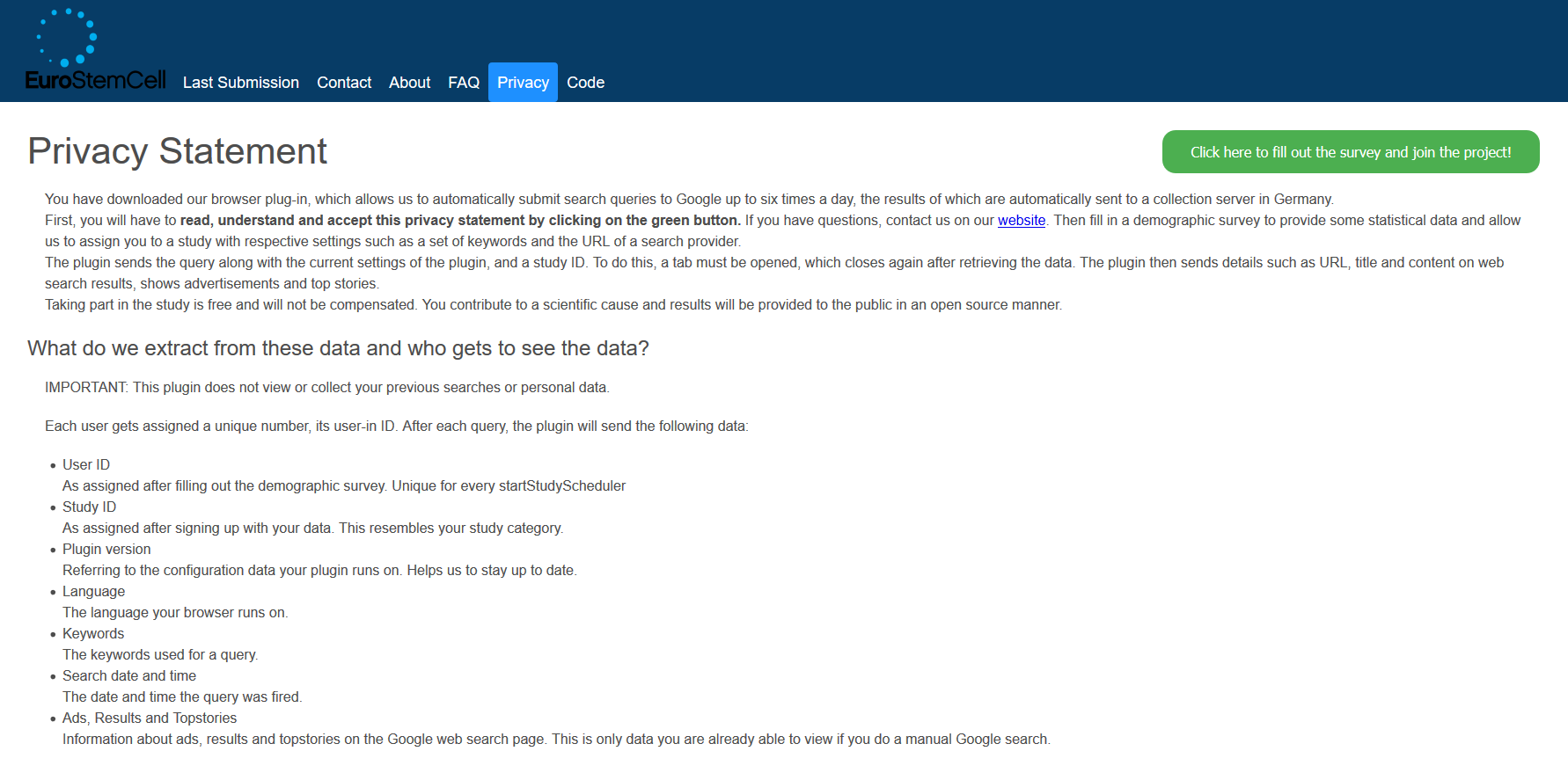}
		\caption{Privacy statement}
		\label{fig:addonprivacy}
	\end{subfigure}
	\begin{subfigure}[b]{\textwidth}
		\centering
		\includegraphics[width=\linewidth]{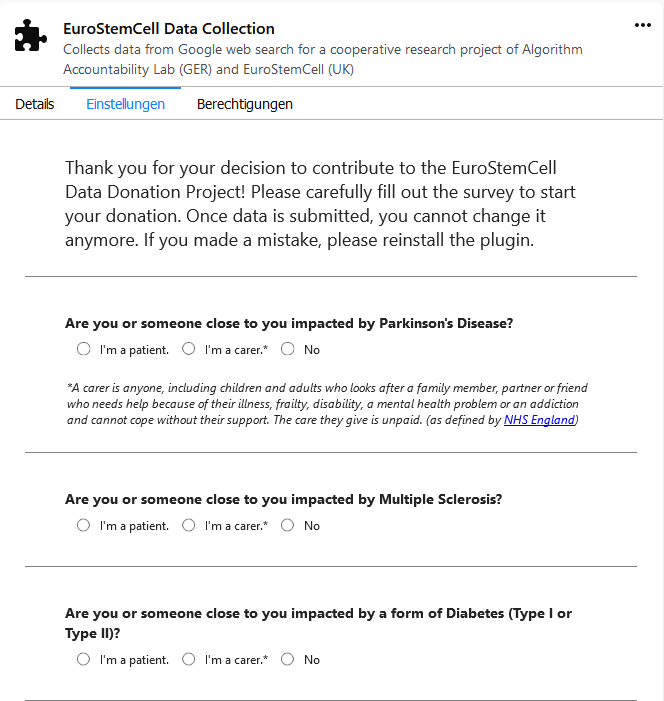}
		\caption{Segment of user survey}
		\label{fig:addonsurevey}
	\end{subfigure}
	\caption[On-bording Process]{Screenshots of the on-boarding process, by author}
\end{figure}

The survey presented in the registration process comprised the following questions and informational footnotes:

\begin{enumerate}
	\item Are you or someone close to you impacted by Parkinson's Disease?
	\begin{itemize}
		\item I'm a patient.
				\item I'm a carer.\footnote{\enquote{A carer is anyone, including children and adults who looks after a family member, partner or friend who needs help because of their illness, frailty, disability, a mental health problem or an addiction and cannot cope without their support. The care they give is unpaid.} (as defined by the NHS,  \url{https://www.england.nhs.uk/commissioning/comm-carers/carers/})}
				\item No
	\end{itemize}
	\item Are you or someone close to you impacted by Multiple Sclerosis?
		\begin{itemize}
		\item I'm a patient.
		\item I'm a carer.
		\item No
	\end{itemize}
	\item Are you or someone close to you impacted by a form of Diabetes (Type I or Type II)?
		\begin{itemize}
				\item I'm a patient.
		\item I'm a carer.
		\item No 
	\end{itemize}
	\item Are you a stem cell researcher or medical professional?
		\begin{itemize}
		\item Yes
		\item No
	\end{itemize}
	\item What is your country of residence?\footnote{Note: At this point we are only studying the impact of Google advertising in the four English speaking countries above. We will consider data from other countries to guide future research.}
		\begin{itemize}
		\item Australia 
		\item Canada
		\item United Kingdom
		\item United States Of America
		\item Other
	\end{itemize}
	\item Your age range
		\begin{itemize}
		\item 18-29
		\item 30-39
		\item 40-49
		\item 50-59
		\item 60-69
		\item 69+
	\end{itemize}
	\item Your gender
		\begin{itemize}
		\item Female
		\item Male
		\item Other
		\item Prefer not to say
	\end{itemize}
	\item How often do you use your computer, laptop, tablet and/or smartphone?
		\begin{itemize}
		\item Daily (More than 2 times a day)
		\item Daily (Less than 2 times a day)
		\item Weekly
		\item Monthly
	\end{itemize}
	\item How often do you use Google Search?
		\begin{itemize}
		\item Daily (More than 2 times a day)
		\item Daily (Less than 2 times a day)
		\item Weekly
		\item Monthly 
	\end{itemize}
	\item Have you ever paid for or inquired about stem cell treatments?\footnote{If Yes: We’d like to hear about your experience. Please contact us.}
		\begin{itemize}
		\item Yes
		\item No
	\end{itemize}
		\item What is the next largest city near you?\footnote{Please enter only letters. If you feel uncomfortable answering this, please choose "Prefer not to say".}
		\begin{itemize}
		\item City: \texttt{textfield}
		\item Prefer not to say
	\end{itemize}

\end{enumerate}

\section{Query composition and crawled HTML elements}
\subsection{Query composition}\label{app:comp}
The following search terms were composed at the project's kick-off meeting. They were meant to formulate popular queries with respect to the field we examined. Thus, we included keywords like \textit{stem cell}, the names of the respective diseases (\textit{parkinsons disease}, \textit{multiple sclerosis}, \textit{diabetes}, denoted by \texttt{disesase} here). Also we included natural language questions as we assumed searchers to query search engines with direct questions if they are not Internet literate in a sense that they understand search engines capabilities and mechanics.
	\begin{itemize}
		\item stem cells
		\item stem cells cost
		\item stem cells treatment
		\item stem cells cure
		\item stem cells therapy
		\item can stem cells help me?
		\item can stem cells cure \texttt{[disease]}?
		\item \texttt{[disease]} cure
		\item \texttt{[disease]} therapy
		\item \texttt{[disease]} treatment
		\item \texttt{[disease]} cells cost
		\item \texttt{[disease]} stem cells treatment
		\item \texttt{[disease]} stem cells cure
		\item \texttt{[disease]} stem cells therapy		
	\end{itemize}

\subsection{Crawled HTML elements}\label{app:crawl}
	\begin{itemize}
		\item 	Ads
			\begin{itemize}
				\item Name
				\item Title
				\item URL
				\item Content
			\end{itemize}
		\item 	Search results
		\begin{itemize}
			\item	Title
			\item	Content
			\item	URL
			\item	Position
		\end{itemize}
		\item	Top Stories
			\begin{itemize}
				\item	Title
				\item	Author
				\item	URL
				\item	Position
			\end{itemize}
	\end{itemize}
\begin{figure}
	\centering
	\includegraphics[width=0.7\linewidth]{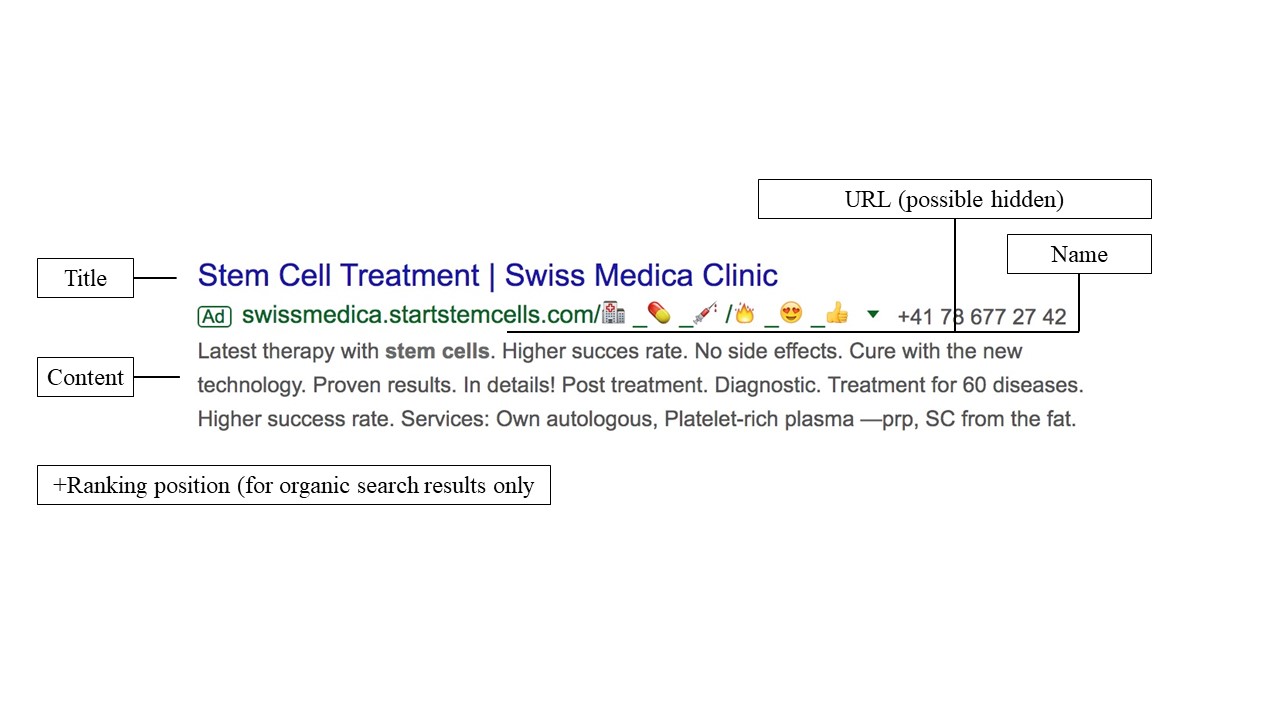}
	\caption[Description of crawled elements]{Detailed example of crawled elements (here: ad and organic results), Screenshot by author}
	\label{fig:crawlelements}
\end{figure}

\chapter{EuroStemCell Data Donation: Data Analysis and Visualizations}
\section{Downloads}\label{app:download_stats}
\begin{figure}[H]
	\centering
	\includegraphics[width=\linewidth]{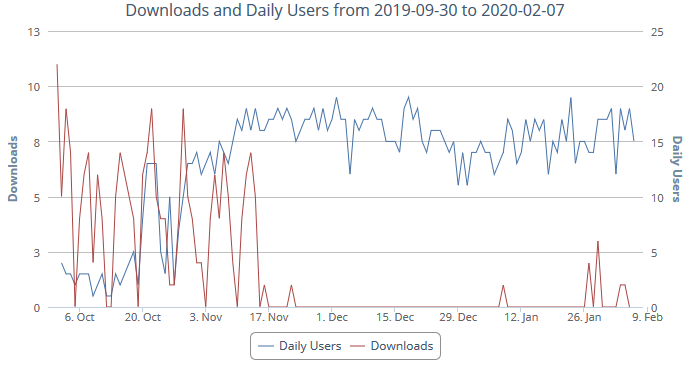}
	\caption[Firefox plugin statistics]{Daily users and downloads of the Firefox addon as documented on the Mozilla Developer Hub statistics, screenshot by author}
	\label{fig:statisticsffdownloadsusersovertime}
\end{figure}

\begin{figure}[H]
	\centering
	\includegraphics[width=\linewidth]{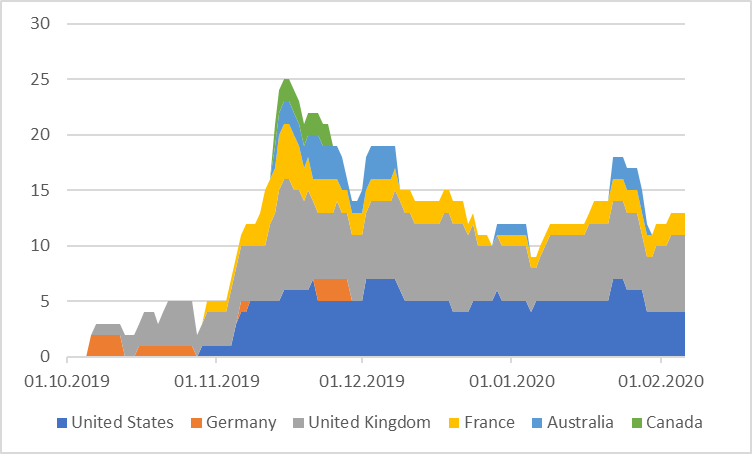}
	\caption[Daily chrome users]{Daily Users from 2019-10-01 to 2020-02-07}
	\label{fig:statisticschromeweekly}
\end{figure}

\begin{figure}[H]
	\centering
	\includegraphics[width=\linewidth]{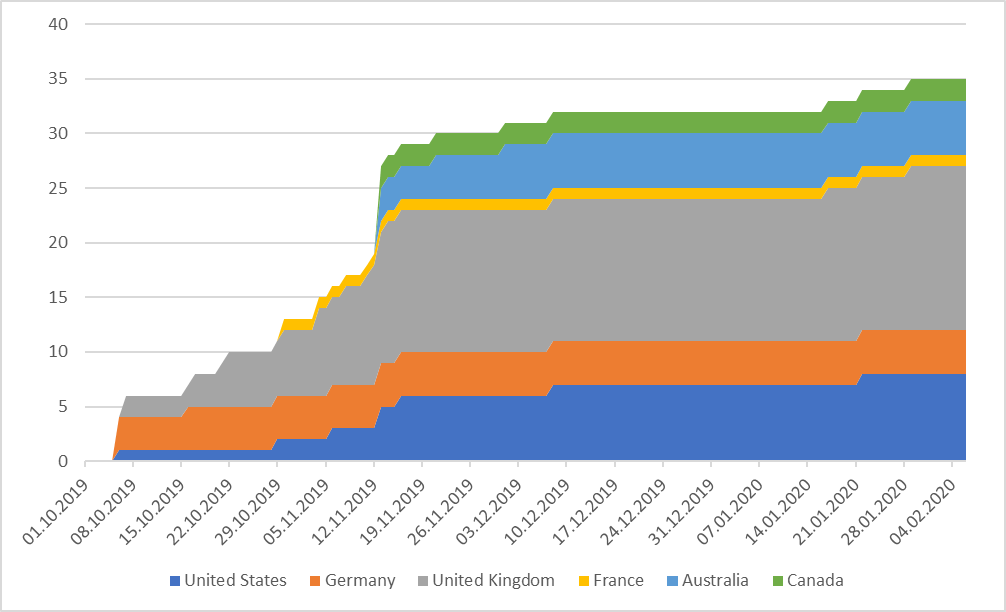}
	\caption[Cumulative Chrome registrations]{Cumulative registrations via the Chrome plugin from 1.10.2019 to 7.2.2020}
	\label{fig:statisticschromeregs}
\end{figure}

\section{Participants}\label{sec:participants}

\begin{figure}[H]
	\centering
	\includegraphics[width=\linewidth]{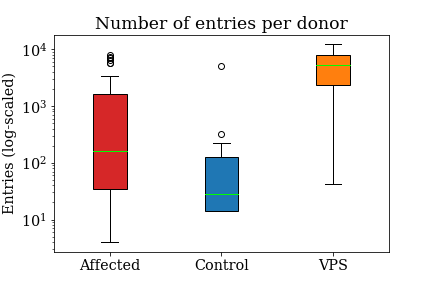}
	\caption{Donations by group}
	\label{fig:entrycountper-group}
\end{figure}

\section{Advertisements and Advertisers}

\begin{figure}[H]
	\centering
	\includegraphics[width=0.7\linewidth]{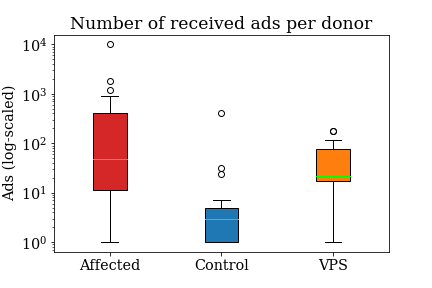}
	\caption{Absolute Number of advertisements per group}
	\label{fig:adcountpergroup2}
\end{figure}

\begin{figure}[H]
	\centering
	\includegraphics[width=0.7\linewidth]{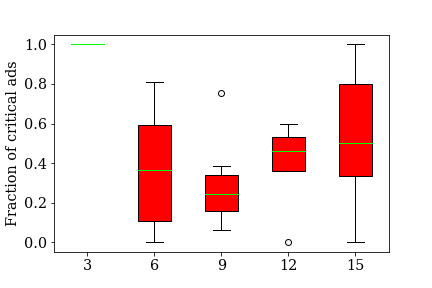}
	\caption{Fraction of Prescription Treatment Advertisements per group}
	\label{fig:probpresccountpergroup}
\end{figure}

\chapter{Functionality of a Search Engine}\label{app:functsearch}
At first, the original mechanism of Google will be portrayed by reference to \autocite{Brin.1999}, the initial paper of the Google founders and the company blog at \autocite{Google.2019f}. Then, these insights will be enriched with observations of search engine researchers, tech observers and industrial professionals. Later on, patents provide a possible outlook.

Web search engine operate in a three-stepped process of crawling the WWW, indexing web pages and serving results\footnote{For more details, see \cref{app:functsearch}}. Additionally, they generally display advertisements along with organic search results to fund their operations.
Search engines developed from merely using on-page data, link-analysis and other web-specific data (anchor text, e.g.)~\autocite{Brin.1999} to leveraging manifold sources to determine a searcher's intentions.
The factors that contribute to a ranking are unknown to the public. Google only gives implicit advice on how to design websites and what they think is \enquote{high quality} that leads to an appropriate ranking with respect to a user query~\autocite{Google.2019u,Google.2020d}.
Due to this publishers might anxiously avoid anything that could possibly be a black hat SEO technique, scholars criticize~\autocite{Pasquale.2008}.

While most search engine operate on well-known principles, the specific details of their algorithms remain undisclosed trade secrets, mainly to sustain search quality and remain competitive \autocite{Granka.2010}.
The fundamental tasks of a search engine are as follows:
\begin{description}
	\item[crawling] Google uses a web crawler\footnote{Definition Crawler: \enquote{Automated software that crawls (fetches) pages from the web and indexes them.}\autocite{Google.2019f}} (or robot / spider) that operates from many computers and collects publicly available web pages on the storeserver. The algorithm behind it receives a list of URLs\footnote{Uniform Resource Locator, or Internet address, see https://en.wikipedia.org/wiki/URL} of prior crawls from the storeserver and sitemap data. Then, crawlers collect those website, send them to the storeserver and follow links recursively. Eventually, newly created pages, changes and deletions are added to the index. It does not crawl blocked\footnote{Websites will not be crawled if a file named \textit{robots.txt} is located on the host. Through inbound hyperlinks it might still be indexed, though.\autocite{Google.2019g}} website, restricted areas and sites that are already known\footnote{\enquote{Pages that have already been crawled and are considered duplicates of another page, are crawled less frequently.}\autocite{Google.2019g}}\autocite{Google.2019g}.

	\item[indexing] The indexer parses the pages it receives from the storeserver's repository and creates an index. All significant words and their position on a website, key content tags and attributes are stored. 
	According to an in-memory has table (the \textit{lexicon}) the words are transformed into word IDs. Their occurrences on a website are recorded on a hit list that is stored in \textit{barrels} sorted  by document ID. Then, the content of the barrels is used to create an inverted index. Additionally, the indexer derives a database of linked documents from anchor files to assess meaning of linked content (web pages and media)\autocite{Brin.1999}. If a page is inaccessible due to a \textit{robots.txt} file, authorization measures or another device, it is not indexed\autocite{Google.2019g}. According to Google, the index \enquote{contains hundreds of billions of webpages and is well over 100,000,000 gigabytes in size}\autocite{Google.2019e}. Observers estimate that Google only indexes a marginal part of the WWW\footnote{Grey literature: The estimates range from 0.004\% to 4\%, depending on source. What they call \textit{Deep Web} consists of website without inbound links, password protected areas, databases that only respond to certain input, subscription services.\autocite{Rosen.2014}}
	Grimmelmann distinguishes \textit{general} from \textit{vertical} search engines. While general index the whole web, vertical ones specialize in a particular category (like news, travel, shopping, e.g.). Over the years of its development, Google has added vertical search capabilities to its existing general search.\autocite{Grimmelmann.2013b}
	
	\item[serving] Upon a user query, words from the parsed query are converted to word IDs and searched for in the barrels. For the documents that include those words, a weighed rank is computed based on a multitude of parameters. Today, Google uses an unknown number of signals or variables to determine relevance\footnote{Brin and Page state in their initial paper how \enquote{[F]iguring out the right values for these parameters is something of a black art}\autocite{Brin.1999}.}. Linguistic cues (website content), user cues (feedback loop) and web structure (Page Rank) all contribute to a final score \autocite{Granka.2010}. The relevance is assessed using an unknown amount of \textit{signals} and factors ranging from context variables (location, time, current situation) to semantic information of the search query all the way to very personalized factors (search history, profiling)\autocite{Google.2019i}.
	In 2010, they used to amount to about 200 \autocite{Google.2010}. 
	Google itself provides assistance to Search Engine Optimization (SEO) and qualitatively describes how publishers should design their websites in order to receive an accurate ranking without penalties. From this advice, one could infer the nature of signals contributing to the measurement like in ~\autocite{Google.2019f}.  The $k$ highest ranked results are presented to the user in descending order of relevance.\autocite{Brin.1999}
	Today, hundreds of \textit{signals} count towards the rank calculation\autocite{Sullivan.2016b}, links, content and \textit{RankBrain} being the most significant ones\autocite{Schwartz.2016}.
	
\end{description}

Originally, Google used an algorithm called \textit{Page Rank} to assess the importance of a website by the web's link structure. The creators intended to compute the measure in accordance with people's subjective idea of importance. They argued that a source which received many citations is probably credible, important or relevant. The more important those referrers are, the higher the PageRank of the respective site. Hence, they used the normalized links of other pages that direct users to a particular website to caclulate the measurement iteratively. A damping factor was included to simulate a surfer that randomly jumps to a different website to avoid dead ends. Then, search results were prioritized based on their respective weight. 
Along with link structure, the link text was considered in assessing a page's relevance. The authors argue that it usually describes the webpage it points to more accurately than the page it is located on. On top of that, websites without text content can thus be crawled (media, databases or other non-textual objects).\autocite{Brin.1999}

\begin{equation}\label{equ:pr}
PR(A)=(1-d) + d \Bigl( \frac{PR(T_1)}{C(T_1)}+\cdots+\frac{PR(T_n)}{C(T_n)} \Bigr)
\end{equation}

The PageRank algorithm above calculates the PageRank $PR(A)$ of website $A$ by summing up the PageRanks of websites pointed at it, normalized by the respective sites total number of outgoing links. The damping factor $d$ is used to allow for personalization. This iterative algorithm computes a probability distribution over all websites, so 
$\sigma_i=1^n PR(A_i)$~\autocite{Brin.1999}.

\textbf{Disclaimer:} Below, some prominent features and most recent developments are discussed by observers\footnote{The discussions are mostly based on the personal opinions by authors of \textit{SEO by the sea} (\url{http://www.seobythesea.com/}), especially Bill Slawski} and \textit{Search Engine Land}\footnote{\url{https://searchengineland.com/}, especially Danny Sullivan}. Note that a patented feature is not necessarily part of the actual search algorithm. For most of the patents reviewed, there is no evidence of their clear implementation. However, it can \enquote{offer an interesting perspective on where [Google] is steering search and how it’s thinking about evolution of search.}\autocite{Nguyen.2019} Also, the observations and argumentations below are documented by industry professionals outside of the Google universe. They stem from original interviews, research, experience and conferences and are published on their website. Thus, they do not ensure that Google uses these technologies.\\

The search engine enhances queries using query expansion\autocite{Smarty.2008}\footnote{Gray literature: Techniques include word stemming, acronyms, synonyms, translations, spelling corrections and removal of stop words.}. This allows to broaden the search horizon and rely less on a user's distinct input.

In 2008, Google introduced auto suggestion, a feature that showed numeral possible text completions when users started to type their query\autocite{Sullivan.2008}.

The introduction of the \textit{Knowledge Graph}\autocite{Henry.3.8.2012}\footnote{
	Critics argue that the concept of a knowledge graph (KG) is not properly defined yet. Research work dealing with KGs cite Google's blog even though it does not explain what constitutes a KG. In \autocite{Ehrlinger.2016}, Ehrlinger and Wöß criticize the wide variety of interpretations of the concept. They propose to define KG as follows: \enquote{A knowledge graph acquires and integrates information into an ontology and applies a reasoner to derive new knowledge.}
	\autocite{Ehrlinger.2016}}
indicates a paradigm shift from things to strings, as the Official Google Blog puts it\autocite{Singhal.2012}. Now, search on their platform is no longer about connecting keywords, but finding semantically correct results. The development was kickstarted through acquisition of Metaweb, a company maintaining \enquote{an open database of things in the world}\autocite{Menzel.2010}.
Pages will not only be indexed with respect to keywords but they will also be crawled for entities their attributes, classes and relationships between them to create  ontologies~\autocite{Semturs.6.6.2015}.
Google files numerous patents to bridge the \enquote{semantic gap}\autocite{GoogleAIBlog.2013} and further develop its KG\autocite{Pasca.16.6.2010}\autocite{Gubin.12.5.2014}\autocite{Gupta.15.3.2013} up the point where it grows and matures self-sufficiently from query input \autocite{Halevy.28.10.2014}\autocite{Pasca.2.11.2012} and understands conversational queries\autocite{Sullivan.2013}\autocite{Slawski.2018}. 

The KG displays information from different sources in an infobox next to the search results to enrich the search experience through contextual and diverse information\autocite{Singhal.2012}. Google possibly adapts this method to the individual users' background\autocite{Balog.2019} and enriches it with signals from their social network\autocite{Wu.7.5.2013}. Then, it may order result based on a user profile \autocite{Zamir.13.7.2004}.
Google expands their concept of finding \enquote{things} in \autocite{Huynh.12.12.12} where they discusses how entity metrics can be used to rank results in a way that considers semantics and context\footnote{Examples from \autocite{Starr.2015} include \textit{relatedness} (co-occurence of entities),\textit{notable entity type} (multi-categorization of entities),\textit{contribution} (content generated by an entity, like social media posts or published works) and \textit{prize} (awards and prizes)}.

In \autocite{Lu.2019}, for example question-answer-relationships on Q\&A-websites are identified as well as how question concerning these relationships can be parsed.
They filed numerous knowledge-oriented patents that tried to grasp a user's individual context and understand semantic relationships.

In \autocite{Starr.2015}, Starr refers to \autocite{Huynh.12.12.12} and points out how different regions on a search results page may be computed by different kinds of algorithms. She implies that \enquote{different algorithms apply at different times}\autocite{Starr.2015} and results might be of mixed origin to allow optimal presentation and information to users.

Another leap in Google's search engine design was the introduction of a a Natural Language Processor \textit{RankBrain} in 2015\autocite{Schachinger.2017}. According to \autocite{Clark.2015}, it is an AI-driven addition to the algorithm affecting a large fraction of searches that are ambiguous in their meaning or have never been asked before\footnote{The latter amounting to 15\% of all searches.\autocite{Farber.2013}}.
If a query cannot be confidently answered, RankBrain tries to guess the searcher's intentions. Its goal is to come up with a sufficiently good answer by inferring associations from the input and then find similarities to queries in the past\autocite{Sullivan.2016b}\autocite{Clark.2015}. This guesswork is made possible by the semantic network of entities and their attributes and relationships mentioned above\autocite{Schachinger.2017}.

Recent advances include predictive computing that tries to guess user intent and guide them through search and decision processes.
Davies reviews two patents (\autocite{Peddinti.15.9.2015}\autocite{Foerster.21.9.2015})that support that development in \autocite{Davies.2017} and \autocite{Davies.2017b}. The patents describe how a search engine can include various information to infer future behavior or intent. Davies points out, that \enquote{[B]asically, the patent is built on the idea that all data from virtually any source can be used to determine expected actions a user is likely to take.}\autocite{Davies.2017}\footnote{
	According to Davies, this includes social media, motion, purchase history, weather, network account data, data from third-party applications and services all sorts of communication processed on the device.\autocite{Davies.2017}}
With this knowledge, the patented system tries to estimate future behavior and indicate to the user if an action is jeopardizing the expected outcome\autocite{Peddinti.15.9.2015}.
The second patent allows to inject suggestive steps into the purchasing process and enables highly targeted bidding on advertisements\autocite{Foerster.21.9.2015}.
Davies raises awareness to how these two inventions have massive impact on search behavior, advertisement bidding, purchasing processes.
Consequently, this allows nudging the user in a third party's interest which is critical in terms of the practices introduced in \cref{sec:digihealth}.

Google assesses search quality with feedback from third-party services and users \autocite{Google.2019k}\autocite{Levy.2010}. Additionally they are constantly testing and reviewing algorithm prototypes through A/B testing\footnote{\enquote{Every time engineers want to test a tweak, they run the new algorithm on a tiny percentage of random users, letting the rest of the site’s searchers serve as a massive control group.}\autocite{Levy.2010}} \autocite{Levy.2010}.

\typeout{get arXiv to do 4 passes: Label(s) may have changed. Rerun}

\end{document}